\documentclass[fleqn,usenatbib]{mnras}

\usepackage{newtxtext,newtxmath}
\usepackage{lscape}
\usepackage{float}
\usepackage{subcaption}
\usepackage{multicol}
\usepackage{multirow}
\usepackage[T1]{fontenc}

\DeclareRobustCommand{\VAN}[3]{#2}
\let\VANthebibliography\thebibliography
\def\thebibliography{\DeclareRobustCommand{\VAN}[3]{##3}\VANthebibliography}

\usepackage{graphicx}	% Including figure files
\usepackage{amsmath}	% Advanced maths commands
\title[Library of sMILES SSPs With $\alpha$ Abundance Variations]{sMILES SSPs: A Library of Semi-Empirical MILES Stellar Population Models with Variable [$\alpha$/Fe] Abundances}
\author[A. T. Knowles et al.]{
Adam T. Knowles$^{1,2,3}$\thanks{E-mail: adam.knowles@iac.es}, A. E. Sansom$^{3}$, A.Vazdekis$^{1,2}$, C. Allende Prieto$^{1,2}$
\\
$^{1}$Instituto de Astrof\'isica de Canarias, V\'ia L\'actea, 38205 La Laguna, Tenerife, Spain\\
$^{2}$Universidad de La Laguna, Departamento de Astrof\'isica, 38206 La Laguna, Tenerife, Spain\\
$^{3}$Jeremiah Horrocks Institute, School of Natural Sciences, University of Central Lancashire, Preston, PR1 2HE, UK
}

\date{Accepted XXX. Received YYY; in original form ZZZ}

\pubyear{2023}

\defcitealias{LaBarbera13}{LB13} % AES added alias 22/7/22
\defcitealias{Vaz2015}{V+15}

\begin{document}
\label{firstpage}
\pagerange{\pageref{firstpage}--\pageref{lastpage}}
\maketitle

\begin{abstract}
We present a new library of semi-empirical stellar population models that are based on the empirical MILES and semi-empirical sMILES stellar libraries. The models span a large range of age and metallicity, in addition to an [$\alpha$/Fe] coverage from $-$0.2 to $+$0.6 dex, at MILES resolution (FWHM=$2.5\,${\AA}) and wavelength coverage ($3540.5-7409.6\,${\AA}). These models are aimed at exploring abundance ratios in the integrated light from stellar populations in star clusters and galaxies. Our approach is to build SSPs from semi-empirical stars at particular [$\alpha$/Fe] values, thus producing new SSPs at a range of [$\alpha$/Fe] values from sub-solar to super-solar. We compare these new SSPs with previously published and well-used models and find similar abundance pattern predictions, but with some differences in age indicators. We illustrate a potential application of our new SSPs, by fitting them to the high signal-to-noise data of stacked SDSS galaxy spectra. Age, metallicity and [$\alpha$/Fe] trends were measured for galaxy stacks with different stellar velocity dispersions and show systematic changes, in agreement with previous analyses of subsets of those data. These new SSPs are made publicly available.
\end{abstract}


%%%%%%%%%%%%%%%%%%%%%%%%%%%%%%%%%%%%%%%%%%%%%%%%%%

%%%%%%%%%%%%%%%%% BODY OF PAPER %%%%%%%%%%%%%%%%%%
\section{Introduction}
A powerful method in the analysis of the unresolved stellar content of galaxies is the use of stellar population models. Through matches of such models to observed galaxy spectra, fundamental properties such as population age, metallicity, initial mass function (IMF) and abundance patterns can be estimated; properties that hold key information in understanding the formation history of the host galaxy. Fitting of spectral indices or full spectral predictions of stellar population models to unresolved populations in external galaxies are common place and have been for some time for various applications (e.g. \citealt{Bruzual83}; \citealt{Worthey94}; \citealt{Vaz2010}). In particular, the elemental abundance patterns of galaxies provide good indicators of the time-scales in which their constituent stellar populations were formed and can be used to constrain models of galaxy formation. A historical example of this is the well-known measurement of the overabundance of [Mg/Fe]\footnote{[A/B]=$\log[{n(A)/n(B)}]_{*}$ - $\log[{n(A)/n(B)}]_{\odot}$, where $n(X)/n(Y)$ is the number density ratio of element A, relative to element B.} compared to the solar neighbourhood observed in early-type galaxies (ETGs); a property that is usually attributed to short star formation time-scales (e.g. see the review of \citealt{Trager98} and references therein).

The computation of a Single Stellar Population (SSP) model, defined as a model of a stellar population with a single age, metallicity and abundance pattern, requires an input stellar library; a collection of stellar spectra; to translate the evolutionary predictions of how stars age into predictable observables of indices or full spectra. SSP models have been generated with stellar libraries that have been fully theoretical (e.g. \citealt{Maraston05}; \citealt{Coelho07} or empirical (e.g. \citealt{Vaz99Model,Vaz2010}).

With increasingly powerful and wide-field spectroscopic surveys and instrumentation (e.g. WEAVE \citealt{Dalton12}, 4MOST \citealt{deJong19} and X-Shooter \citealt{Vernet11}), it is becoming possible to generate ever improving SSP models based on high-quality empirical libraries that cover large portions of the Hertzsprung–Russell diagram and large wavelength ranges. Recent works that have incorporated large empirical stellar libraries include the E-MILES (\citealt{Rock16}; \citealt{Vazdekis16}), E-IRTF (\citealt{Conroy18}), MaStar (\citealt{Maraston20}) and X-Shooter (\citealt{Verro21}) models. However, a fundamental shortcoming of these models is the limited parameter space coverage in terms of abundance patterns;  an unavoidable limitation because spectra are typically taken from stars in the vicinity of the solar neighbourhood and therefore represent the chemical evolution of the Milky Way. Small samples of bright star spectra with differing chemical compositions can be obtained from other galaxies, but long exposure times limit the number of these observations.

With this limitation in mind, the use of theoretical stellar predictions is required to build SSP models with abundance ratios that differ from the typical solar neighbourhood pattern. An approach to account for non-solar abundances in SSP models is to use differential predictions of theoretical spectra in unison with empirical spectra on either the star or SSP level. Using only differential predictions from theoretical spectra has been shown to reproduce observations of abundance pattern effects more accurately than fully theoretical spectra, particularly for wavelengths below the $\textrm{Mg}_{\textrm{b}}$ index (e.g see figure 11 of \citealt{Knowles19} or \citealt{Martins07}; \citealt{Bertone08}; \citealt{Coelho2014}; \citealt{Villaume17}). \cite{Coelho20} produce an thorough analysis of the impact of using theoretical or empirical stellar spectra in the generation of stellar population models. Using a quantification of how stars are affected by changes in atmospheric abundances, it is possible to modify empirical spectra to generate models with different element compositions. These modifications, known as differential corrections, were originally performed on individual spectral line indices (e.g. \citealt{Tripicco1995}; \citealt{Thomas03Mod}; \citealt{Korn2005}), but have since been done for full spectral predictions (\citealt{Coelho07}; \citealt{Prugniel07}; \citealt{Cervantes07}; \citealt{Walcher09}; \citealt{Conroy2012a}; \citealt{Vaz2015} - hereafter \citetalias{Vaz2015}; \citealt{LaBarbera17}; \citealt{Conroy18}). Recent modelling has also started to consider abundance effects on isochrones more widely (\citealt{Worthey22}).

Abundance pattern predictions (differential corrections) can be applied at different levels of the SSP calculation. SSP models generated using empirical stellar spectra can be corrected from predictions of SSP models generated using theoretical stellar spectra. Using this approach, for example, \citetalias{Vaz2015} computed SSPs for variations in [$\alpha$/Fe]. Alternatively, abundance pattern predictions from theoretical stellar spectra can be applied to individual stellar spectra to create semi-empirical stars, that are then incorporated into the SSP calculation (e.g. \citealt{LaBarbera17} for [Na/Fe] variations).

In this work we compute a new library of semi-empirical SSP models based on the publicly available\footnote{\url{http://miles.iac.es/}} semi-empirical MILES (sMILES) spectral library (\citealt{Knowles21}). Using families of sMILES stellar spectra with [$\alpha$/Fe] values we generate sMILES SSP models for different [$\alpha$/Fe] abundances over a larger range and at finer sampling than previously computed SSP models (e.g. \citetalias{Vaz2015}), for a number of population ages and metallicities with different IMF prescriptions. The range of [$\alpha$/Fe] is chosen to better match the abundance pattern measurements of various extragalactic environments (e.g. see \citealt{WortheyTangServen2014}; \citealt{Sen2018}). We make the sMILES SSP models available for public use.

The structure for this paper is as follows. Section~\ref{sec:SSPCalc} describes the generation and calculation of the sMILES SSP models. Section~\ref{sec:sMILESTests} presents then tests these new models through comparisons to other published models and observations. Section~\ref{sec:SSPEtypeGalaxies} shows an application of the sMILES SSPs to external galaxy data and demonstrates the use of [$\alpha$/Fe] variations available for future work. Section~\ref{sec:SummaryConclusions} presents our summary and conclusions.

\section{Building Stellar Population Models}
\label{sec:SSPCalc}
With sMILES stars generated in \cite{Knowles21}, we now incorporate them into new SSPs, building on the previous methods of \cite{LaBarbera17}. Using semi-empirical stars, SSPs are computed with varying [$\alpha/$Fe] abundances, for a range of ages and metallicities.

Section~\ref{SSP_Calc} details the SSP calculations, including a description of parameter conversions that allow for translation of the stellar library component into locations on pre-computed isochrones.

\subsection{SSP calculation}
\label{SSP_Calc}
For the calculation of SSP spectra, we follow the general methodology of \citetalias{Vaz2015}, using families of sMILES stars to compute SSP spectra of varying [$\alpha$/Fe] abundances. The difference in methodology between sMILES SSPs here and those of \citetalias{Vaz2015} is that the differential corrections are performed on individual MILES star spectra, rather than on MILES empirical SSP spectra. SSPs are computed for the same range of [$\alpha$/Fe] as the sMILES stars, from $-$0.2 to $+$0.6 dex in steps of 0.2 dex. The availability of [$\alpha$/Fe] estimates for MILES stars (from \citealt{Milone2011}) allowed for differentially correcting their spectra to other [$\alpha$/Fe] values, and hence to compile consistent SSPs spectra at different [$\alpha$/Fe] values, taking the $\alpha$-elements as a group. This is a different approach than others who have modelled individual elements, but for Lick indices rather than full spectra (e.g. \citealt{Johansson12}) or varying individual elements one-by-one, relative to an assumed element abundance (e.g. \citealt{Worthey2011}), or fitting spectra but still treating individual elements as trace element changes relative to an assumed abundance pattern (e.g. \citealt{Conroy2012a}), rather than the more self-consistent approach taken here, where more is known about the base stars. We discuss the individual components of the SSPs and then describe the calculation.

\subsubsection{IMF}
Several IMF parameterisations can be considered in the computation of SSPs, with recent applications of published models including the investigation of IMF variations within early-type galaxies (e.g. \citealt{LaBarbera16,LaBarbera17,LaBarbera21}). We compute models with five IMF variations, starting with the multipart power-law universal and revised Kroupa IMFs, described in \cite{Kroupa01}. The revised version, which removed estimated effects of unresolved binary stars, adopts $\alpha_{1}$ and $\alpha_{2}$ values of 1.8 and 2.7, respectively, compared to the 1.3 and 2.3 values of the universal Kroupa IMF, from equations 1 and 2 in \cite{Kroupa01}. We provide SSPs described by a \cite{Chabrier03} IMF, with a massive star segment logarithmic slope of 1.3.

We also compute SSPs using the unimodal and bimodal IMF described in \cite{Vaz96} and in appendix A of \cite{Vaz03}, parameterised by logarithmic slopes $\Gamma$ and $\Gamma_\textrm{b}$, respectively. A bimodal IMF of $\Gamma_\textrm{b}$=1.3 is close to the universal Kroupa IMF. We compute SSPs for thirteen values of $\Gamma$ and $\Gamma_\textrm{b}$, ranging from 0.3 to 3.5; the same range and values as those provided previously in \citetalias{Vaz2015}\footnote{\url{http://research.iac.es/proyecto/miles/pages/ssp-models.php}}. We set the lower and upper mass cutoffs at 0.1 and 100 $\textrm{M}_\odot$, respectively.

\subsubsection{Isochrones}
We adopt two sets of theoretical isochrones in the SSP calculations. For the [$\alpha$/Fe]=$-$0.20, 0.0 and $+$0.20 SSPs we adopt the scaled-solar isochrones from \cite{Pietrinferni04} and for the [$\alpha$/Fe]=$+$0.40 and $+$0.60 SSPs we use the $\alpha$-enhanced isochrones from \cite{Pietrinferni06}. We refer to these sets of isochrones as the BaSTI models. The $\alpha$-enhanced isochrones are computed at [$\alpha$/Fe]=0.40. Both sets of isochrones, and therefore the resulting sMILES SSPs, are computed for 53 different ages in the range 0.03-14 Gyr, with the coverage given in Table~\ref{sMILES_SSPmodels_Tab}. Total metallicities, defined on the \cite{Grevesse93} solar abundance scale, were computed for 10 steps in total metal mass fraction (Z) for Z=0.0003, 0.0006, 0.0010, 0.0020, 0.0040, 0.0080, 0.0100, 0.0198, 0.0240, 0.0300. On this scale, the solar metallicity at birth is given as Z$_\odot$=0.0198. We note that although this solar abundance reference is deemed obsolete by the original authors (\citealt{Grevesse13}), we are tied to this scale because the isochrones we implement are calculated adopting this value. The BaSTI models include a consistent prescription for atomic diffusion of helium and metals in the solar metallicity models, in order to match helioseismological constraints of the depth of the convective envelope, the present helium abundance of the solar envelope and current (Z/X) ratio. These isochrones have been constrained by various observations, such as eclipsing binaries, cluster colour-magnitude diagrams and unresolved stellar populations (\citealt{Pietrinferni04}; \citealt{Percival09}). We use the isochrones that include convective overshooting with a mass loss rate given by $\eta=0.4$. $\eta$ is the free parameter in Reimers law (\citealt{Reimers75}), describing the mass loss of a star depending on its luminosity, surface gravity and radius. The value of 0.4 is a commonly used value as this provides good matches to observations of horizontal branch colours in globular clusters. The thermally-pulsing asymptotic giant branch is included in the isochrones, through models described in \cite{Marigo96} based on methods from \cite{Iben78}. We acknowledge that there are updated versions of these isochrones (\citealt{Hidalgo2018}; \citealt{Pietrinferni21}), however implementation, testing and comparisons of an updated set of isochrones is out of the scope of this current work. The aim of this work is to provide a set of models that have larger range of varying abundance ratios, but are based on well-established models of \citetalias{Vaz2015}. Details of these techniques and extensive tests of the isochrones used here are described in \citetalias{Vaz2015} and \cite{Pietrinferni04,Pietrinferni06,Pietrinferni09,Pietrinferni13}.

\subsubsection{Stellar Spectral Library}
 The sMILES spectral library is based on the widely-used Medium-resolution Isaac Newton Telescope Library of Empirical Spectra (MILES) (\citealt{SanchezBlazquez2006}; \citealt{Falcon2011}) with differential corrections made from predictions of ATLAS9 (\citealt{Kurucz1993}) model atmospheres, opacity distribution functions presented in \cite{Mezaros2012} and ASS$\epsilon$T (\citealt{Koesterke2009}) radiative transfer. The empirical spectra have good signal-to-noise that is typically above 100 and were carefully flux calibrated (\citealt{Falcon2011}). Details of sMILES stellar spectra generation and the underlying theoretical stellar spectra are described in \cite{Knowles21}. Both sMILES and MILES are used in the SSP calculations.
The stellar parameters of effective temperature ($\textrm{T}_\textrm{eff}$), surface gravity (log g), metallicity ([Fe/H]) and [Mg/Fe] values adopted were those of \cite{Cenarro07} and \cite{Milone2011}. In \cite{Knowles21} we used [Mg/Fe] as a proxy for [$\alpha$/Fe] in the MILES stars. For 75 stars without [Mg/Fe] estimates, we made approximate estimates ([Mg/Fe] values of 0.0, 0.2 or 0.4) using measurements from both \cite{Milone2011} (their figure 10) and a Milky Way pattern based on \cite{Bensby2014} (their figure 15).
A subsample of empirical MILES stars, that were found not to be representative of their tagged stellar parameters, were removed prior to the SSP calculation. These inspections are described and presented in sections 2.2 of \cite{Vaz2010} and 2.3.1 of \citetalias{Vaz2015} and resulted in a final empirical MILES library of 925 stars.

The sMILES stellar library was created through differential corrections to the 925 empirical MILES spectra mentioned above. These differential corrections were calculated, using interpolated fully theoretical stellar spectra, and applied to empirical MILES stars through equations 7 and 8 of \cite{Knowles21}, respectively, to produce semi-empirical stellar spectra. This final sMILES library consists of families of 801 spectra for five [$\alpha$/Fe] abundances of $-$0.2, 0.0, $+$0.2, $+$0.4 and $+$0.6 dex. In the stellar models of \cite{Knowles21} and therefore the resulting sMILES stars and SSPs we vary O, Ne, Mg, Si, S, Ca and Ti as the $\alpha$-elements in lock-step, to be consistent with the underlying stellar atmospheres we use from \cite{Mezaros2012}. The 124 stars that could not be differentially corrected, due to their stellar parameters falling outside the range of the theoretical stellar grid (mainly at T$_\textrm{eff}>$10000K or T$_\textrm{eff}<$3500K), were used only empirically in each family of stars and corresponding SSP calculation. Therefore, our corrections can be considered conservative. We note that this upper limit of T$_\textrm{eff}$ will reduce the $\alpha$ sensitivity and accuracy of modelling for stellar populations with ages less than approximately 2 Gyr.

\subsubsection{Calculation}
An SSP can be represented as a probability distribution described by a mean and variance (\citealt{Cervino2006}; \citealt{Vaz2020}). For this work, the final products that we make publicly available are the mean spectra of the stellar populations and therefore we provide details for this calculation.

\begin{figure}
\centering
 \includegraphics[width=\linewidth, angle=0]{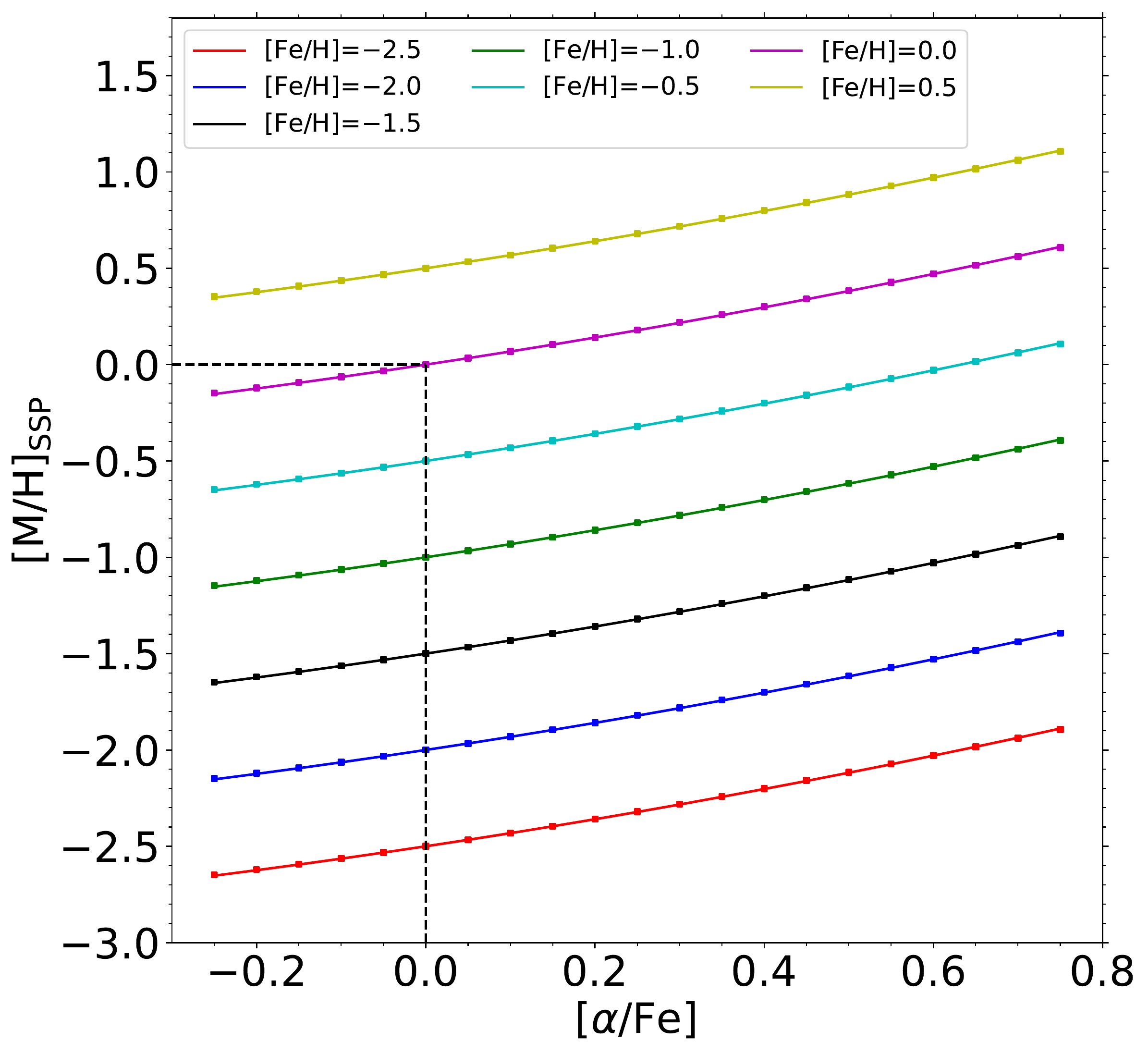}
\caption{Total metallicity ([M/H]$_{\textrm{SSP}}$) as a function of [$\alpha$/Fe], for the range of metallicities ([Fe/H]) in our stellar spectral models. The coloured points represent the full calculation of metallicity and the solid lines represent the fitted relation, given in Equation~\ref{MetallicityFitEqn}. For the full range of [Fe/H] and [$\alpha$/Fe] of the models, the calculations are well fitted by the relation. }
\label{MetallicityFit}
\end{figure}

In SSP calculations, translations between the different parameter planes of the isochrone and stellar library are necessary, because the resultant population spectrum is calculated through integrations of star contributions at different locations on the isochrone. Therefore, a relation between the observed parameters of the stellar library and theoretical isochrone parameters is required. In this work, the underlying BaSTI isochrones are computed with T$_\textrm{eff}$, log g and total metallicity, whereas stellar spectra are usually tagged with T$_\textrm{eff}$, log g and [Fe/H]\footnote{In \cite{Knowles21}, we computed theoretical spectra with a metallicity tag of [M/H] that is defined the same as [Fe/H] here. We differentiate between the isochrone and stellar definitions, by defining the isochrone symbol as [M/H]$_{\textrm{SSP}}$.}. The total metallicity, as defined in isochrone parameters ([M/H]$_{\textrm{SSP}}$) is given by:
\begin{equation}
\textrm{[M/H]}_{\textrm{SSP}}=\log_{10}(\textrm{Z/X})_{*} - \log_{10}(\textrm{Z/X})_{\odot},
\label{MetallicitySSPDef}
\end{equation}
where Z and X are defined as mass fractions of metals and hydrogen, respectively. The spectroscopic metallicity ([Fe/H]) is usually defined for stellar spectra as a scaled-metallicity\footnote{In which all metals, apart from the $\alpha$-elements and carbon if they are also non-solar, are scaled by the same factor from the solar mixture (e.g. [M/H]=0.2=[Fe/H]=[Li/H]).}. For the case of scaled-solar abundances, the total metallicity and [Fe/H] are equivalent, however, in the case where [$\alpha$/Fe] abundance ratios are non-solar a conversion is needed, which requires a relation between the two metallicity definitions, similar to that done in equation 4 of \citetalias{Vaz2015}. Therefore, a calculation of the total metallicity for various [Fe/H] and [$\alpha$/Fe] values was made, assuming \cite{Asplund2005} solar abundances to be consistent with the adoption made in the stellar model calculations used in \cite{Knowles21}. We highlight here that there is an inconsistency between the solar abundances in the isochrones and those used in our theoretical stellar spectra that are the basis of the differential corrections. The stellar models are computed assuming \cite{Asplund2005} abundances whereas BaSTI isochrones are calculated with \cite{Grevesse93} abundances. The solar metallicity, Z$_\odot$, defined by \cite{Grevesse93} is given as 0.0198 compared to the value of 0.0122 found by \cite{Asplund2005}. The calculation was performed for a range of [Fe/H] from $-$2.5 to $+$0.5, in steps of 0.05 dex, and a range of [$\alpha$/Fe] from $-$0.25 to $+$0.75 in steps of 0.05 dex (i.e. the range of the stellar models generated in \citealt{Knowles21}). A relation between total and spectroscopic metallicities, as well as [$\alpha$/Fe], was estimated using the SciPy routine `curvefit' (\citealt{Scipy2020}), of the form:
\begin{equation}
\textrm{[M/H]}_{\textrm{SSP}}\textrm{=[Fe/H]+a}[\alpha\textrm{/Fe]+b}[\alpha\textrm{/Fe}]^{2},
\label{MetallicityFitEqn}
\end{equation}
The coefficients a and b were found to be 0.66154$\pm$0.00128 and 0.20465$\pm$0.00218, respectively. In Figure~\ref{MetallicityFit}, we compare the results of the full calculation and fitted relation for a range of [Fe/H] and varying [$\alpha$/Fe] abundances. For the full range of stellar models, the fit is good.
We are accounting for the difference in solar compositions in isochrone and stellar library by calculating this conversion for the \cite{Asplund2005} mixture so that the interpolation within the sMILES library produces stars at the approximately the correct $\textrm{[M/H]}_{\textrm{SSP}}$, and therefore approximately the correct location on the isochrone, for the required SSP. However, we acknowledge that even with this conversion our models will not be fully consistent between stellar library and isochrone components of the calculation, as there are differences in the $\alpha$ element abundances between the two mixtures. For example, the oxygen abundances differ by $\sim$ 0.2 dex. We note that the impact of isochrone choice on the resulting SSP spectrum is secondary to the impact of the stellar spectra used, particularly in older stellar populations (e.g. see figure 9  of \citetalias{Vaz2015}).

To convert theoretical isochrone parameters into observables (e.g. colours and fluxes), we use relations between fundamental stellar parameters (T$_\textrm{eff}$, log g and [Fe/H]) and colours from extensive empirical photometric libraries, rather than solely using predictions of theoretical atmosphere calculations. The empirical relations used are those of \cite{Alonso96,Alonso99} that are metallicity-dependent relations for dwarfs and  giants. Note that these relations do still have a marginal dependence on theoretical atmospheres. Metal-dependent Bolometric corrections from \cite{Alonso95,Alonso99} are used.

Computations of SSPs are performed through methods described in detail in \cite{Vaz2010} and \citetalias{Vaz2015}. We summarise their method below.
The calculation involves the integration of stellar spectra along isochrones, with the adopted IMF providing the number of stars per mass bin. SSPs are computed for fixed ages and total metallicities for various [$\alpha$/Fe] values through:
\begin{multline}
S_{\lambda}(t,\mathrm{[M/H]}_{\mathrm{SSP}},[\alpha/\mathrm{Fe}],\Phi,\mathrm{I}_{\alpha})=\\\int_{m_{l}}^{m_{t}} S_{\lambda V}(m,t,[\mathrm{Fe/H}],[\alpha/\mathrm{Fe}]) \times\\ F_{V}(m,t, \mathrm{[Fe/H]}, [\alpha/\mathrm{Fe}]) \times N_{\Phi}(m, t) dm,
\end{multline}
where $S_{\lambda}(t,\textrm{[M/H]}_{\textrm{SSP}},[\alpha/\textrm{Fe}],\Phi,\textrm{I}_{\alpha})$, gives the SSP spectrum at time $t$, with total metallicity $\textrm{[M/H]}_{\textrm{SSP}}$ (defined in Equation~\ref{MetallicitySSPDef}), $[\alpha$/Fe] abundance, with a specific IMF ($\Phi$) and isochrone with an [$\alpha$/Fe] abundance $\textrm{I}_{\alpha}$ (either 0.0 or 0.4 dex depending on the [$\alpha$/Fe] value of the desired SSP). $S_{\lambda V}(m,t,[\textrm{Fe/H}],[\alpha/\textrm{Fe}])$ is a star spectrum (in units of erg s$^{-1}${\AA}$^{-1}$), normalised by its V-band flux for each sMILES (or MILES) star, for a given star mass (m), spectroscopic metallicity ([Fe/H]) and $[\alpha$/Fe] abundance, which is alive at time $t$. The $[\alpha$/Fe] abundances here are the values discussed in previous sections, which are made up of sMILES stars. $F_{V}(m,t, \textrm{[Fe/H]}, [\alpha/\textrm{Fe}])$ is the absolute flux of the star in the V-band and is predicted by the method described in \cite{Falcon2011}, based on relations from \cite{Alonso96,Alonso99}, for the atmospheric parameters of the star. $N_{\Phi}(m, t)$ is the number fraction of stars in a mass interval ($m+dm$). ${m_{l}}$ and ${m_{t}}$ represent the lowest and highest mass stars alive at time $t$, which is provided by the isochrone. The product of $S_{\lambda V}$ and $F_{V}$ is a monochromatic luminosity that when integrated with respect to mass (using the adopted IMF) gives a monochromatic luminosity of the SSP. V-band normalisation is used so that absolute magnitudes can be found from the calculated SSPs and will be fully consistent with absolute V-band magnitudes found from the photometric libraries used in the isochrone parameter conversions. In other words, the photometric and spectroscopic predictions of the SSPs will be consistent. We normalise to solar luminosity and mass and therefore the resulting SSP spectra have units of $\frac{L_{\lambda}}{L_{\odot}}$\AA$^{-1}$M$_{\odot}^{-1}$, where L$_{\odot}$=$3.826\times10^{33}$ erg s$^{-1}$ (\citetalias{Vaz2015}).

%The adopted total initial SSP mass is 1M$_{\odot}$.
To obtain stellar spectra that match the required T$_\textrm{eff}$, log g and [Fe/H] for locations on isochrones,  the 3{\small D} interpolator described in \cite{Vaz03,Vaz2010} was used. This interpolator follows a local interpolation scheme in which the routine locates stars in the stellar library within a cube around the required location. We direct interested readers to those works for further details. We note here that because we have computed families of 801 sMILES stars that all have the same $[\alpha$/Fe] abundance and treat the remaining 124 empirical stars as if they had the same $[\alpha$/Fe] abundance as the sMILES stars, we only interpolate in the three dimensions of T$_\mathrm{eff}$, log g and [Fe/H] to sample individual stars of known parameters, at a particular [$\alpha$/Fe].

In summary, we computed five libraries of SSPs that adopt different IMFs; Universal Kroupa, Revised Kroupa, thirteen Unimodal and Bimodal, and Chabrier IMFs; for a wide range of isochrone ages and total metallicities, for [$\alpha$/Fe] values of $-$0.2, 0, 0.2, 0.4 and 0.6. There are 53 steps in age from 0.03-14 Gyr and 10 steps in metallicity from 0.0003-0.030, resulting in 2650 SSPs per IMF variation. The models are produced at MILES wavelength coverage (3450.5-74906\AA), sampling (0.9\AA) and resolution (2.5{\AA} FWHM). We summarise the sMILES SSP parameter coverage in Table~\ref{sMILES_SSPmodels_Tab}.
\begin{table*}
   \caption{Age, metallicity, [$\alpha$/Fe] ranges and IMF variations available for the sMILES SSP models computed in this work.}
    \centering
    \begin{tabular}{|c|p{50mm}|p{30mm}|p{20mm}|p{35mm}|}
    \hline
         {SSP Model} & Age (Gyr) & [M/H]$_{\textrm{SSP}}$ & [$\alpha$/Fe] & IMF  \\
        \hline
       sMILES  &0.03, 0.04, 0.05, 0.06, 0.07, 0.08, 0.09, 0.10,\newline0.15, 0.20, 0.25, 0.30, 0.35, 0.40, 0.45, 0.50,\newline 0.60, 0.70, 0.80, 0.90, 1.00, 1.25, 1.50, 1.75, \newline2.00, 2.25, 2.50, 2.75, 3.00, 3.25, 3.50, 3.75,\newline 4.00, 4.50, 5.00, 5.50, 6.00, 6.50, 7.00, 7.50,\newline8.00, 8.50, 9.00, 9.50, 10.0, 10.5, 11.0, 11.5,\newline 12.0, 12.5, 13.0, 13.5, 14.0 & $-1.79$, $-1.49$, $-1.26$,\newline $-0.96$, $-0.66$, $-0.35$,\newline $-0.25$, 0.06, 0.15,\newline 0.26 & $-$0.2, 0.0, 0.2,\newline 0.4, 0.6 & 13 Unimodal ($\Gamma=0.3-3.5$), \newline 13 Bimodal ($\Gamma_\mathrm{b}=0.3-3.5$),\newline Universal Kroupa,\newline Revised Kroupa,\newline Chabrier  \\
    \end{tabular}
    \label{sMILES_SSPmodels_Tab}
\end{table*}
\section{\protect\lowercase{s}MILES SSP Models}
\label{sec:sMILESTests}
\subsection{Properties}
\subsubsection{Age and Metallicity}
\begin{figure*}
\includegraphics[width=\linewidth]{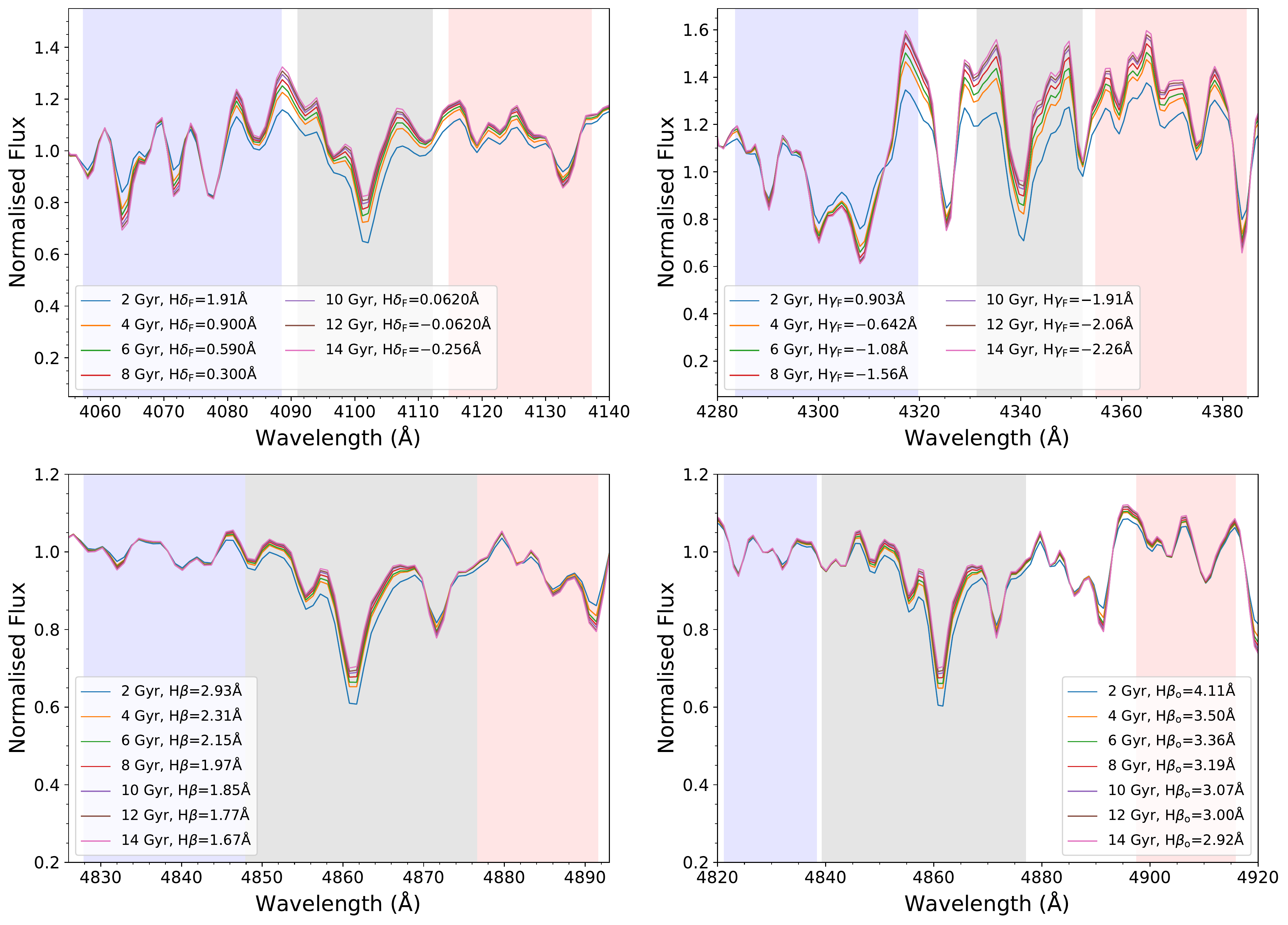}
\caption{Sequence of sMILES SSPs of varying age, for four age-sensitive spectral features for solar metallicity and [$\alpha$/Fe] abundance pattern ([M/H]$_{\textrm{SSP}}$=0.06, [$\alpha$/Fe]=0.0) populations, computed assuming a Universal Kroupa IMF. (Top panel: H$\delta_{\textrm{F}}$ and H$\gamma{\textrm{F}}$. Bottom panel: H$\beta$ and H$\beta_{\textrm{o}}$. Index values are also shown, demonstrating the age-sensitivity of these features. The blue pseudo-continuum, feature and red pseudo-continuum bands definitions (\citealt{Trager98}; \citealt{Cervantes09}) are plotted in blue, grey and red respectively. Spectra are normalised to the average flux within the blue pseudo-continuum band for each index. Index values are measured at MILES FWHM (2.5{\AA}) resolution.}
  \label{sMILES_SSP_Age_Seq}
\end{figure*}

\begin{figure*}
\includegraphics[width=\linewidth]{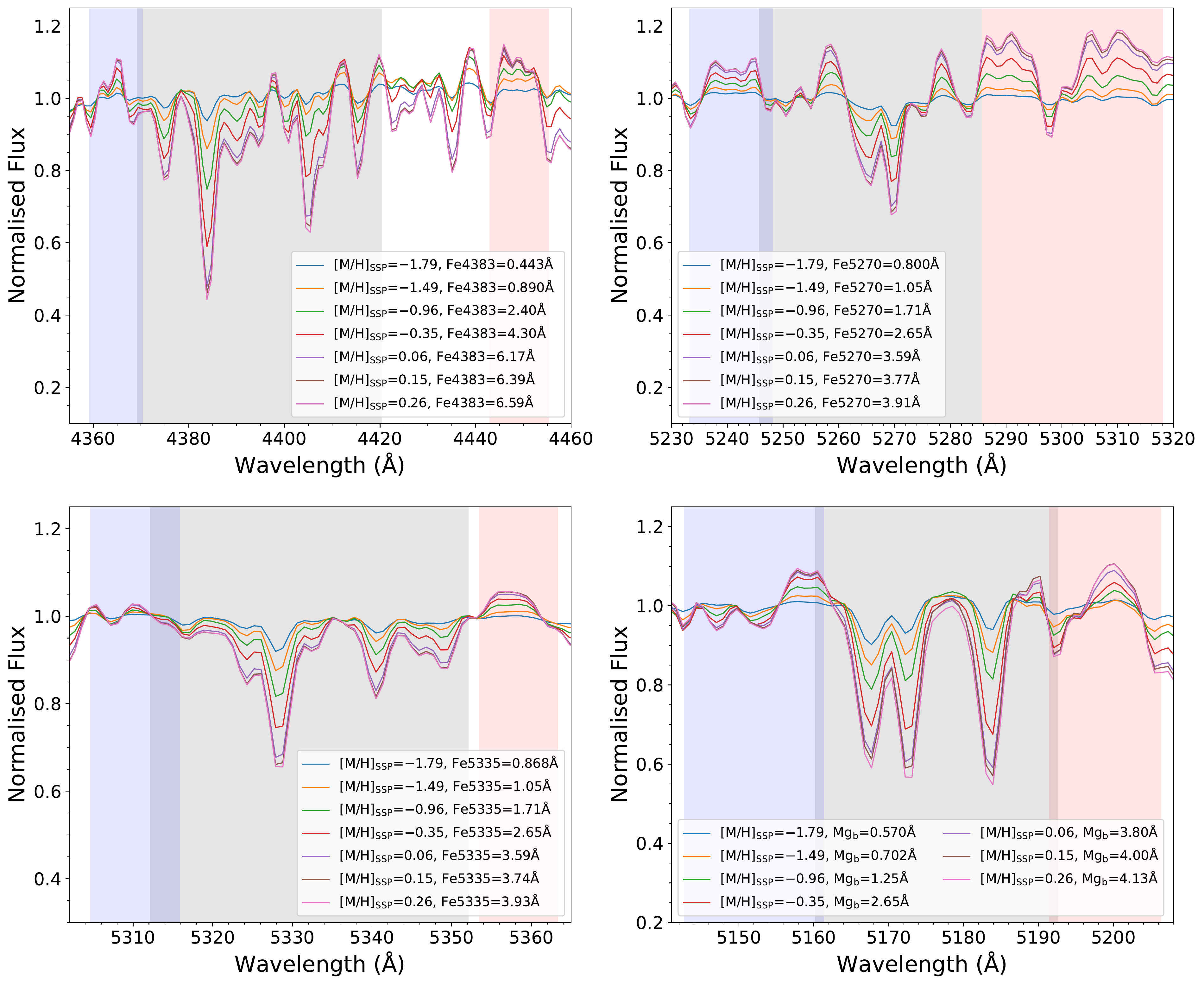}
\caption{sMILES SSP sequences for four metallicity sensitive features for fixed age (10 Gyr) and scaled-solar abundance pattern populations, computed assuming a Universal Kroupa IMF. Top panel: Fe4383 and Fe5270, Bottom panel: Fe5335 and Mg$_{\textrm{b}}$. Index values are also shown, demonstrating the metallicity-sensitivity of these features. The blue pseudo-continuum, feature and red pseudo-continuum bands definitions (\citealt{Trager98}; \citealt{Cervantes09}) are plotted in blue, grey and red respectively. Spectra are normalised to the average flux within the blue pseudo-continuum band for each index. Index values are measured at MILES FWHM (2.5{\AA}) resolution.}
  \label{sMILES_SSP_Metal_Seq}
\end{figure*}

In Figure~\ref{sMILES_SSP_Age_Seq}, we show a sequence of sMILES SSP spectra for varying age, with fixed solar metallicity and $\alpha$ abundance ([M/H]$_{\textrm{SSP}}$=0.06, [$\alpha$/Fe]=0.0) and a universal Kroupa IMF. This figure shows four commonly used age-sensitive Lick indices, H$\delta_{\textrm{F}}$, H$\gamma_{\textrm{F}}$, H$\beta$ and H$\beta_{o}$. Spectra are normalised to their average flux within the blue pseudo-continuum side band for each index. As shown, all indices behave as expected with age, in that the H$\delta_{\textrm{F}}$, H$\gamma_{\textrm{F}}$ H$\beta$ and H$\beta_{o}$ features decrease in strength with increasing SSP age. As expected from \cite{Cervantes09}, H$\beta_{o}$ provides a stronger age indicator than H$\beta$, with larger indices present for all ages tested.
In Figure~\ref{sMILES_SSP_Metal_Seq}, we show a sequence of sMILES SSPs for varying metallicity, with a fixed age (10 Gyr),  scaled-solar abundance ([$\alpha$/Fe]=0.0) and universal Kroupa IMF. We plot four Lick indices commonly used in total metallicity probes, particularly in the [MgFe] and [MgFe]' indices we investigate later in Section~\ref{sec:MgFe}. SSPs are plotted in the wavelength range of Fe4383, Fe5270, Fe5335 and Mg$_{\textrm{b}}$ indices, defined in \cite{Trager98}, with their strengths shown. Spectra are normalised to their average flux within the blue pseudo-continuum side band for each index. As expected, for fixed scaled-solar abundance ([$\alpha$/Fe]=0.0), Fe4383, Fe5270, Fe5335 and Mg$_{\textrm{b}}$ all increase in strength with increasing metallicity.
We highlight the well known age-metallicity degeneracy in Figure S1 of the Supplementary Material, where we show sequences of sMILES SSP spectra of varying age and metallicity. We demonstrate how old, metal-poor populations look like younger, metal-rich populations.
\subsubsection{[$\alpha$/Fe]}
\begin{figure}
    \includegraphics[width=\linewidth,angle=0]{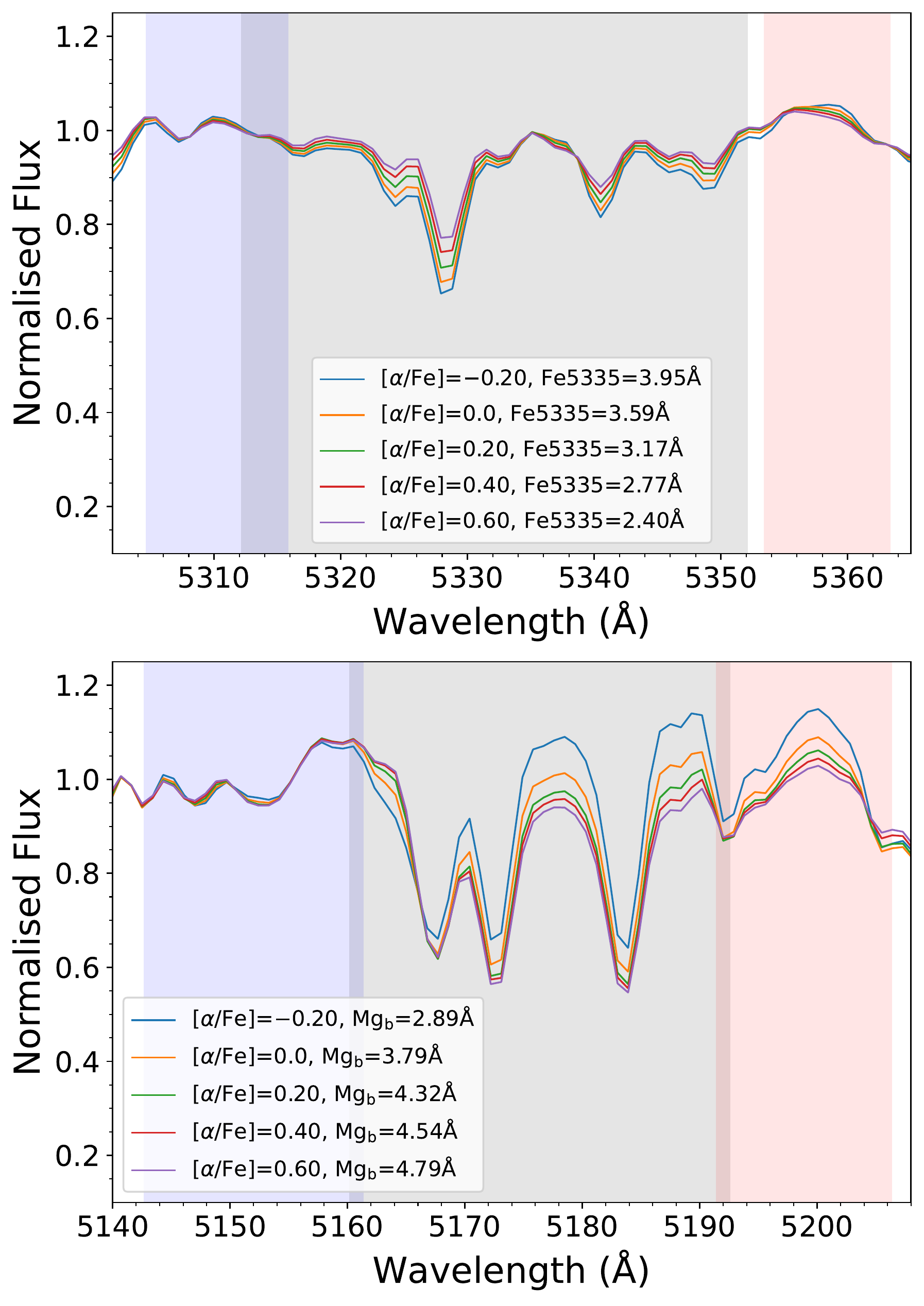}
    \caption{sMILES SSP spectral sequences for two features showing their sensitivity to [$\alpha$/Fe].
 (Top panel: Fe5335, Bottom panel: Mg$_{\textrm{b}}$, for a fixed age (10 Gyr) and solar metallicity ([M/H]$_{\textrm{SSP}}$=0.06), computed assuming a Universal Kroupa IMF. The blue pseudo-continuum, feature and red pseudo-continuum bands definitions (\citealt{Trager98}) are plotted in blue, grey and red respectively. Lick index values are also shown. Spectra are normalised to the average flux within the blue pseudo-continuum band for each index. Index values are measured at MILES FWHM (2.5{\AA}) resolution.}
    \label{sMILES_SSP_Alpha_Seq}
\end{figure}

In Figure~\ref{sMILES_SSP_Alpha_Seq}, we show a sequence of sMILES SSPs for varying [$\alpha$/Fe] abundance ratio with fixed solar metallicity, 10 Gyr age and universal Kroupa IMF. In this figure we focus on two Lick indices, namely Fe5335 and Mg$_{\textrm{b}}$, both of which are used in the total metallicity-sensitive index definitions of [MgFe] and [MgFe]'. Spectra are normalised to their average flux within the blue pseudo-continuum side band for each index. The sense of the change is as expected, in that there is a general decrease and increase of index strength for Fe5335 and Mg$_{\textrm{b}}$ respectively, for increasing [$\alpha$/Fe] abundance at fixed metallicity. We note that other known [$\alpha$/Fe]-sensitive Lick indices, namely Ca4227, TiO$_{1}$, TiO$_{2}$, also follow the general qualitative trend of increasing strength for increasing [$\alpha$/Fe], at this fixed age (10 Gyr) and metallicity ([M/H]$_{\textrm{SSP}}$=0.06). Interestingly, we find in sMILES models that Ca4455 and Mg$_{1}$ indices decrease in strength with increasing [$\alpha$/Fe], in agreement with the \citetalias{Vaz2015} models at the same age and metallicity.

\begin{figure*}
  \includegraphics[width=150mm,angle=0]{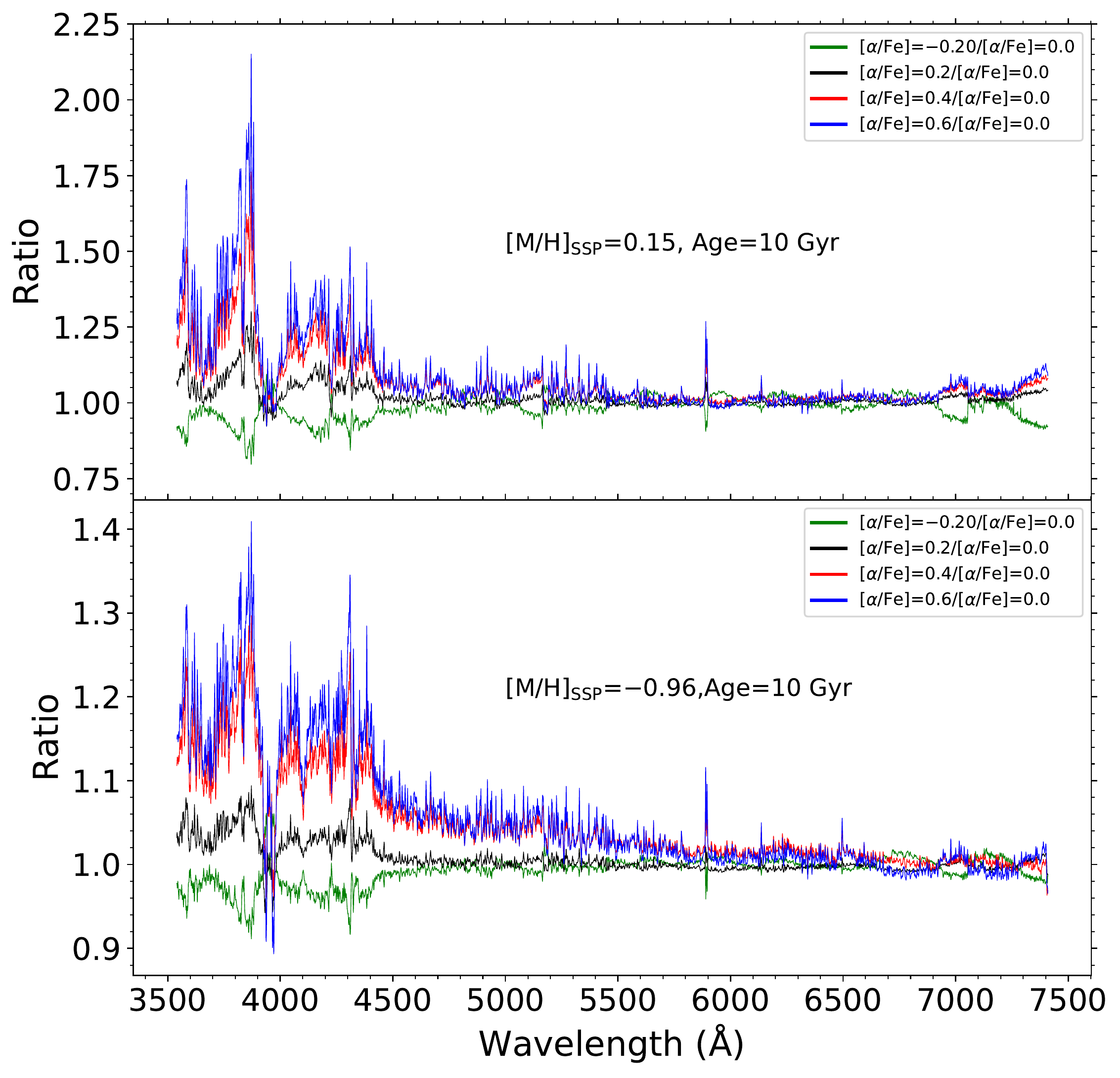}
    \caption{sMILES SSP predictions of differential [$\alpha$/Fe] effects for a super-solar (Top panel) and sub-solar (Bottom panel) metallicity, 10 Gyr old population in the full MILES wavelength range. SSP spectra are calculated adopting a Universal Kroupa IMF and are shown at MILES resolution (2.5{\AA} FWHM).}
    \label{sMILES_SSP_Alpha_Differential}
\end{figure*}

In Figure~\ref{sMILES_SSP_Alpha_Differential}, we show sMILES SSP predictions of spectral changes due to [$\alpha$/Fe] variations for a 10 Gyr old population at two metallicity values; one super-solar ([M/H]$_{\textrm{SSP}}$=0.15) and one sub-solar ([M/H]$_{\textrm{SSP}}$=$-$0.96). We plot ratios of [$\alpha$/Fe] enhanced or deficient and solar abundance pattern SSP spectra to show the impact of [$\alpha$/Fe] changes across the full MILES wavelength range. The effect of enhancing or reducing the [$\alpha$/Fe] abundance ratio is particularly large in the blue. The excess of flux in the blue is largely attributed to the fact that at fixed total metallicity, the [$\alpha$/Fe]-enhanced (deficient) element mixture has a decreased (increased) iron abundance, and therefore lower (higher) opacity, with respect to the scaled-solar model of the same total metallicity. Large changes in flux are found for Ca II H–K lines around $\sim$3950{\AA}. These residuals reflect the [$\alpha$/Fe] changes between SSPs, with calcium included with the $\alpha$-elements in the underlying stellar models of \cite{Knowles21}. TiO band residuals are visible around and above $\sim$6500{\AA}, also reflecting the inclusion of titanium and oxygen in the $\alpha$-elements of our models. The Mg$_\textrm{b}$ and MgH region around 5150{\AA} is also seen to vary, as expected due to changing magnesium abundance.

Carbon, nitrogen and oxygen-related molecular absorption features are also clear, such as CNO, CN and CH which are present around $\sim$3800-4300{\AA} (see \citealt{Tripicco1995}). The changes of flux for these features are most likely due to the differences in C, N, O and individual $\alpha$-element abundances between empirical MILES stars and the theoretical stellar spectra underpinning the sMILES stars in \cite{Knowles21} and corresponding SSP calculations. In that work and this work, C, N and O abundances in the MILES stars have not been accounted for; only the estimates of [Mg/Fe] (as a proxy for [$\alpha$/Fe]) and [Fe/H] have been used in the differential correction process when generating sMILES stars and SSPs. In the theoretical stellar spectra of \cite{Knowles21}, carbon and nitrogen are assumed to be scaled-solar and $\alpha$-elements, of which oxygen is a part, are all assumed to track [Mg/Fe]. CNO contribute significantly to the opacity of stellar photospheres (see the discussions of \citetalias{Vaz2015} and \citealt{Sansom2013}). For Na$\mathrm{_D}$, as was found in \citetalias{Vaz2015}, a clear peak (or trough in the case of the [$\alpha$/Fe]-deficient ratio) is evident at the Na I doublet around $\sim$5895{\AA} for all ratios. \cite{Barbuy03}, \cite{Coelho05,Coelho07} and \citetalias{Vaz2015} discuss the cause of this; an increased electron donation from increased $\alpha$ abundances can cause a lowering of the continuum and therefore a weakening of lines, particularly for iron lines and the Na I doublet.
\subsubsection{sMILES SSP Predictions}
\begin{figure*}
\setlength{\columnsep}{-12pt}
\begin{multicols}{2}
    \includegraphics[width=\linewidth]{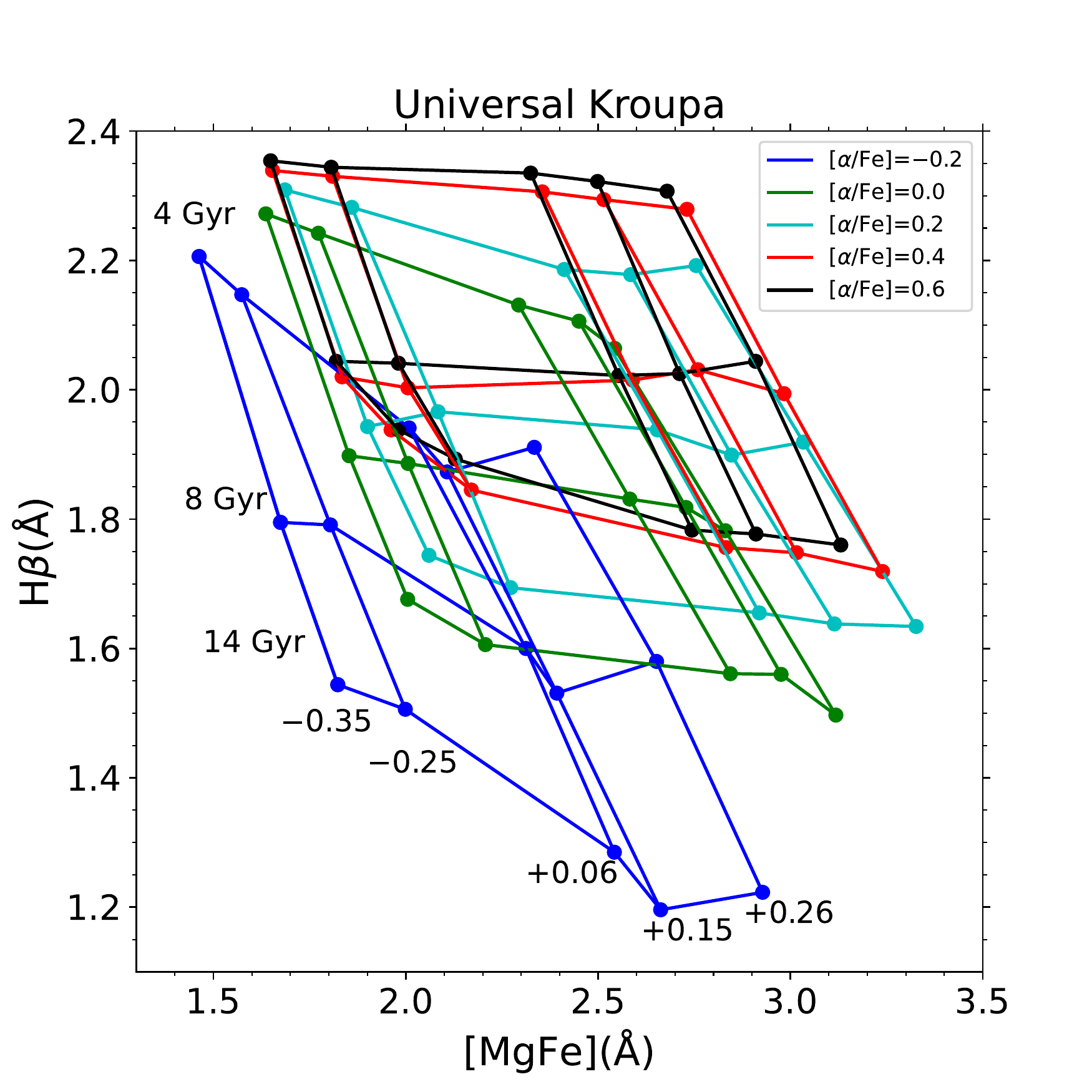} \par
    \includegraphics[width=\linewidth]{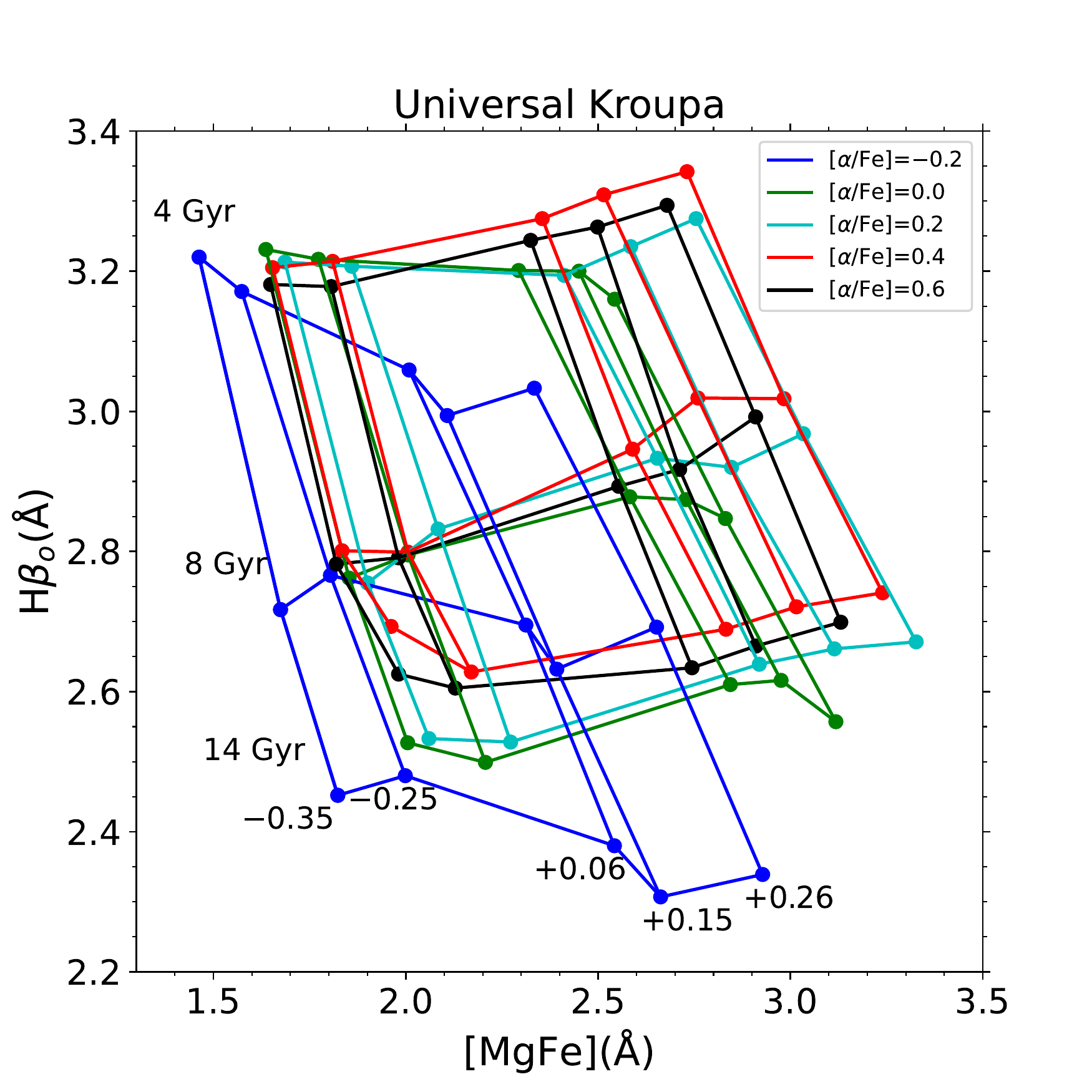}\par
    \end{multicols}
    \vspace{-29pt}
\begin{multicols}{2}
    \includegraphics[width=\linewidth]{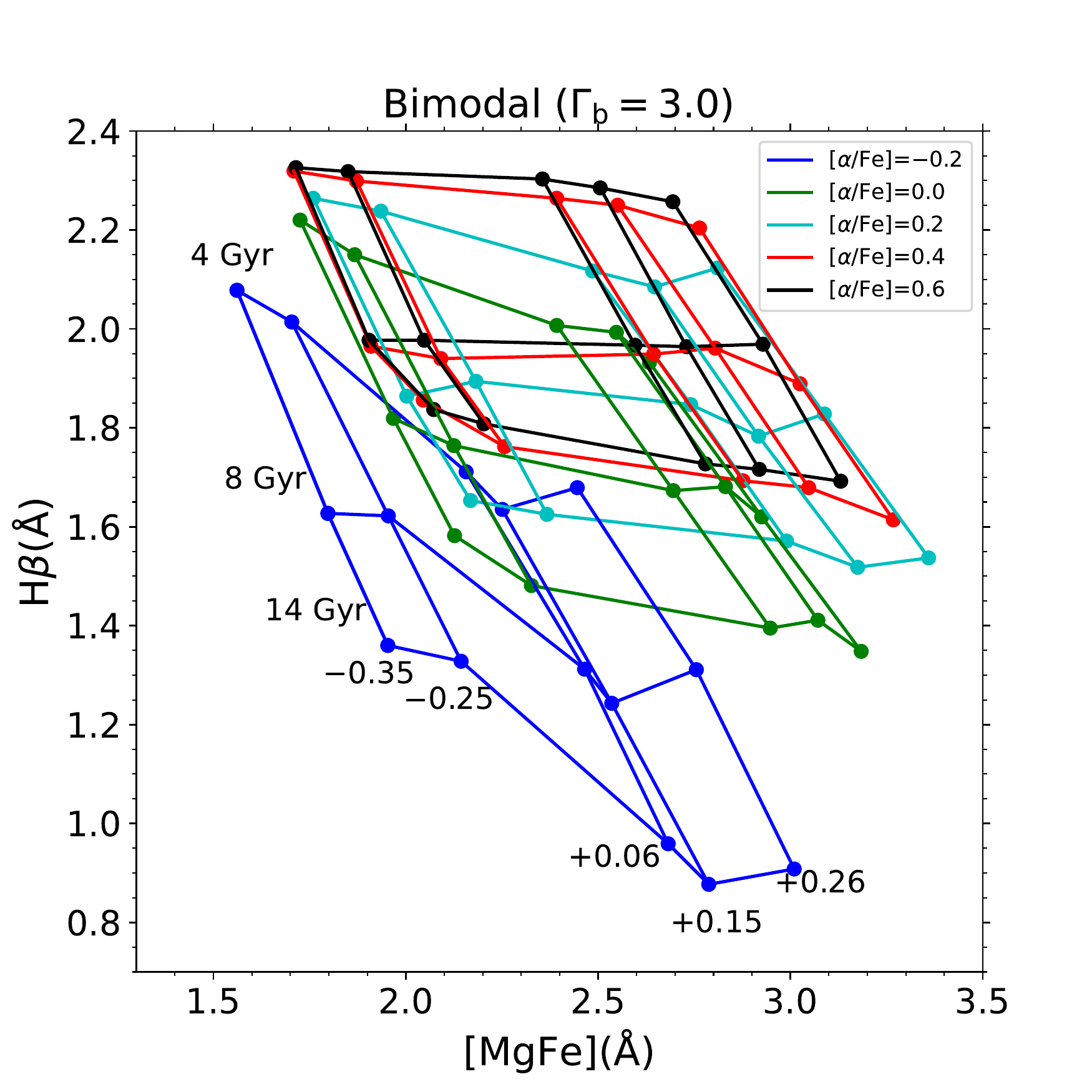}\par
    \includegraphics[width=\linewidth]{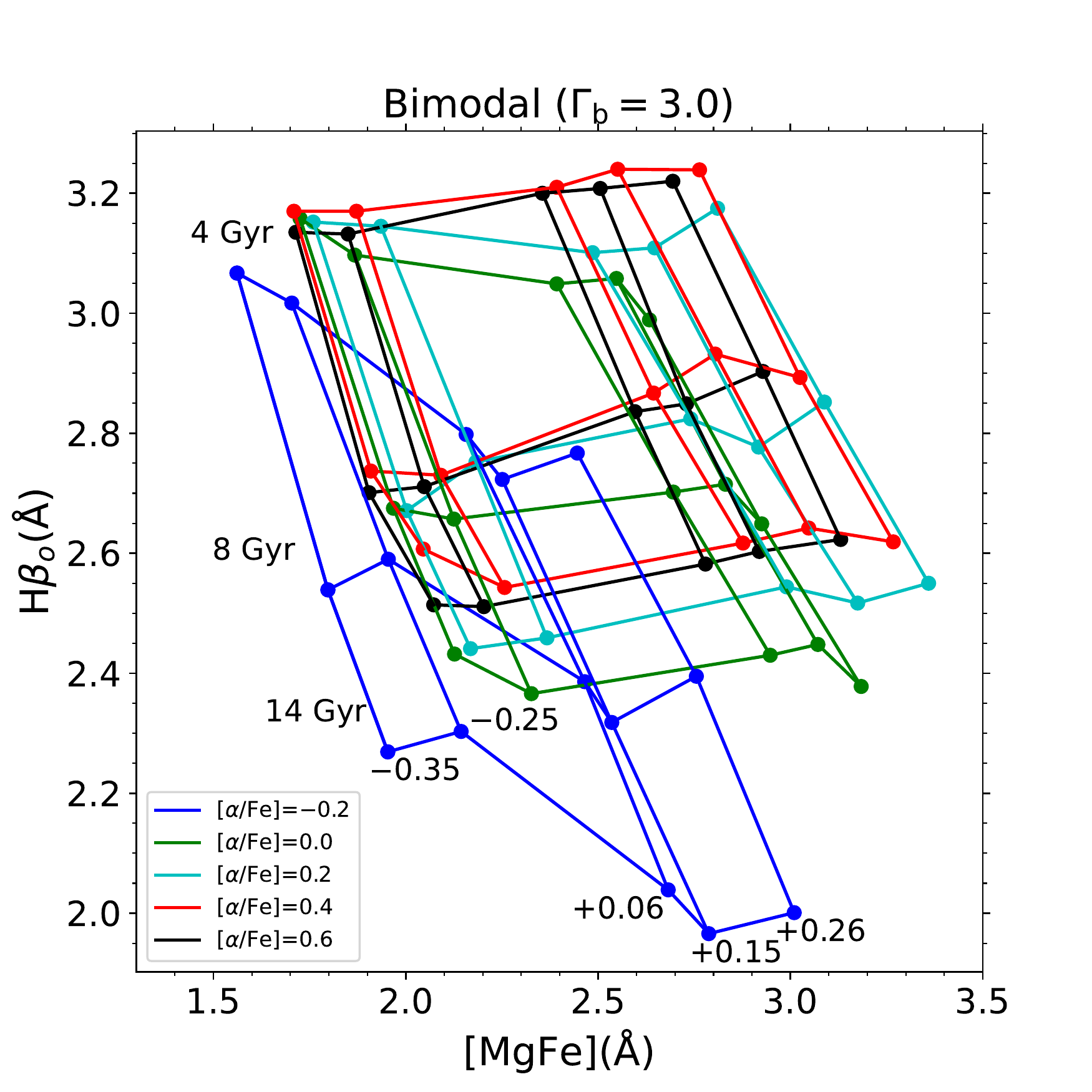}\par
\end{multicols}
\caption{sMILES SSP predictions of H$\beta$, H$\beta_{\textrm{o}}$ and [MgFe] index strengths for varying age, metallicity and [$\alpha$/Fe]. Top panel: For a Universal Kroupa IMF. Bottom panel: For a Bimodal IMF with a slope $\Gamma_\textrm{b}=3.0$. Grids show the effect of increasing total metallicity ($\mathrm{[M/H]}_{\mathrm{SSP}}$, moving from left to right in each grid) and
age (increasing from top to bottom in each grid), respectively, as labelled. Index values are measured at MILES FWHM (2.5{\AA}) resolution.}
\label{sMILES_Grid_HbetaHbetao_MgFe}
\end{figure*}

\begin{figure*}
\setlength{\columnsep}{-7pt}
\begin{multicols}{2}
    \includegraphics[width=\linewidth]{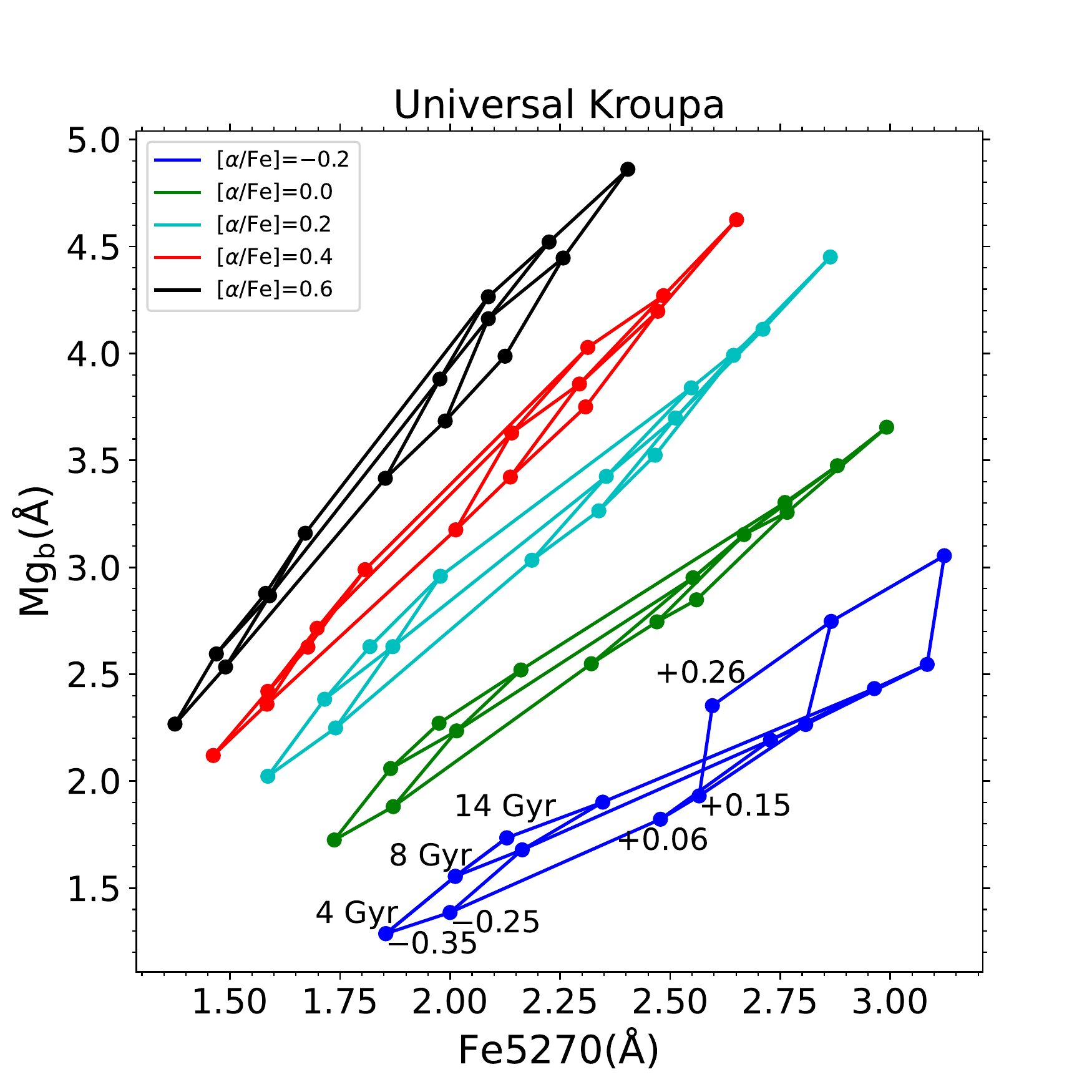}\par
    \includegraphics[width=\linewidth]{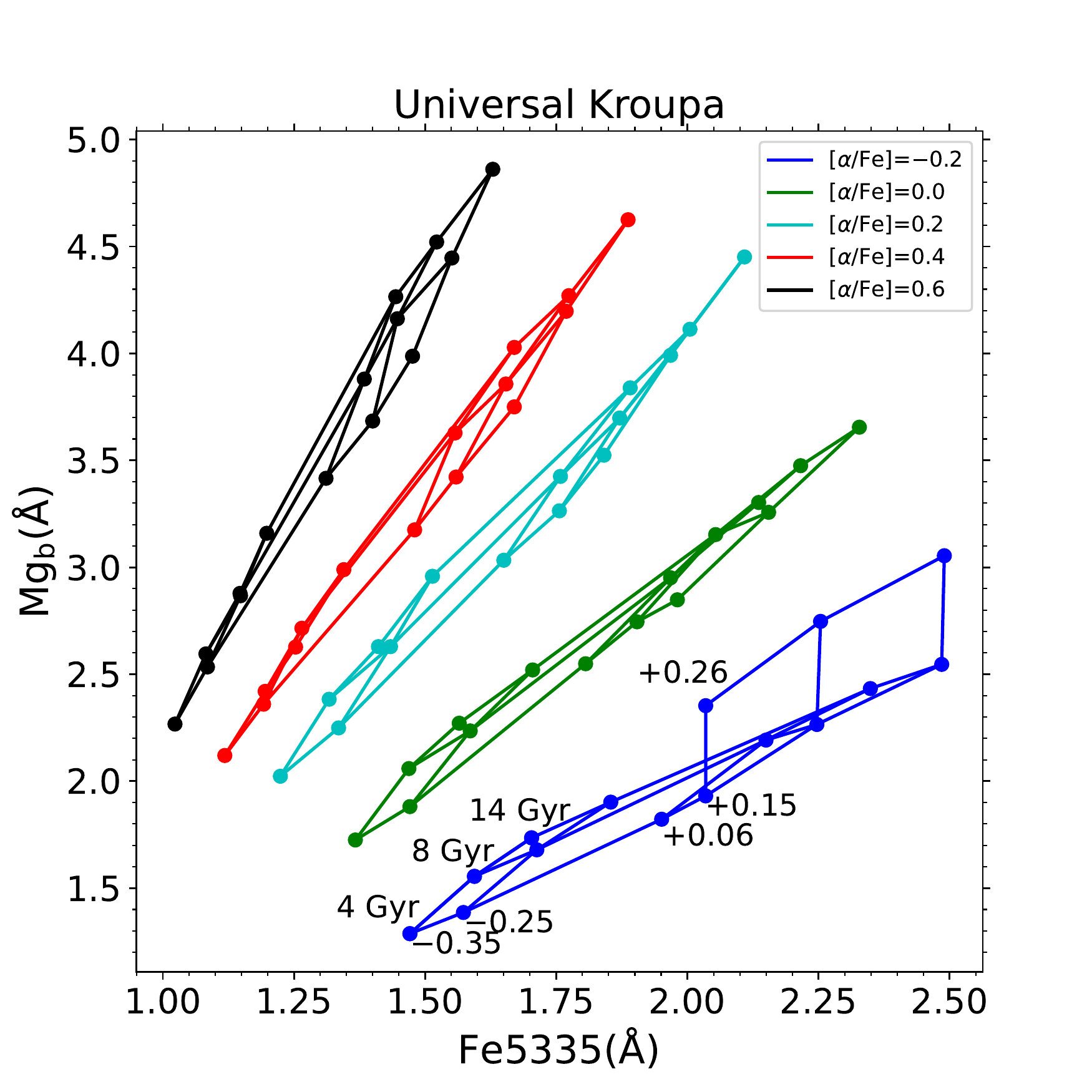}\par
    \end{multicols}
\vspace{-31pt}
\begin{multicols}{2}
    \includegraphics[width=\linewidth]{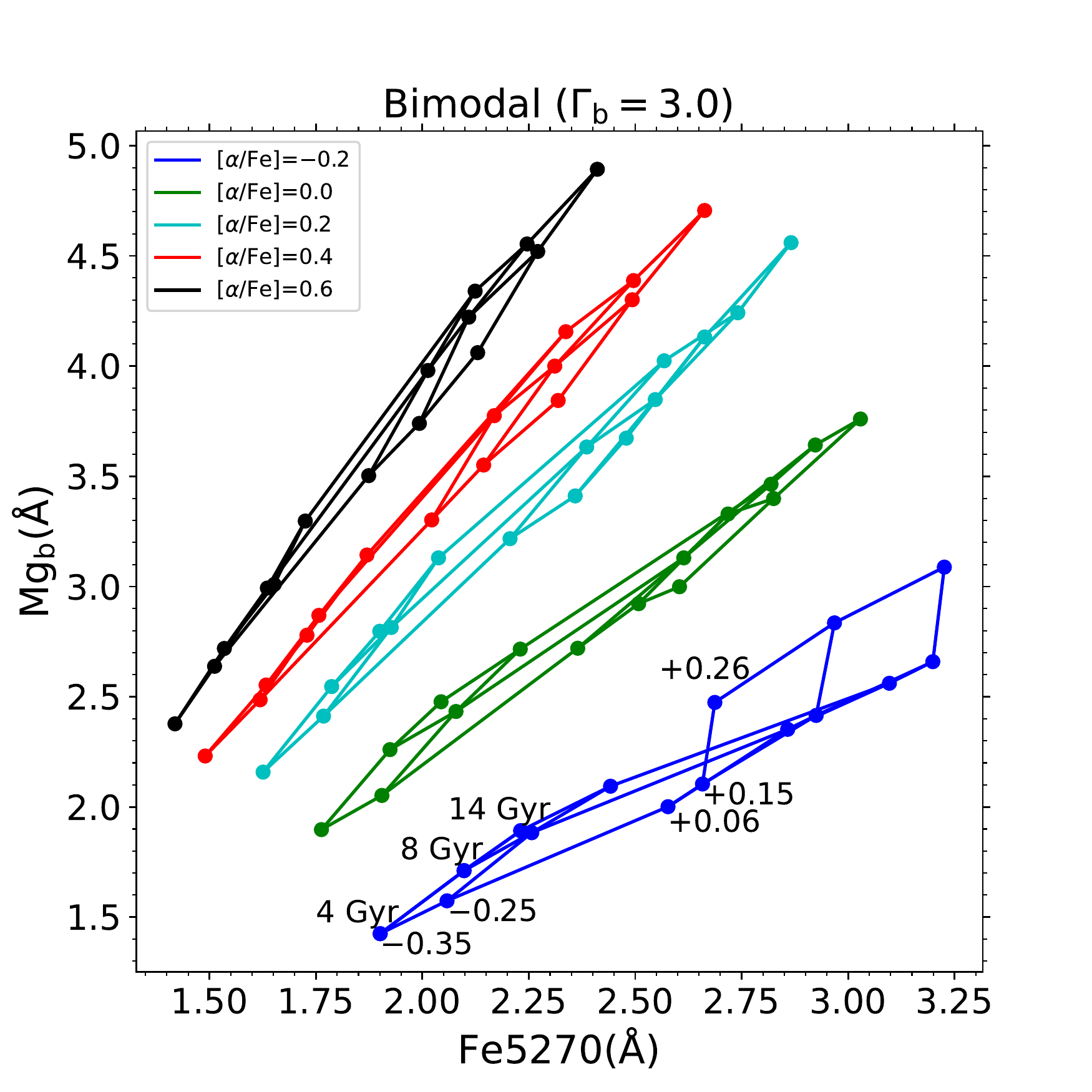}\par
    \includegraphics[width=\linewidth]{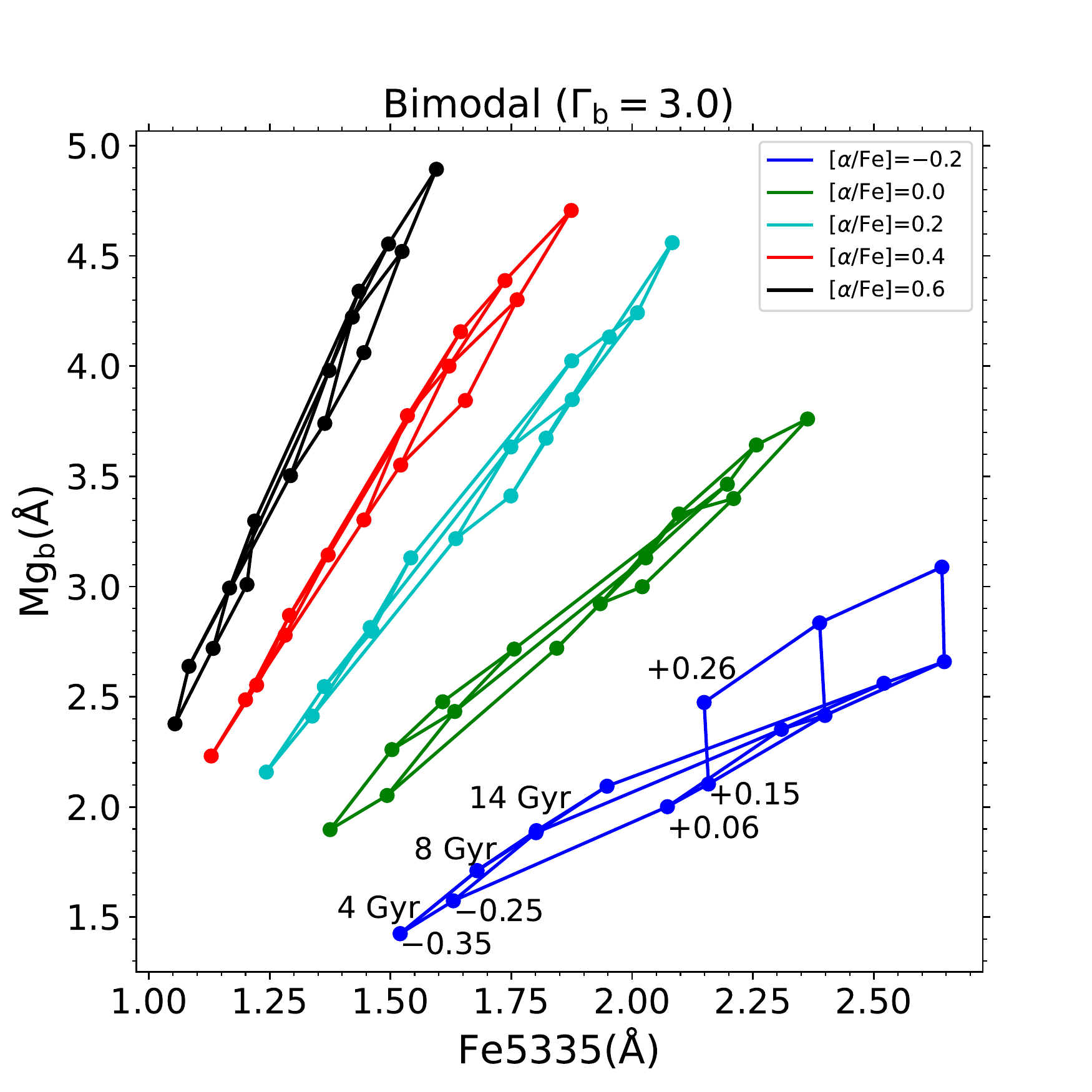}\par
\end{multicols}
\caption{sMILES SSP predictions of Mg$_\mathrm{b}$, Fe5270 and Fe5335 line strengths for varying age, metallicity and [$\alpha$/Fe]. Top panel: For a Universal Kroupa IMF. Bottom panel: For a Bimodal IMF with a slope $\Gamma_\textrm{b}=3.0$. Grids show the effect of changing total metallicity ($\mathrm{[M/H]}_{\mathrm{SSP}}$, moving from left to right in each grid) and
age (increasing from bottom to top in each grid), respectively, as labelled. Index values are measured at MILES FWHM (2.5{\AA}) resolution.}
\label{sMILES_SSP_Mgb_vs_Fe}
\end{figure*}

We now focus on the predictions of sMILES SSPs when varying the parameters of age, metallicity and [$\alpha$/Fe] together. In Figures~\ref{sMILES_Grid_HbetaHbetao_MgFe} and~\ref{sMILES_SSP_Mgb_vs_Fe}, we plot the predictions of Lick line strengths with varying age, metallicity and [$\alpha$/Fe] changes together. Figure~\ref{sMILES_Grid_HbetaHbetao_MgFe} shows how H$\beta$ and H$\beta_{\mathrm{o}}$ varies with [MgFe] for a range of age, metallicity and [$\alpha$/Fe] abundance. Figure~\ref{sMILES_SSP_Mgb_vs_Fe} shows the variation of Mg$_\mathrm{b}$ and Fe5270 for the same range of stellar population properties, illustrating how abundance patterns can be distinguished almost independently from effects of age-metallicity degeneracy and that the choice of IMF does not change these results.
\subsection{Comparisons With Other SSP Models}
\label{ModelComparisons}
The newly computed sMILES SSP models are first compared to previous models of \citetalias{Vaz2015}, with an analysis of indices and spectra.

The two sets of models are based on the same underlying methods and therefore differences between them should originate from the treatment of the differential correction and theoretical stellar models used, rather than differences in SSP calculation. The main differences are that \citetalias{Vaz2015} models differentially correct on the SSP level, rather than star level as in the sMILES SSPs. The differential corrections used in \citetalias{Vaz2015} were calculated using predictions from the theoretical stellar library of \cite{Coelho05,Coelho07} and corresponding fully theoretical SSPs computed using that library. The resulting \citetalias{Vaz2015} have consistent solar abundance references in both stellar library and isochrone components of the calculations.

Another widely-used set of models is that of \citealt{Conroy18}. These models are an update of the \cite{Conroy2012a} models, calculated for a larger range of metallicites than those original models. The three sets of models differ in several ways, particularly in the adopted stellar libraries, isochrones and solar abundance reference as summarised in Table~\ref{SSP_Differences}.

In addition to this, as in the work of \citetalias{Vaz2015}, \cite{Conroy18} perform differential corrections on an SSP level and the theoretical stellar spectra adopted were computed with a larger number of molecules included in the line lists than the models computed in \cite{Knowles21}, with the inclusion of FeH, H$_2$O, MgO, AlO, NaH, VO, SiH, CrH and CaH. The differential corrections were also computed differently, in that \cite{Conroy18} calculate the responses of individual elements and then combine them, through multiplications, to obtain arbitrary abundance patterns. However, effects on an SSP spectrum of changing several elements at once is not necessarily the same as multiplying by individual element responses. This individual element approach is a good approximation for changes in trace elements that do not significantly affect the atmospheric structure of the stars. Difference between adding up responses of individual elements, compared to more global changes was illustrated in \cite{Proctor2002}, their table 9, which showed that adding up effects from individual $\alpha$-elements can lead to large discrepancies compared to overall $\alpha$ changes in the stellar atmospheres. \cite{Conroy18} calculate SSP responses for 18 elements, for [X/H]=$-$0.3 and $+$0.3 apart from C, which is computed at [C/H]=0.15 to avoid the generation of carbon stars. sMILES models have the stellar spectral responses for total [$\alpha$/Fe] changes, computed with fully consistent model atmospheres and spectral synthesis calculations, and those responses are used to differentially correct empirical MILES stars that are used in the SSP calculations. sMILES SSP calculations are inconsistent in solar abundance scales, in that the model stellar fluxes (both atmospheric structures and the spectral synthesis calculations) are computed assuming \cite{Asplund2005} abundances whereas BaSTI isochrones are calculated with \cite{Grevesse93} abundances.

With these differences in mind, sMILES model predictions are mainly compared to the models of \citetalias{Vaz2015}, with some limited comparisons made between sMILES, \citetalias{Vaz2015} and \cite{Conroy18} models later in Section~\ref{sMILES_Vaz_Conroy}.
\begin{table*}
	\centering
\caption{Adopted components used in the generation of stellar populations models, for this work (sMILES), \citetalias{Vaz2015} and \protect\cite{Conroy18}.}
	\begin{tabular}{cp{50mm}p{40mm}p{40mm}}
		\hline
		Model & Stellar Libraries & Isochrones & Solar Abundance Reference\\
		\hline
		sMILES  & sMILES (\citealt{Knowles21}), - Empirical MILES library (with [Fe/H] and [Mg/Fe] measures) + stellar corrections from theoretical stellar spectra based on ATLAS9 model atmospheres (\citealt{Kurucz1993}, \citealt{Mezaros2012}) & Scaled-solar isochrones (\citealt{Pietrinferni04}) for [$\alpha$/Fe]=$-$0.20, 0.0 and $+$0.20 SSPs.\newline $\alpha$-enhanced isochrones (0.4) (\citealt{Pietrinferni06}) for [$\alpha$/Fe]=$+$0.40 and $+$0.60 SSPs & Stellar model component - \cite{Asplund2005} \newline  Isochrone Component - \cite{Grevesse93}\\
    &  & &  \\
		\citetalias{Vaz2015}  & Empirical MILES library (with [Fe/H] and [Mg/Fe] measures) and corresponding SSPs + SSP corrections from \cite{Coelho05,Coelho07} based on \cite{Castelli03} and \cite{Plez92} model atmospheres & Scaled-solar isochrones (\citealt{Pietrinferni04}) for [$\alpha$/Fe]= 0.0 SSPs.\newline  $\alpha$-enhanced isochrones (0.4) (\citealt{Pietrinferni06}) for [$\alpha$/Fe]=$+$0.40 SSPs & Stellar model component - \cite{Grevesse98} \newline Isochrone component - \cite{Grevesse93} \\
  &  & & \\
  \cite{Conroy18}  & Empirical MILES  and Extended IRTF libraries (with [Fe/H] measures and adopted abundance patterns) and corresponding SSPs + SSP corrections based on Kurucz model atmosphere and stellar spectral predictions (\citealt{Kurucz1979}; \citealt{Kurucz1981}; \citealt{Kurucz1993})  & MIST scaled-solar isochrones (\citealt{Choi16}; \citealt{Dotter16}) & Stellar model component - \cite{Asplund2009} \newline Isochrone component - \cite{Asplund2009}\\
		\hline
	\end{tabular}
 \label{SSP_Differences}
\end{table*}

\subsubsection{Age}
\begin{figure}
 \includegraphics[width=\linewidth, angle=0]{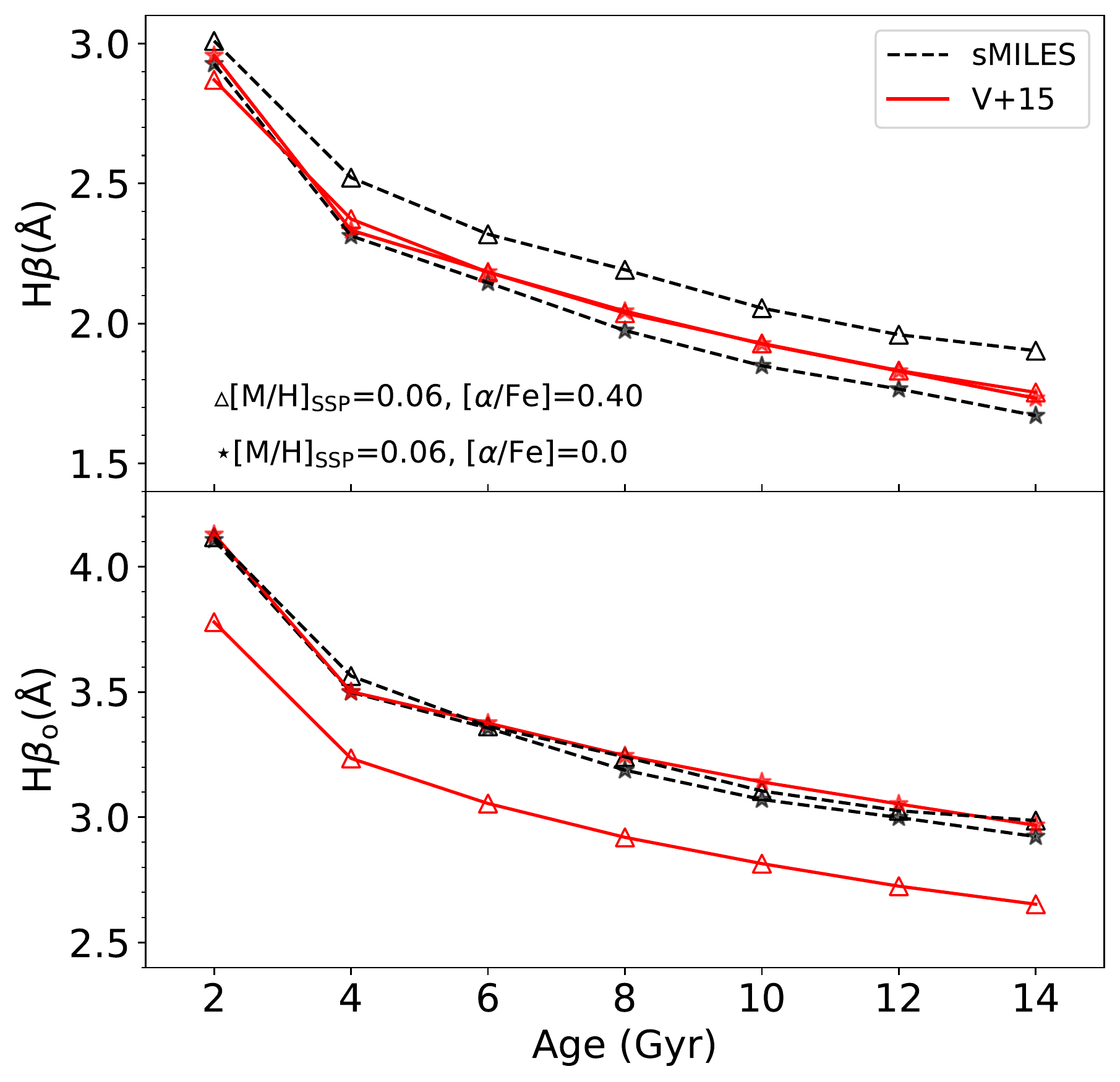}
 \centering
 \caption{sMILES (black points and dashed-lines) and \citetalias{Vaz2015} (red points and soild-lines) SSP model predictions of H$\beta$ (Top panel) and H$\beta_{\textrm{o}}$ (Bottom panel) index variations with age for solar metallicity, Universal Kroupa IMF SSPs at different [$\alpha$/Fe] values. The star and triangular points represent scaled-solar and $\alpha$-enhanced populations, respectively. Index values are measured at MILES FWHM (2.5{\AA}) resolution.}
\label{sMILES_Vaz_SSP_Age}
\end{figure}

In Figure~\ref{sMILES_Vaz_SSP_Age} we show sequences of both sMILES and \citetalias{Vaz2015} SSP predictions for H$\beta$ and H$\beta_{o}$ indices, for varying age and [$\alpha$/Fe] abundance, with fixed solar metallicity ([M/H]$_{\textrm{SSP}}$=0.06) and universal Kroupa IMF.

For [$\alpha$/Fe]=0.0 and [M/H]$_{\textrm{SSP}}$]=0.06 populations (star points), sMILES SSPs show approximately the same decrease with age for these features, compared with the models of \citetalias{Vaz2015}. For the same parameters, \citetalias{Vaz2015} and sMILES models predict a decrease of 1.25{\AA} and 1.26{\AA} respectively, in H$\beta$ with a change of 2 to 14 Gyr. For H$\beta_{o}$, \citetalias{Vaz2015} and sMILES models predict a decrease of 1.16{\AA} and 1.19{\AA} respectively for the same age increase. The similarity in index strength and strength change with age presented gives confidence in the sMILES models in this part of parameter space. It is as expected because around [$\alpha$/Fe]=0.0 and [M/H]$_{\textrm{SSP}}$]=0.06, the SSP predictions are constructed mainly from empirical stars, for both sMILES and \citetalias{Vaz2015} models.

We now test a region of parameter space in which the SSP construction becomes more reliant on the underlying differential corrections to empirical stars and SSPs. Also in Figure~\ref{sMILES_Vaz_SSP_Age}, we plot sMILES and \citetalias{Vaz2015} SSP predictions of H$\beta$ and H$\beta_{o}$ changes with age, for [$\alpha$/Fe]=0.4 populations (triangular points). The [$\alpha$/Fe]-enhancement requires differential corrections, which are performed on individual MILES stars in sMILES models and on an SSP level in \citetalias{Vaz2015} models. At [$\alpha$/Fe]=0.4, the sMILES models predict stronger index strengths than \citetalias{Vaz2015} models, at all ages for both H$\beta$ and H$\beta_{o}$. Despite this offset, the change of index strength with age is similar in both models. For H$\beta$, sMILES models predict a change of index strength from 3.01 to 1.90{\AA} for a change in age from 2 to 14 Gyr, whereas \citetalias{Vaz2015} models predict a change of index strength from 2.87 to 1.75 for the same change in age. For H$\beta_{o}$, sMILES models predict a change of index strength from 4.12 to 2.99{\AA} and \citetalias{Vaz2015} models predict a change of index strength from 3.87 to 2.65, for the same change in age. sMILES and \citetalias{Vaz2015} models therefore have similar predictions of the effect of age changes on SSPs, in this region of parameter space, albeit with an offset in absolute predictions. This offset highlights a difference in model behaviour. A change in [$\alpha$/Fe] causes a significant increase in H$\beta$ and a minor increase in H$\beta{o}$ for sMILES models, whereas \citetalias{Vaz2015} models predict a significant decrease of H$\beta{o}$ with increasing [$\alpha$/Fe] and only a minor change (an increase or decrease for different ages) in H$\beta$. Interestingly, sMILES and \citetalias{Vaz2015} predict the same effect of [$\alpha$/Fe] to other Balmer line index predictions (H$\delta_{\mathrm{A\&F}}$ and H$\gamma_{\mathrm{A\&F}}$), with indices increasing for an $\alpha$-enhancement.

In the Supplementary Materials (Figure S2), we investigate these differences of index predictions further by showing the ratio of [$\alpha$/Fe]-enhanced and solar abundance SSP spectra in the region of H$\beta$ and H$\beta_{o}$ indices and over the broader region of Balmer features. Differences are particularly seen in spectral features found in the red pseudocontinuum bands of H$\beta$ and H$\beta_{o}$ indices.
\begin{figure}
 \includegraphics[width=\linewidth, angle=0]{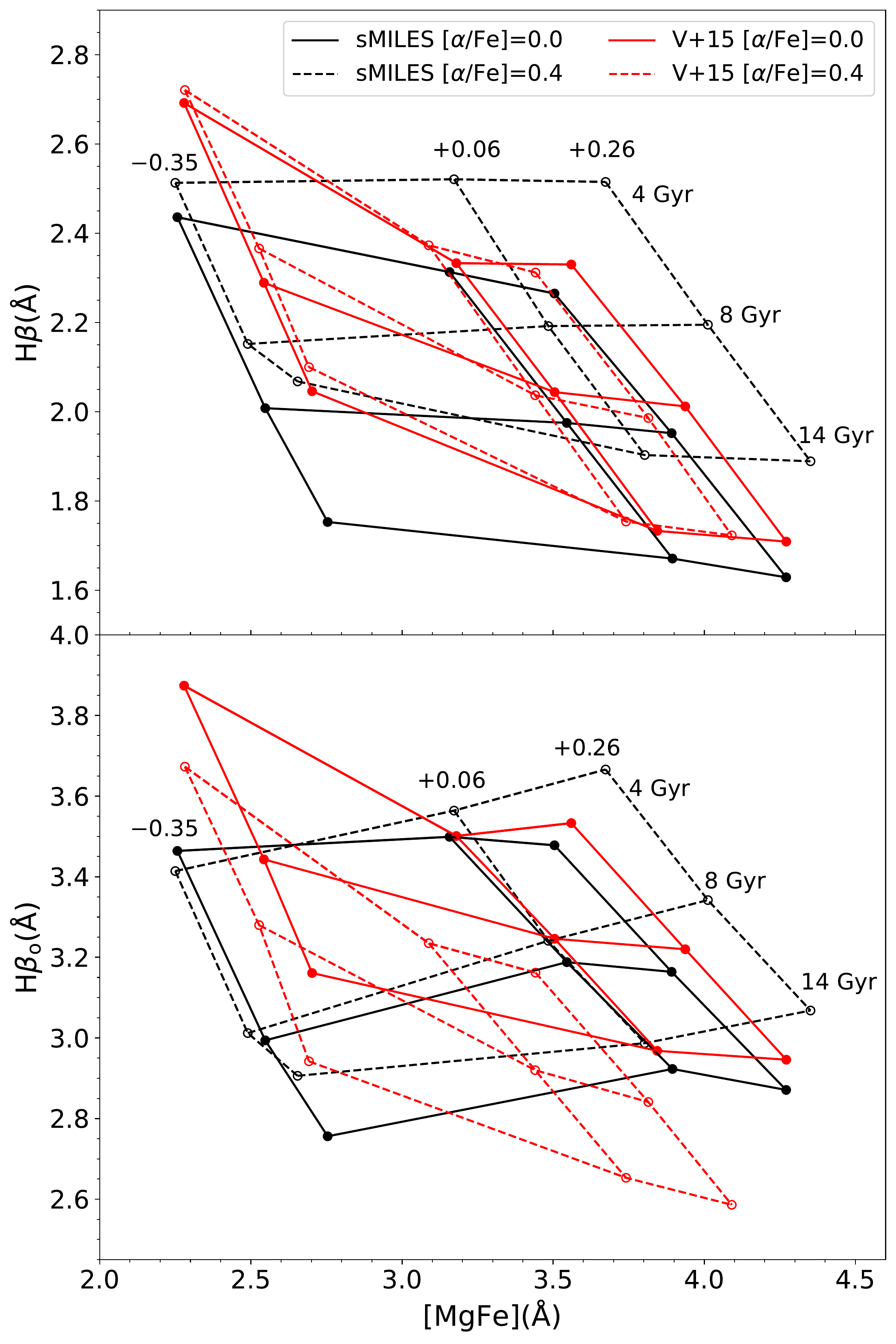}
 \centering
\caption{sMILES and \citetalias{Vaz2015} SSP predictions of H$\beta$, H$\beta_{\textrm{o}}$ and [MgFe] line strengths for varying age, metallicity and [$\alpha$/Fe]. Grids show the effect of changing total metallicity  ($\mathrm{[M/H]}_{\mathrm{SSP}}$, moving from left to right in each grid) and
age (increasing from top to bottom in each grid), respectively, as labelled. SSPs are calculated assuming a Universal Kroupa IMF. Index values are measured at MILES FWHM (2.5{\AA}) resolution.}
 \label{sMILES_V15_HbetaHbetao_Grid}
\end{figure}
To compare the sMILES and \citetalias{Vaz2015} models across a wider parameter space, we plot their predictions of H$\beta$ and H$\beta_{\textrm{o}}$ when varying the parameters of age, metallicity and [$\alpha$/Fe] together in Figure~\ref{sMILES_V15_HbetaHbetao_Grid}. We show the variation of H$\beta$ (top panel) and H$\beta_{\textrm{o}}$ (bottom panel) with [MgFe] (defined in Section~\ref{sec:MgFe}) for changes in age, total metallicity and [$\alpha$/Fe] abundance. The same results as shown in Figure~\ref{sMILES_Vaz_SSP_Age} are found for all ages and metallicities shown, with sMILES models showing a greater change to H$\beta$ with [$\alpha$/Fe] than H$\beta_{o}$, highlighted by the larger separation of [$\alpha$/Fe]=0.0 and 0.4 grids between indices. \citetalias{Vaz2015} models show the opposite behaviour; a greater sensitivity of H$\beta_{o}$ than H$\beta$ to [$\alpha$/Fe] changes.

As the main input distinction between SSPs computed in this work and those of \citetalias{Vaz2015} is the stellar models adopted in the differential correction, the cause of the H$\beta$ and H$\beta_{\textrm{o}}$ differences is anticipated to be found by comparing the \cite{Knowles21} and \cite{Coelho05,Coelho07} predictions. In Section~\ref{sMILES_Coelho_StellarMods}, we compare the Balmer line spectral and index predictions of these stellar model sets.
\subsubsection{Stellar Models}
\label{sMILES_Coelho_StellarMods}
To investigate causes of the different predictions from \citetalias{Vaz2015} and sMILES SSP models in the H$\beta$ region, we compare the underlying theoretical stellar models used in the SSP calculations. \citetalias{Vaz2015} models are constructed using predictions of \cite{Coelho05,Coelho07} stellar models and sMILES SSPs are computed from predictions of the theoretical stellar library presented in \cite{Knowles21}. It has been previously noted that different stellar models can affect predictions of H$\beta$ and H$\beta_{\textrm{o}}$ changes with [$\alpha$/Fe] (e.g. \citealt{Cervantes09}; \citetalias{Vaz2015}), possibly due to differences in the line lists adopted for those computations. In Appendix~\ref{sec:HbetaAppendix} we show the change of H$\beta$ and H$\beta_{\textrm{o}}$ with an [$\alpha$/Fe] enhancement for five typical stars present within a stellar population for \cite{Coelho05,Coelho07} and \cite{Knowles21} models. Both sets of models were degraded and rebinned to the MILES FWHM and sampling of 2.5{\AA} and 0.9{{\AA}}, respectively. \cite{Knowles21} models predict an increase of both indices with [$\alpha$/Fe] for all stars, resulting in a net increase of indices in the sMILES stars and resulting SSPs, whereas \cite{Coelho05,Coelho07} models predict a mixture of increasing and decreasing index strength for different stars, resulting in a net increase or decrease on the SSP level depending on the weighting of those stars in the isochrone integration. In the Supplementary Materials (Figure S3), we also show these differences on the spectral level by plotting the ratios of an [$\alpha$/Fe] enhanced and a scaled-solar abundance theoretical stellar spectrum for both \cite{Coelho05,Coelho07} and \cite{Knowles21} models in the region of H$\beta$ and H$\beta_{\textrm{o}}$ and for a broader region of 4000-5000{\AA}. Differences of $\alpha$-enhancement predictions can also be seen at the star level for three star types tested in those materials.

\subsubsection{Metallicity}
Similar behaviours between model sets (sMILES and V+15 SSPs) are found in terms of changes of index strength with varying metallicity. This is shown in Figure S4 of the Supplementary Materials for Mgb, Fe5335 and Fe5270 indices for 10 Gyr old populations at [$\alpha$/Fe]=0.0 and at [$\alpha$/Fe]=0.4. Differences found are always less than 0.3{\AA}. This demonstrates the similarity of the two differential correction methods, and underlying stellar models that predict the correction, in these regions of parameter space.

\subsubsection{[$\alpha$/Fe]}
\begin{figure}
 \includegraphics[width=\linewidth, angle=0]{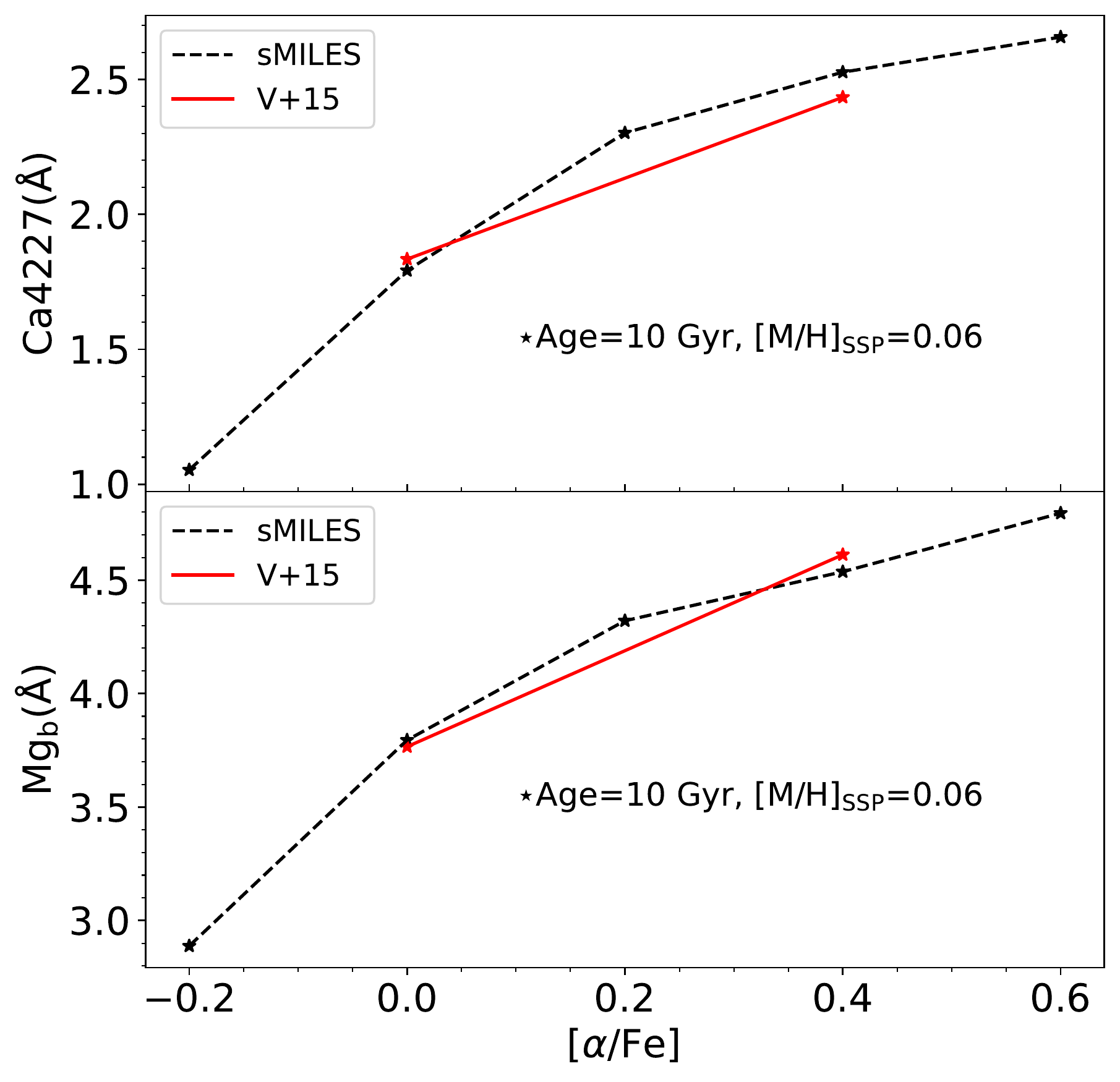}
     \caption{sMILES and \citetalias{Vaz2015} SSP predictions of Mg$_{\textrm{b}}$ (Top panel) and Ca4227 (Bottom panel) index variations with changing [$\alpha$/Fe], for a solar metallicity population. The SSPs are 10 Gyr old, with a Universal Kroupa IMF. Index values are measured at MILES FWHM (2.5{\AA}) resolution.}
 \label{sMILES_Vaz_SSP_Alpha_Compare}
\end{figure}

 sMILES SSP models cover a wider range of [$\alpha$/Fe] ($-$0.2 to $+$0.6) than the \citetalias{Vaz2015} models (0.0 to 0.4), therefore comparisons can only be made for the two [$\alpha$/Fe] points sampled in the latter. In Figure~\ref{sMILES_Vaz_SSP_Alpha_Compare} we show model predictions of changes in Ca4227 and Mg$_{\textrm{b}}$ indices with an increase in [$\alpha$/Fe] abundance, for 10 Gyr, solar metallicity populations. The sMILES models reveal a non-linear increase in these line strengths with increasing [$\alpha$/Fe]. For [$\alpha$/Fe] enhancements from scaled-solar to 0.4, the change of both indices is similar in both sMILES and\citetalias{Vaz2015} model sets. sMILES models predict an increase of 1.79 to 2.53{\AA} and 3.79 to 4.54{\AA} in Ca4227 and Mg$_{\textrm{b}}$, respectively, whereas \citetalias{Vaz2015} models predict changes of 1.83 to 2.43{\AA} for Ca4227 and 3.77 to 4.61{\AA} for Mg$_{\textrm{b}}$. This demonstrates that in this part of parameter space, the methods of differential corrections on SSPs compared with corrections on individual stars are similar, in addition to the similarity between stellar model predictions of \cite{Coelho05,Coelho07} and \cite{Knowles21} for Ca4227 and Mg$_{\textrm{b}}$ variations with changes in [$\alpha$/Fe]. The sensitivity of Mg$_{\textrm{b}}$ to [$\alpha$/Fe] abundances may appear smaller than expected. This can be explained due to the total metallicity ([M/H]$_{\textrm{SSP}}$) being fixed for changing abundance patterns in the SSP calculations. To fix [M/H]$_{\textrm{SSP}}$, the [$\alpha$/Fe] enhanced models are deficient in other element abundances, such as iron, that can also have an impact on the  Mg$_{\textrm{b}}$ index (e.g. \citealt{Korn2005}). We discuss this further in Section~\ref{Mg1}, where we show carbon abundance effects on the $\mathrm{Mg}_1$ index.
\subsubsection{[MgFe] and [MgFe]'}
\label{sec:MgFe}
Two index combinations widely used in the study of integrated stellar populations are the total metallicity-sensitive indices of [MgFe] and [MgFe]'. These indices, defined in \cite{Gonzalez93} and \cite{Thomas03Mod}, respectively, are combinations of $\textrm{Mg}_\textrm{b}$, Fe5270 and Fe5335.

\begin{figure*}
\centering
 \includegraphics[width=174mm,angle=0]{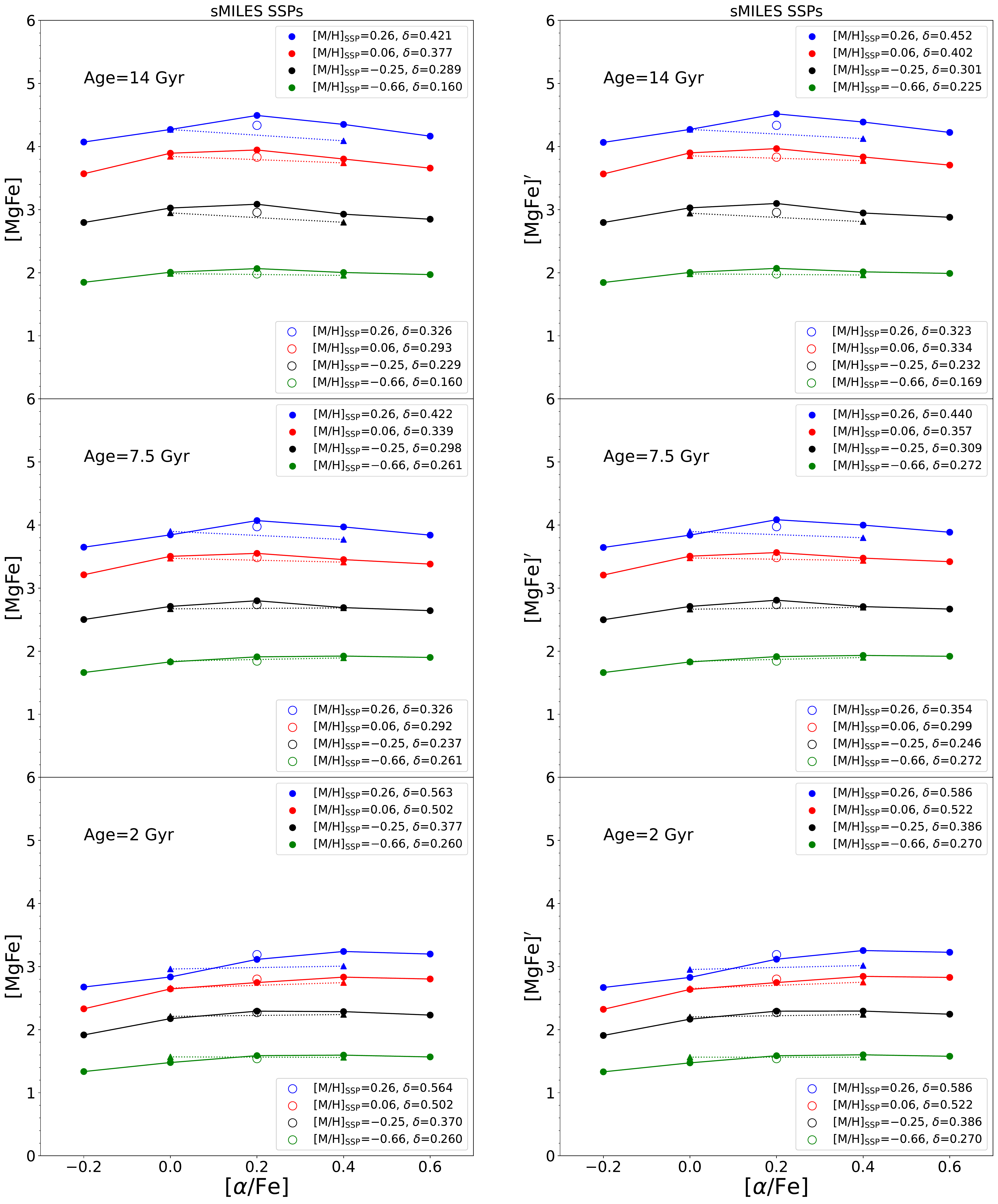}
\caption{Difference in [MgFe] and [MgFe]' index values between sMILES SSP models that compute the [$\alpha/$Fe]=0.2 model with a scaled-solar or $\alpha$-enhanced ([$\alpha/$Fe]=0.4) isochrone, for three age and four metallicity bins. Left and right panels show the [MgFe] and [MgFe]' values, respectively. Closed symbols represent sMILES SSPs models that have [$\alpha/$Fe]=$-$0.2, 0.0 and 0.2 computed with scaled-solar isochrones, and [$\alpha/$Fe]=0.4 and 0.6 computed with $\alpha$-enhanced ([$\alpha/$Fe]=0.4) isochrones. Open symbols represent [$\alpha/$Fe]=0.2 models computed with the $\alpha$-enhanced ([$\alpha$/Fe]=0.4) isochrones.  The difference between the maximum and minimum values of these indices ($\delta$) for the range of [$\alpha$/Fe] in the metallicity and age bin is given in units of {\AA}. Index values are measured at MILES FWHM (2.5{\AA}) resolution. For comparison we also show points from \citetalias{Vaz2015} models as triangles (see Section 4 of the Supplementary Materials) and Table~\ref{sMILES_vs_Vaz_MgFeMgFeDash_Tab} for additional comparisons).}
\label{sMILES_SSP_alpha02_isochrone_choice}
\end{figure*}

We compare sMILES SSP predictions of [MgFe] and [MgFe]' to those previously calculated with \citetalias{Vaz2015} models. Both indices were found to be almost insensitive to [$\alpha$/Fe] abundance in \citetalias{Vaz2015} (their figure 14) and \cite{Thomas03Mod} (their figure 7 for an old, solar metallicity model) SSP models. Both of these models took a semi-empirical approach to account for [$\alpha$/Fe] variations, with \citetalias{Vaz2015} performing differential corrections through ratios of theoretical SSP spectra, whereas \cite{Thomas03Mod} modified Lick indices through response functions. We test for the full range of [$\alpha$/Fe] values computed in this work in Figure~\ref{sMILES_SSP_alpha02_isochrone_choice} and include the two [$\alpha$/Fe] points computed in \citetalias{Vaz2015} (triangular points). We show the differences in [MgFe] and [MgFe]' indices between sMILES and \citetalias{Vaz2015} models for 2, 7.5 and 14 Gyr old stellar populations at various metallicities. In the Supplementary Material (Figures S5 and S6) we show this comparison for just the two [$\alpha$/Fe] variations of \citetalias{Vaz2015}. Figure~\ref{sMILES_SSP_alpha02_isochrone_choice} illustrates that, for either choice of SSP models, the sensitivity of [MgFe] index to [$\alpha$/Fe] variations is generally much weaker than their sensitivity to total metallicity. A similar result is found for the [MgFe]' index.

In summary, the sMILES SSP predictions of [MgFe] and [MgFe]' changes with [$\alpha$/Fe] agree well with \citetalias{Vaz2015} models, for intermediate and old SSP ages over a wide range of total metallicities. Differences exist at the youngest ages tested (2 Gyr), with sMILES models predicting larger changes in [MgFe] and [MgFe]' indices with changing [$\alpha$/Fe], compared to \citetalias{Vaz2015} models. These differences are largest at the highest metallicities tested (see Section 4 of the Supplementary Materials).

Further work to understand the origin of these differences is required, as well as comparisons to observations to determine the true sensitivity of [MgFe] and [MgFe]' indices to abundance pattern. Observations at the star level would help. If [MgFe] and [MgFe]' is measured as a function of [Fe/H] and [$\alpha$/Fe] for a large number of Milky Way stars, this would help define a correlation that in principle the SSP models should also follow. In regards to the type of stars required, AGB, main-sequence and RGB stars all have significant contributions to the total SSP light at 2 Gyr, however at the wavelength regime of [MgFe] and [MgFe]', main-sequence and RGB stars are the dominant source. Due to the chemical history of the Milky Way, trends would only be available in the age, metallicity and [$\alpha$/Fe] regimes where stars currently reside (e.g. young, metal-rich with $\sim$ solar [$\alpha$/Fe] or old, metal-poor with high [$\alpha$/Fe]. Observations of stars in other nearby systems (like those highlighted in \citealt{Sen2018}), would allow for [MgFe] and [MgFe]' trends to be obtained in other metallicity and abundance pattern regimes.

These trends with [MgFe] and [MgFe]' are explored for our wider range of [$\alpha$/Fe] computed in Figure~\ref{sMILES_SSP_alpha02_isochrone_choice} compared to \citetalias{Vaz2015} SSPs. For the highest metallicity bins in 7.5 and 14 Gyr old populations, sMILES models predict a non-linear dependence of [MgFe] and [MgFe]' to [$\alpha$/Fe] variations, such that there is an increase of line strength for increasing [$\alpha$/Fe] until a peak at [$\alpha$/Fe]=0.2, followed by a decrease in strength for increasing [$\alpha$/Fe]. This behaviour flattens to an approximately linear dependence or to no dependence at the lowest metallicity bins as well as the youngest ages. The increased sampling and range of [$\alpha$/Fe] in this work highlights these trends. The $\delta$ values show that these sMILES SSP models do have some dependence on [$\alpha$/Fe], in these overall metallicity sensitive indices, but that it may not always be a monotonic behaviour.

Given that the BaSTI isochrones available  were calculated at either scaled-solar or one $\alpha$-enhanced (0.4) abundance, a choice for our SSP calculations was the treatment of $[\alpha$/Fe]=0.2 models. In Figure~\ref{sMILES_SSP_alpha02_isochrone_choice}, we also investigate the effect of choosing a scaled-solar or $\alpha$-enhanced isochrone on these intermediate models. We show [MgFe] and [MgFe]' predictions of the sMILES SSP models for the full range of [$\alpha$/Fe] sampled, for the same age and metallicity bins as tested in Section~\ref{sec:MgFe}. The closed and open symbols in each panel shows the difference in values for models that included scaled-solar or $\alpha$=0.4 isochrones for $[\alpha$/Fe]=0.2 SSPs, respectively. Also shown are the differences ($\delta$) of maximum and minimum values of [MgFe] and [MgFe]' from the range of [$\alpha$/Fe] for each metallicity and age. The effect of using the $\alpha$=0.4 isochrone for the $[\alpha$/Fe]=0.2 SSP is a reduction in the range of [MgFe] and [MgFe]' for varying [$\alpha$/Fe] at different metallicities. This is a particularly strong effect at 14 Gyr and [M/H]$_{\textrm{SSP}}$=0.26, where the difference between the maximum and minimum value of [MgFe] is reduced from 0.421 to 0.326{\AA} and the difference in [MgFe]' is reduced from 0.452 to 0.354{\AA}. This effect is much smaller at younger ages of SSP. Thus we estimate a difference of $\sim$0.1{\AA} in [MgFe] and [MgFe]' indices can arise from using isochrones at [$\alpha$/Fe]=0.0 or $+$0.4 for the $[\alpha$/Fe]=0.2 sMILES models.

\begin{table*}
   \caption{Comparisons between sMILES and \citetalias{Vaz2015} SSP predictions for changes in Mg$_{\textrm{b}}$, Fe5270, Fe5330, [MgFe] and [MgFe]' indices for a change in [$\alpha$/Fe] of 0.4 dex, at the highest metallicity modelled in sMILES SSPs. Units of Mg$_{\textrm{b}}$, Fe5270, Fe5330, [MgFe] and [MgFe]' are given in {\AA}. The bold rows of \textbf{$\Delta$sMILES} and \textbf{$\Delta$Vazdekis} represent the sMILES and \citetalias{Vaz2015} model predictions of changes in index for a change of [$\alpha$/Fe] from 0.0 to 0.4. Index values are measured at MILES FWHM (2.5{\AA}) resolution.}
    \centering
    \begin{tabular}{|c|c|c|c|c|c|c|c|c|c|}
    \hline
         {SSP Model} & {Age (Gyr)} & [M/H]$_{\textrm{SSP}}$ & [$\alpha$/Fe] & Mg$_{\textrm{b}}$ & Fe5720 & Fe5335 & [MgFe] & [MgFe]'\\
        \hline
       sMILES ($\alpha$=0.0)  &  2.0 & 0.26 & 0.0 & 2.63 & 3.03 & 3.10 & 2.84 & 2.83 \\
       sMILES ($\alpha$=0.40) & 2.0 & 0.26 & 0.40 & 3.76 & 2.85 & 2.73 & 3.24 & 3.25 \\
      \textbf{$\Delta$sMILES=sMILES ($\alpha$=0.40) $-$ sMILES ($\alpha$=0.0)}  & \textbf{2.0} & \textbf{0.26} &  & \textbf{1.14} & \textbf{$-$0.18} & \textbf{$-$0.37} & \textbf{0.40} &  \textbf{0.42} \\
       &  &  &  &  &  & & \\
      \citetalias{Vaz2015} ($\alpha$=0.0) & 2.0 & 0.26 & 0.0 & 2.85 & 3.04 & 3.12 & 2.96 & 2.95\\
      \citetalias{Vaz2015} ($\alpha$=0.4) & 2.0 & 0.26 & 0.40 & 3.71 & 2.47 & 2.40 & 3.01 & 3.02 \\
     \textbf{$\Delta$Vazdekis=\citetalias{Vaz2015} ($\mathbf{\alpha}$=0.4) $-$ \citetalias{Vaz2015} ($\alpha$=0.0)} & \textbf{2.0} & \textbf{0.26} &  & \textbf{0.86} & \textbf{$-$0.57} & \textbf{$-$0.72} & \textbf{0.05} & \textbf{0.07} \\
         \hline
       sMILES ($\alpha$=0.0) & 7.5 &  0.26& 0.0& 3.98 & 3.79 & 3.82 & 3.84 & 3.84  \\
       sMILES ($\alpha$=0.4) & 7.5 & 0.26 & 0.4&5.02&3.32&3.09 &3.97 & 4.00  \\
       \textbf{$\Delta$sMILES=sMILES ($\alpha$=0.40) $-$ sMILES ($\alpha$=0.0)} & \textbf{7.5} & \textbf{0.26} &  & \textbf{1.04} &  \textbf{$-$0.47}& \textbf{$-$0.73} & \textbf{0.13} & \textbf{0.16}\\
        &  & &  &  & \\
        \citetalias{Vaz2015} ($\alpha$=0.0) & 7.5 & 0.26 & 0.0 & 4.09 & 3.78 & 3.80 &3.90 & 3.90\\
        \citetalias{Vaz2015} ($\alpha$=0.40) & 7.5 & 0.26 & 0.4 & 4.99 & 3.02 & 2.82 & 3.77 & 3.80\\
        \textbf{$\Delta$Vazdekis=\citetalias{Vaz2015} ($\alpha$=0.40) $-$ \citetalias{Vaz2015} ($\alpha$=0.0)} & \textbf{7.5} & \textbf{0.26} &  & \textbf{0.90} &  \textbf{$-$0.76}& \textbf{$-$0.98}&\textbf{$-$0.13} & \textbf{$-$0.10} \\
        \hline
        sMILES ($\alpha$=0.0) & 14.0 &  0.26& 0.0& 4.46 & 4.10 & 4.09& 4.27 & 4.27\\
        sMILES ($\alpha$=0.40) & 14.0 &  0.26& 0.4& 5.53 & 3.56 & 3.29& 4.35 & 4.39\\
        \textbf{$\Delta$sMILES=sMILES ($\alpha$=0.40) $-$ sMILES ($\alpha$=0.0)} & \textbf{14.0} & \textbf{0.26} &  & \textbf{1.07} &  \textbf{$-$0.54}& \textbf{$-$0.80}&\textbf{0.08} & \textbf{0.12} \\
        &  & &  &  & \\
        \citetalias{Vaz2015} ($\alpha$=0.0) & 14.0 & 0.26 & 0.0 & 4.51 & 4.05 & 4.04 & 4.27 & 4.27\\
        \citetalias{Vaz2015} ($\alpha$=0.40) & 14.0 & 0.26 & 0.4 & 5.41 & 3.21 & 2.97 & 4.09 & 4.13\\
        \textbf{$\Delta$Vazdekis=\citetalias{Vaz2015} ($\alpha$=0.40) $-$ \citetalias{Vaz2015} ($\alpha$=0.0)} & \textbf{14.0} & \textbf{0.26} &  & \textbf{0.90} &  \textbf{$-$0.84}& \textbf{$-$1.07} & \textbf{$-$0.18} & \textbf{$-$0.14}\\
        \hline
    \end{tabular}
    \label{sMILES_vs_Vaz_MgFeMgFeDash_Tab}
\end{table*}

\subsubsection{$\mathrm{Mg}_1$}
\label{Mg1}
\citetalias{Vaz2015} previously found that $\mathrm{Mg}_1$ indices are stronger in their scaled-solar than $\alpha$-enhanced (0.4) models at all metallicities. This was also mentioned in an application of these models to SDSS MaNGA early-type galaxies by \cite{Liu2020}, who found that the models were unable to match the observations. The reason for this behaviour was attributed the greater sensitivity to carbon than magnesium for $\mathrm{Mg}_1$, shown previously (e.g. \citealt{Korn2005}). If the total metallicity is fixed then $\alpha$-enhanced models are deficient in carbon compared with scaled-solar models, resulting in a decrease in $\mathrm{Mg}_1$ strength.

\begin{figure}
 \includegraphics[width=\linewidth, angle=0]{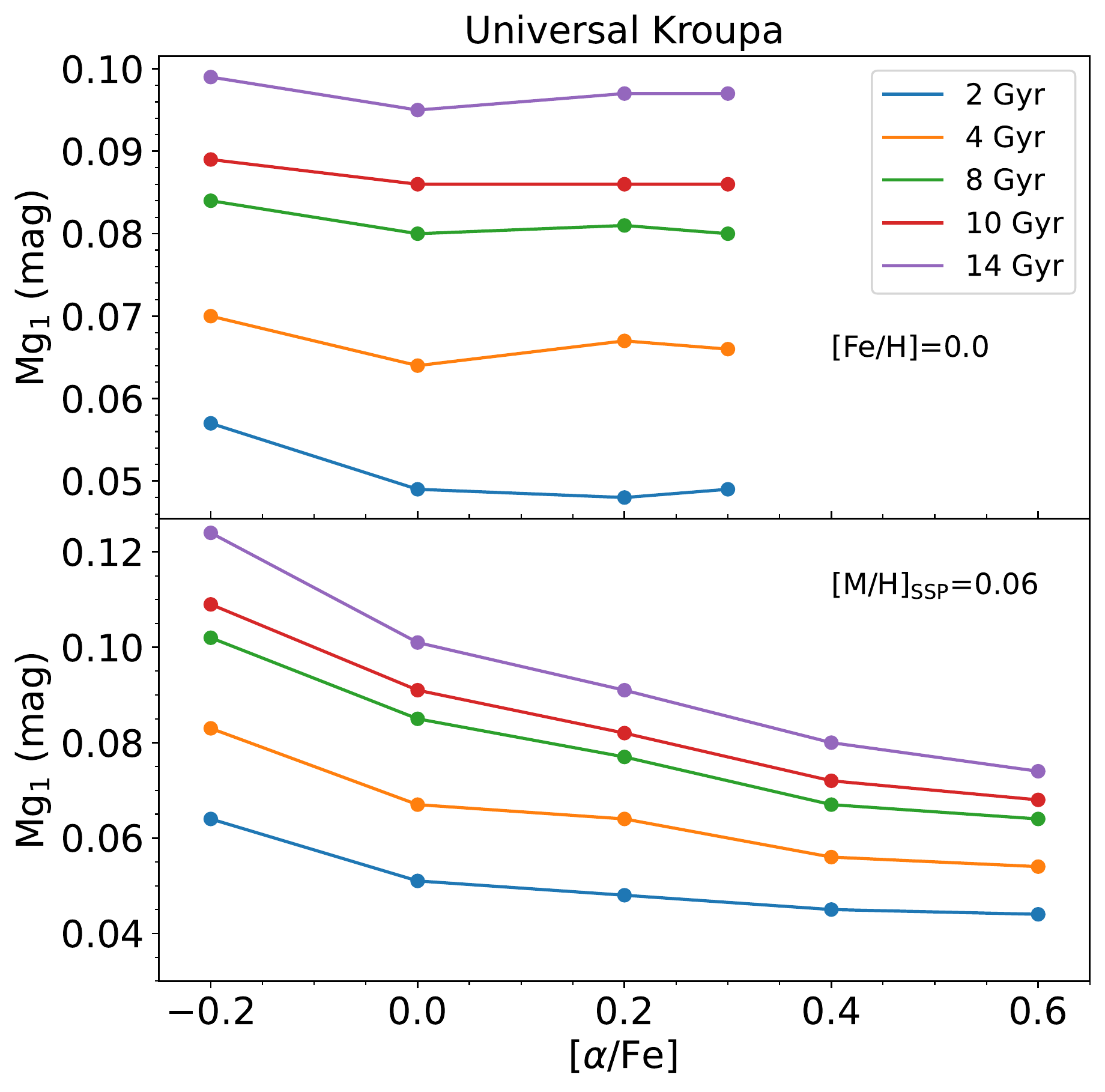}
 \caption{sMILES SSP model predictions of $\mathrm{Mg}_1$ index variations with changing [$\alpha$/Fe] at solar [Fe/H] (Top panel) and [M/H]$_{\textrm{SSP}}$ (Bottom panel), for 2, 4, 8, 10 and 14 Gyr old populations. SSPs are computed adopting a Universal Kroupa IMF and indices are measured at MILES FWHM (2.5{\AA}) resolution.}
\label{Mg1_Plot}
\end{figure}

Here, we investigate $\mathrm{Mg}_1$ index variations for the full range of [$\alpha$/Fe] available in sMILES SSP models. We find that there is a decrease in $\mathrm{Mg}_1$ index when increasing [$\alpha$/Fe] from $-$0.2 to 0.6 for ages greater than 2 Gyr, apart from the lowest metallicity bins ([M/H]$_{\textrm{SSP}}$=$-$1.79 and $-$1.49) that show a minor increase with [$\alpha$/Fe]. Using interpolations within the SSP grid, we calculate models with fixed [Fe/H]=0.0 at [$\alpha$/Fe]=$-$0.2, 0.0, 0.2 and 0.3 (as shown by Equation~\ref{MetallicityFitEqn}, [$\alpha$/Fe] abundances greater than this would require an extrapolation to [M/H]$_{\textrm{SSP}}$ values larger than 0.26). For 2, 4, 8, 10 and 14 Gyr old populations at [Fe/H]=0.0, we find a decrease in index strength from [$\alpha$/Fe]=$-$0.2 to 0.0 and then approximately a constant value for [$\alpha$/Fe] abundances from 0.0 to 0.3. We show an example of how $\mathrm{Mg}_1$ varies with [$\alpha$/Fe] for populations at fixed solar [Fe/H] and [M/H]$_{\textrm{SSP}}$ in Figure~\ref{Mg1_Plot}.

\begin{figure}
 \includegraphics[width=\linewidth, angle=0]{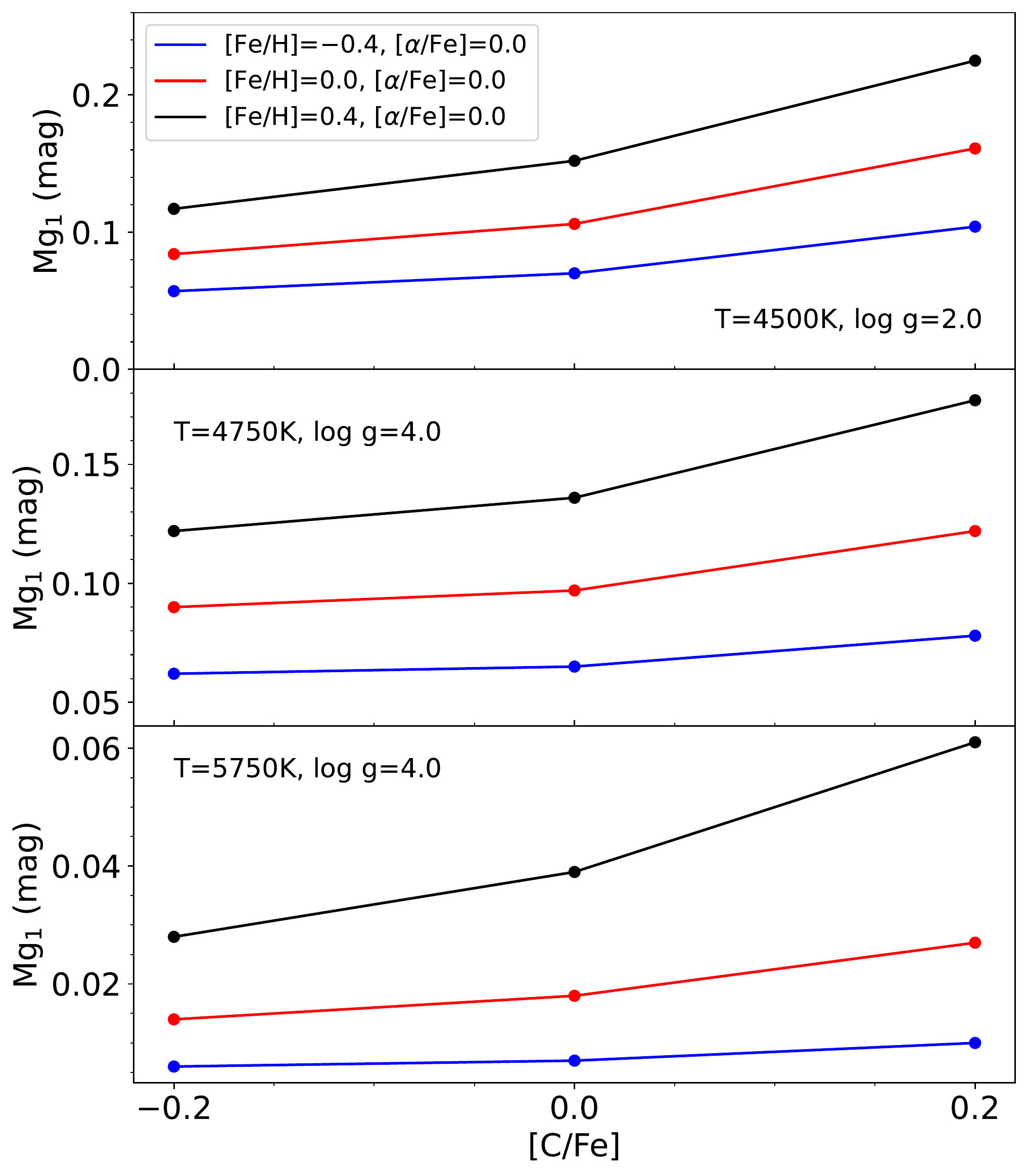}
 \caption{sMILES stellar model (\citealt{Knowles21}) predictions of $\mathrm{Mg}_1$ index variations with changing [C/Fe] at fixed [$\alpha$/Fe]=0.0 and three values of [Fe/H] for three star types. Indices are measured at MILES FWHM (2.5{\AA}) resolution.}
\label{Mg1_Plot_Star}
\end{figure}

To test this behaviour further, we investigate the change of $\mathrm{Mg}_1$ with changing both carbon and $\alpha$ abundances in the underlying stellar models presented in \cite{Knowles21}, for the same star types as Section~\ref{sMILES_Coelho_StellarMods}. We find that at three fixed values of [Fe/H] ($-$0.4 ,0, 0.4) and [$\alpha$/Fe] ($-$0.2, 0.0, 0.4), an increase in [C/Fe] from $-$0.2 to 0.2 results in an increase in $\mathrm{Mg}_1$ for the three star types. We demonstrate this effect in Figure~\ref{Mg1_Plot_Star}, where we show the $\mathrm{Mg}_1$ index with varying [C/Fe] at [$\alpha$/Fe]=0.0 and three fixed values of [Fe/H] for the three star types.  When fixing the total metallicity, we find that the T=4500{\small K}, log g=2.0 and the T=5750{\small K}, log g=4.0 stars show a decrease in $\mathrm{Mg}_1$ for an [$\alpha$/Fe] increase at all metallicities provided, whereas the T=4750{\small K}, log g=4.0 star shows a minor increase in index strength for each metallicity bin.
Fully consistent calculations of semi-empirical SSPs with variable [C/Fe] and [$\alpha$/Fe] would shed light on the relationship between carbon, $\alpha$ and the strength of the $\mathrm{Mg}_1$ at different total metallicites on an SSP level, but is beyond the scope of this work. Despite this, here we can confirm the previous findings of the \citetalias{Vaz2015} SSP models, that the strength of $\mathrm{Mg}_1$ generally decreases with increasing [$\alpha$/Fe], and extend this to the wider [$\alpha$/Fe] range modelled in sMILES SSPs. We find that [C/Fe] abundance has a greater impact on the strength of $\mathrm{Mg}_1$ than the [$\alpha$/Fe] abundance on the star level, in agreement with previous studies of this index (e.g. \citealt{Korn2005} and the response functions presented in \citealt{Knowles19}).
\subsection{sMILES, V+15, Conroy et al. Comparisons}
\label{sMILES_Vaz_Conroy}
Here, we show a limited comparison between the models of sMILES, \citetalias{Vaz2015} and \cite{Conroy18}. We compare the ratio of an $\alpha$-enhanced ([$\alpha$/Fe]=0.3) and scaled-solar SSP spectra for a solar metallicity and 9 Gyr old population.

We use the public distribution of \cite{Conroy18} models and combine their element response functions following the methodology described in \cite{Conroy2012a} (their equation 2) and \cite{Conroy18} to obtain an overall $\alpha$-enhancement of 0.3 dex. Their method to obtain an arbitrary abundance pattern is through multiplication of individual element response functions. For this work we combine the calcium, silicon, magnesium, titanium responses with a response that included an enhancement of oxygen, neon and sulphur together. We use the responses calculated at solar metallicity to obtain an 9 Gyr old, [Fe/H]=0.0 and [$\alpha$/Fe]=0.3 SSP, that is then divided by a scaled-solar abundance pattern SSP at the same metallicity and age.

With sMILES and \citetalias{Vaz2015} models computed at fixed [M/H]$_{\textrm{SSP}}$, whilst \cite{Conroy18} compute models at fixed [Fe/H], interpolations in the model sets were required to make a fair comparison for a similar [$\alpha$/Fe] enhancement. Equation~\ref{MetallicityFitEqn} here and equation 4 of \citetalias{Vaz2015} were used to calculate the required [M/H]$_{\textrm{SSP}}$ values to produce an SSP at [Fe/H]=0.0 and [$\alpha$/Fe]=0.3 to compare to the equivalent \cite{Conroy18} SSP. Hence sMILES models were interpolated to produce an SSP at [M/H]$_{\textrm{SSP}}$=0.217 and [$\alpha$/Fe]=0.3 ([Fe/H]=0.0) and \citetalias{Vaz2015} were interpolated to produce an SSP at [M/H]$_{\textrm{SSP}}$=0.225 and and [$\alpha$/Fe]=0.3 ([Fe/H]=0.0). Both of these $\alpha$-enhanced models were divided by their equivalent scaled-solar abundance pattern model to produce the same ratio to compare with \cite{Conroy18} models. \cite{Conroy18} models were converted to air wavelengths using the \cite{Ciddor96} relation. sMILES and \citetalias{Vaz2015} SSPs were degraded to match the spectral resolution of \cite{Conroy18} models ($\sigma=$100$\mathrm{km\,s^{-1}}$).

\begin{figure*}
 \includegraphics[width=\linewidth, angle=0]{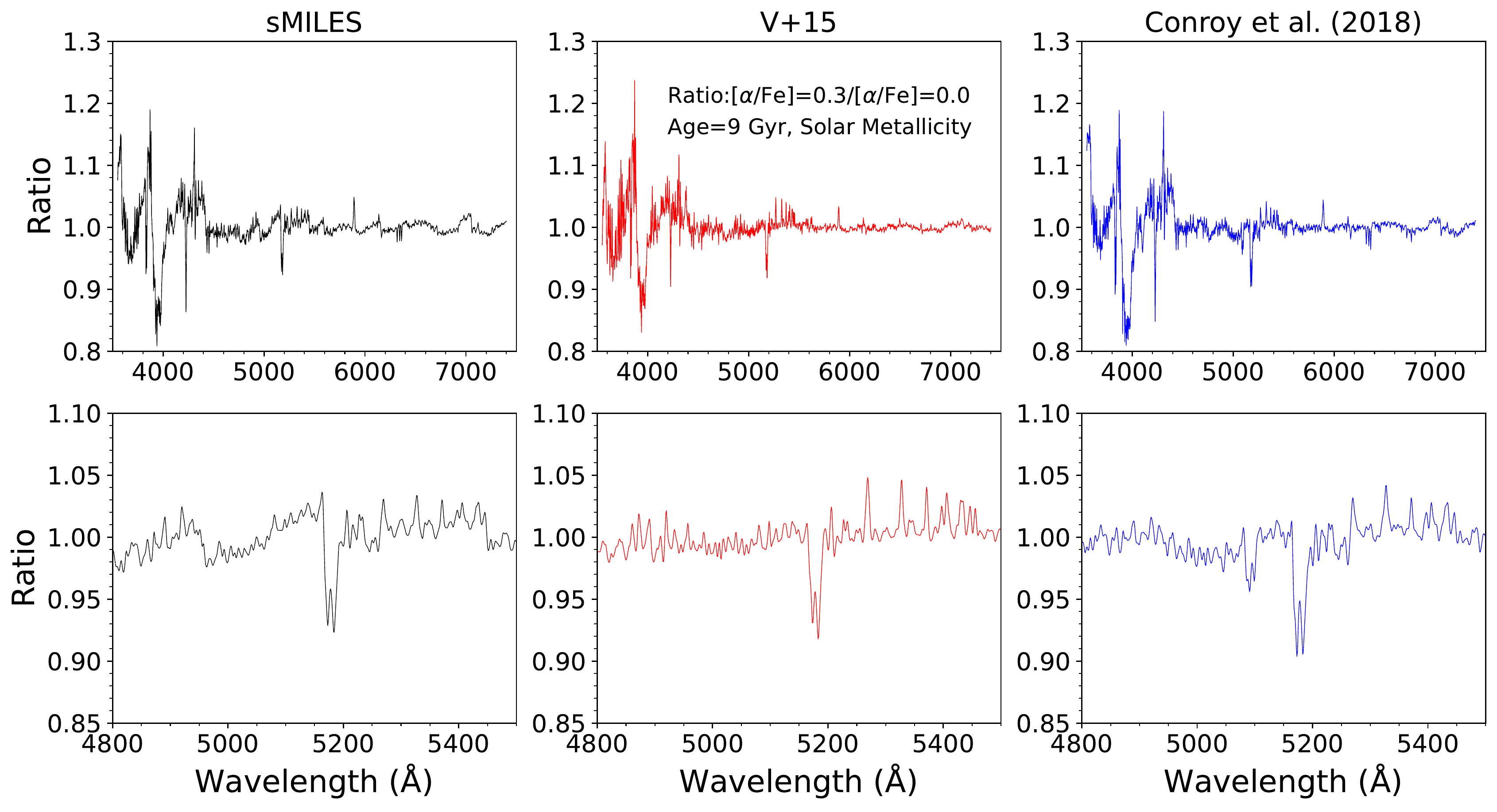}
 \caption{Ratio of an $\alpha$-enhanced to scaled-solar populations at fixed [Fe/H]=
0.0, for our models (black, left panels), \citetalias{Vaz2015} (red, middle panels) and \protect\cite{Conroy18} (blue, right panels) SSP models with an age of 9 Gyr and a Universal Kroupa IMF for the full MILES wavelength range (top panels) and a spectral region containing Mg and Fe lines (bottom panels).}
\label{sMILES_Vaz_Conroy_alpha_ratio}
\end{figure*}

Figure~\ref{sMILES_Vaz_Conroy_alpha_ratio} shows predictions for a ratio of [$\alpha$/Fe] enhanced (0.3 dex) to scaled-solar stellar population for sMILES, \citetalias{Vaz2015} and \cite{Conroy18} models at a fixed age of 9 Gyr, [Fe/H]=0.0 and Universal Kroupa IMF, for both the full optical MILES wavelength range (top panel) and for a spectral region containing magnesium and iron-sensitive features (bottom panel). Despite the numerous differences in assumptions and approaches, discussed in Section~\ref{ModelComparisons}, the overall effect of increasing [$\alpha$/Fe] on the resulting spectrum is similar and in the same sense for the three sets of models. Ratios show the same general behaviour across the full MILES wavelength range and for Mg and Fe lines, similar to what was found in \citetalias{Vaz2015} (their figure 20).

Some differences are found, originating not only from differences in the input isochrone, stellar libraries and differential correction approach but also from the required interpolations to match the sMILES and \citetalias{Vaz2015} SSPs to the fixed [Fe/H] and $\alpha$-enhancement of the \cite{Conroy18} models. In particular, the sMILES SSPs have a finer and larger sampling of [$\alpha$/Fe] compared to that of \citetalias{Vaz2015}, which affect the interpolation of SSPs to the 0.3 dex value of \cite{Conroy18} models.

 We note that for this stellar population (9 Gyr, [Fe/H]=0.0) \cite{Conroy18} and sMILES models predict increasing H$\beta$ and H$\beta_{o}$ with increasing [$\alpha$/Fe], whereas \citetalias{Vaz2015} models predict a decrease for both indices. More detailed investigations of the cause of offsets between the models is required, but is beyond the scope of this work. Next we compare sMILES SSPs to real galaxy data that has previously been analysed by the models of \cite{Vaz2010}. Such comparisons provide further tests of the models, in addition to potential future applications of sMILES SSPs.
\section{Application to Early-type Galaxies}
\label{sec:SSPEtypeGalaxies}
\subsection{SDSS Stacks of ETGs}
\begin{figure}
\centering
 \includegraphics[width=\linewidth, angle=0]{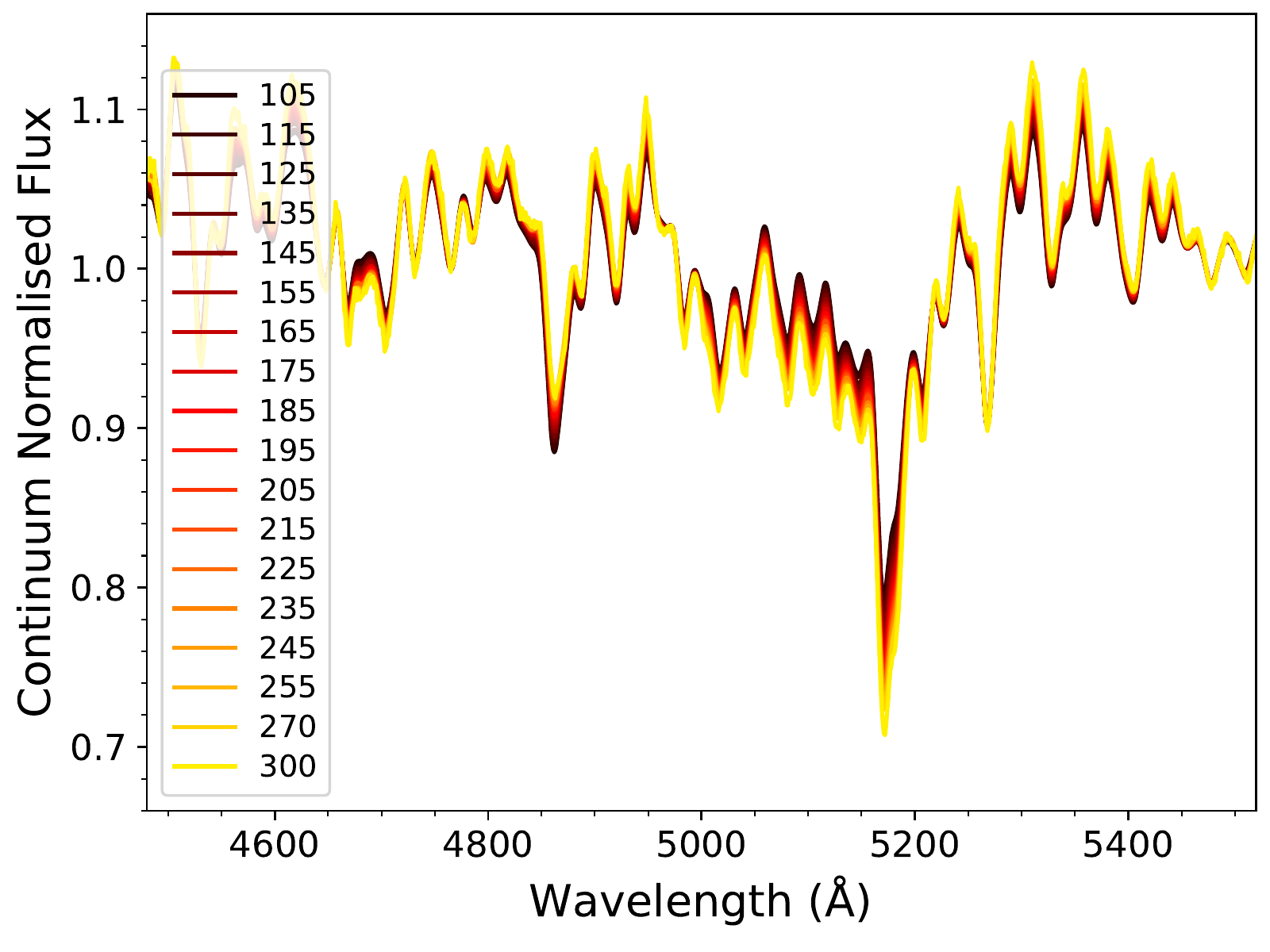}
 \caption{Continuum normalised stacked SDSS spectra, showing smooth changes with $\sigma_0$ (colour key shows $\sigma_0$ in 18 bins, in $\mathrm{km\,s^{-1}}$). See \citetalias{LaBarbera13} for description of these spectral stacks. The spectra shown are all degraded to the same resolution, corresponding to $\sigma_0$ of 300 $\mathrm{km\,s^{-1}}$. The spectral range plotted here shows H$\beta$ strongest in low $\sigma_0$ galaxies and metal lines increasing in strength with $\sigma_0$, with largest increases around Mgb and MgH.}
\label{LB13stacksNormalised}
\end{figure}

To illustrate an application of these new semi-empirical SSPs we use high signal-to-noise stacked SDSS spectra of ETGs from \citet{LaBarbera13} (hereafter - \citetalias{LaBarbera13}). Spectral data and error arrays are available at rest wavelengths, for 18 bins in central stellar velocity dispersion ($\sigma_0$), from 100 to 320 $\mathrm{km\,s^{-1}}$ (see table 1 of \citetalias{LaBarbera13} for bin definitions). These stacks include galaxies along sight lines with low Galactic extinction. See \citetalias{LaBarbera13}, for how these ETGs were selected and processed into stacks. In Figure~\ref{LB13stacksNormalised} we show versions of these spectra, degraded to 300 $\mathrm{km\,s^{-1}}$  and continuum normalised, which qualitatively illustrates the relative changes in feature strengths with changes in $\sigma_0$. Using multiple line indices we measure average stellar population (SSP) properties of the three main parameters; age, metallicity ([M/H]$_{\textrm{SSP}}$) and [$\alpha$/Fe] ratio, for each $\sigma_0$ bin. Whilst it is known that galaxies are not SSPs, we can use SSP fitting to look for relative changes in average parameters (e.g. \citealt{Proctor2002}; \citealt{LaBarbera14}; \citealt{McDermid15}).
\subsection{SSP Fitting Methods}
To measure the best fitting SSPs we developed a python code that degrades and resamples sMILES SSP models to match the SDSS stacked spectral data. The code continuum normalises both data and model spectra, measures Lick line indices and searches for the best fit using Powell minimisation available in scipy.optimize.minimize (\citealt{Scipy2020}). Continuum normalisation is done with a $9^{\mathrm{th}}$ order Chebyshev polynomial, to flatten the spectrum and allow us to focus on absorption lines. Multiple searches were performed to fit three population parameters, starting from different points in the SSP model grid of age, [M/H]$_{\textrm{SSP}}$ and [$\alpha$/Fe]. Uncertainties on these parameters were estimated from runs with these different starting locations, plus perturbations of the spectral flux data by their flux errors. The $2^{\mathrm{nd}}$ and $3^{\mathrm{rd}}$ quartiles of values were taken as the lower and upper error range for each fitted parameter. For instrument resolutions we assumed SDSS resolution as a function of wavelength (see \citetalias{LaBarbera13}, their figure 2) and MILES resolution of 2.5 {\AA} FWHM (\citealt{Falcon2011}).

\begin{figure}
\centering
\includegraphics[scale=0.55,
angle=0]{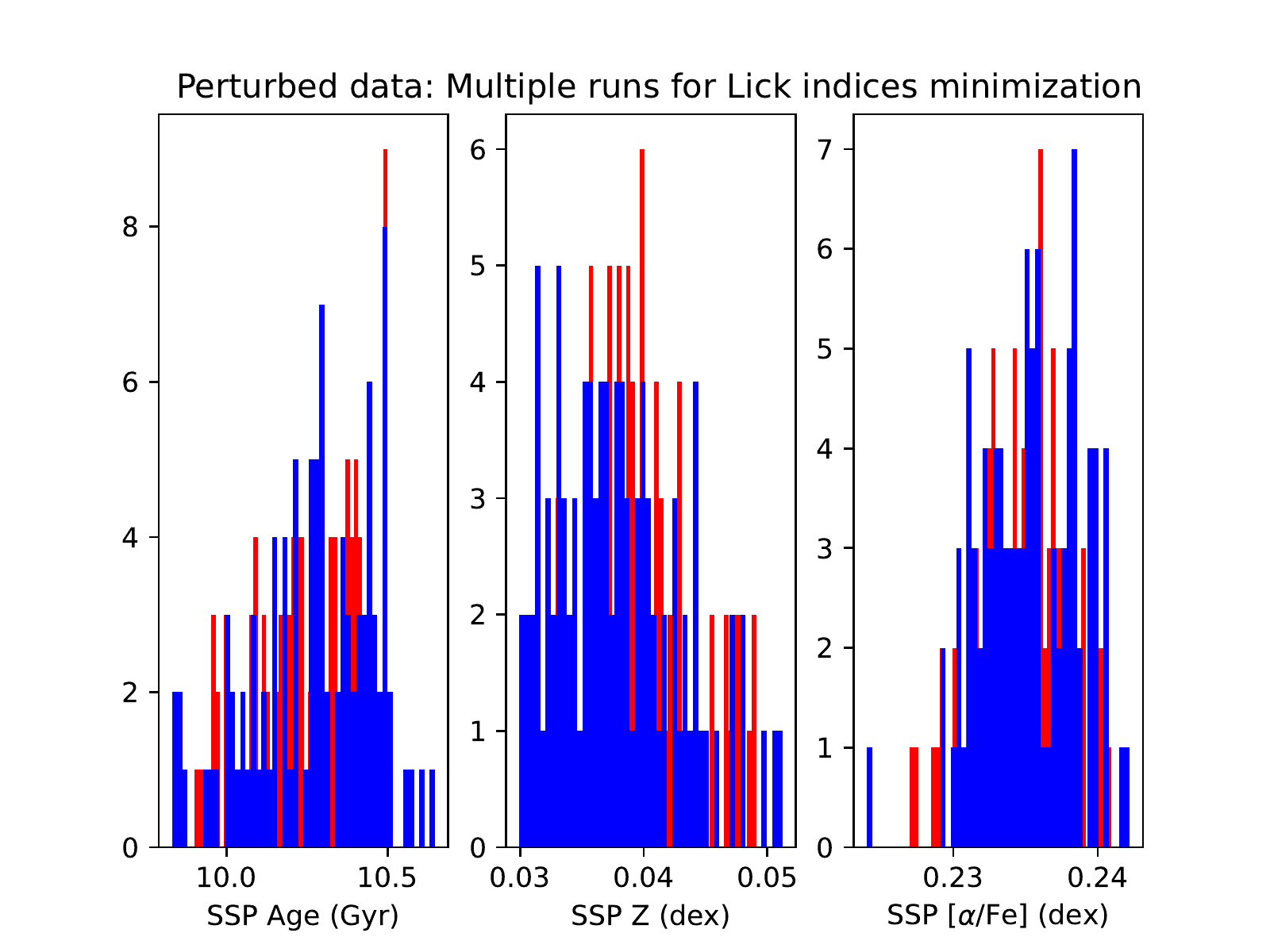}\vspace{-5pt}
\includegraphics[scale=0.55,
  angle=0]{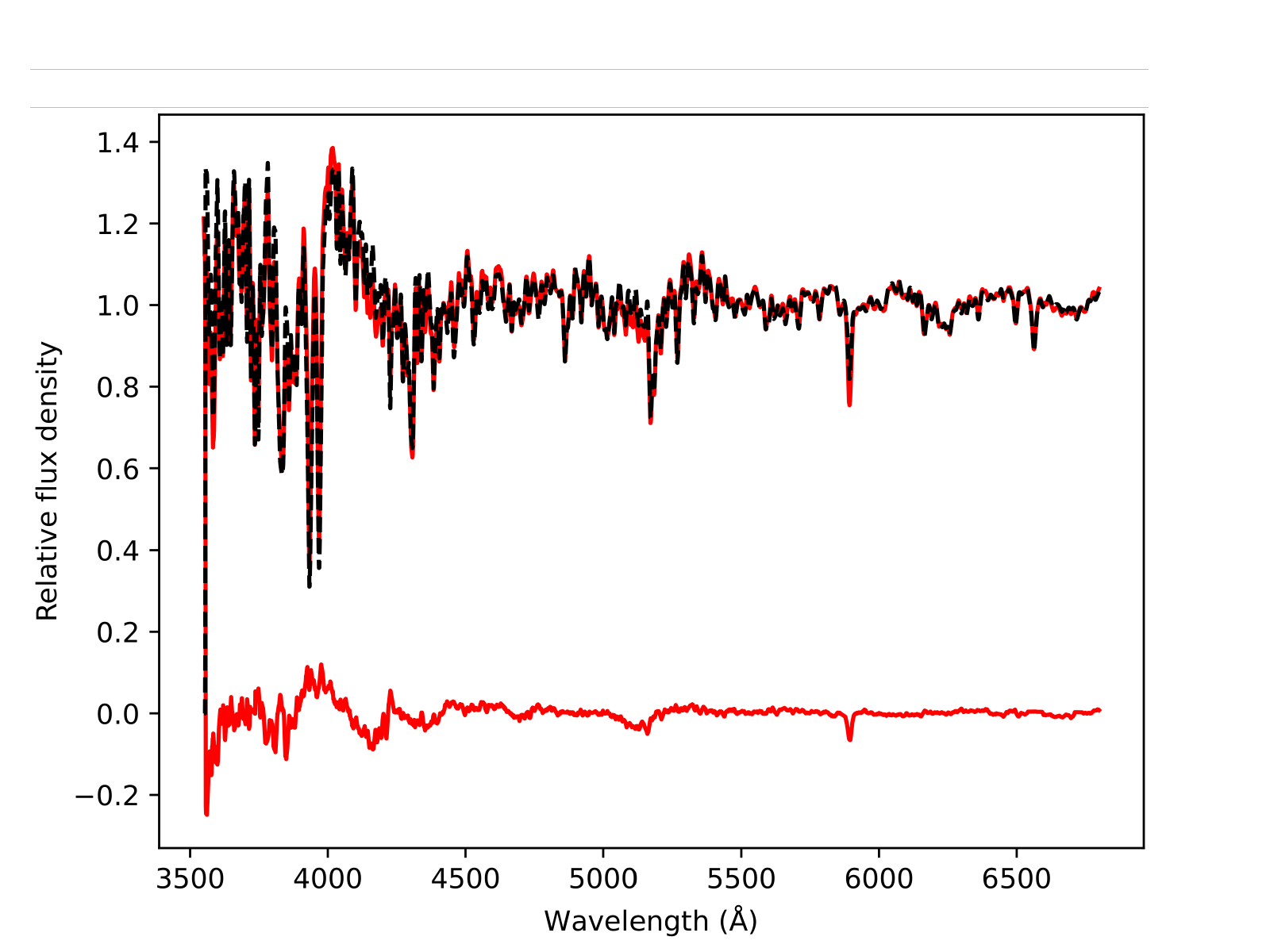}
\caption{Examples from fitting line indices to  $\sigma_0$=190-200 $\mathrm{km\,s^{-1}}$ SDSS stack. Top panel: Parameter fits for 200 perturbations, from which parameter uncertainties are estimated. Blue histograms are from fits starting at a young age (1 Gyr) and red histograms are from fits starting at an older age (7 Gyrs). Errors estimated from quartile ranges in this example are: Age range$=$(10.10 to 10.47 Gyr), Z ($=$[M/H]$_{\textrm{SSP}}$) range$=$(0.0326 to 0.0426 dex), [$\alpha$/Fe] range$=$(0.2322 to 0.2376 dex).
Bottom panel: Black line shows the data and top red line shows the model. Lower red line shows residuals for this example (data-model), which occur mostly in the blue part of the spectrum, affected most by younger contributions, and also in the Ca4227 and NaD line regions not fitted here. This best fit SSP is at Age$=$10.28 Gyr, [M/H]$_{\textrm{SSP}}$$=$0.038 dex and [$\alpha$/Fe]$=$+0.235 dex. The residuals spectrum also illustrates that emission lines were well removed in the initial run.}
\label{PerturbAndFitExample}
\end{figure}

\begin{figure}
\centering
 \includegraphics[scale=0.57, angle=0]{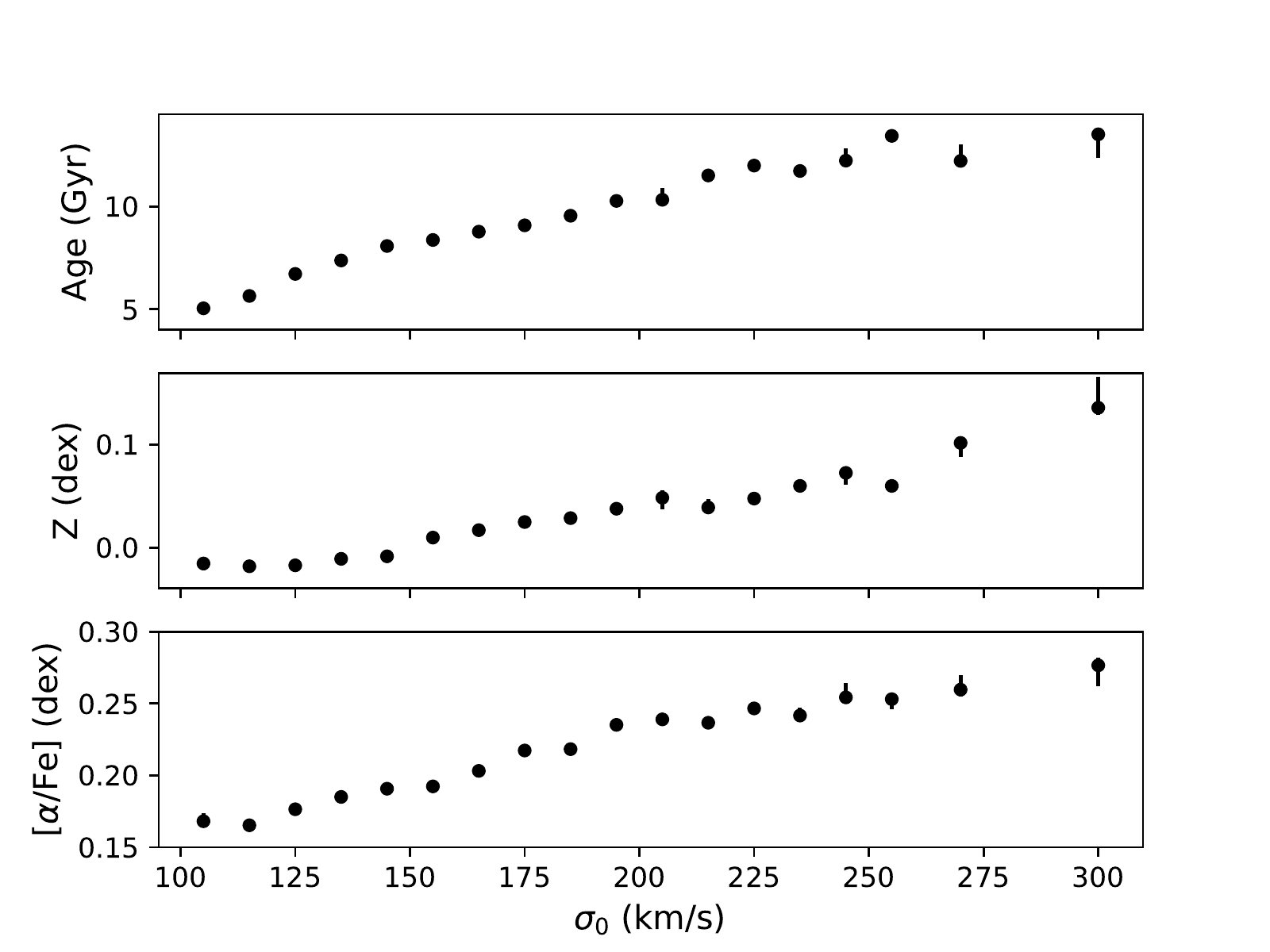}
\caption{Trends of average age, Z($=$[M/H]$_{\textrm{SSP}}$) and [$\alpha$/Fe] from SSP fits to stacked SDSS spectra of ETGs. These plots highlight smooth changes in these parameters, with increasing galaxy velocity
dispersion. Asymmetric error bars are from perturbations.}
\label{SSPfitsToSDSSstacks}
\end{figure}

On a first run through, residuals (data $-$ best-fit model) showed clear, weak emission lines that are not so evident in the original SDSS stacked spectra. Results from this first run were then used to remove emission line flux by subtracting the flux in each line from the original data, scaled to the local continuum. This process only removed flux in positive residuals located in 12 well-known emission lines from star-formation, in this spectral range. SSP fitting was then re-run, using these emission line subtracted spectra. The improvement in SSP fitting was mainly for absorption features most affected by emission line contamination, including Balmer lines. Figure~\ref{PerturbAndFitExample}  shows an example of diagnostic plots for perturbation results and residuals.
SSP fits were carried out using Lick-based line indices. Full spectrum fitting is possible but careful choice of spectral ranges would be needed to avoid biasing the fit. Hence here we start with fitting Lick indices. NaD was excluded from these fits because it is known to be sensitive to the stellar initial mass function in ETGs and can be affected
by interstellar absorption (\citetalias{LaBarbera13}). Ca4227 was also excluded because it is known to be poorly modelled (e.g. \citealt{Vaz97}; \citealt{Proctor04}; \citetalias{LaBarbera13}). Molecular band indices, plus C$_2$4668 and Fe5015 were excluded as being too broad, hence not well modelled by these continuum normalised fits. This left 15 line indices that were being fit (H$\delta_{\mathrm{A\&F}}$, G4300, H$\gamma_{\mathrm{A\&F}}$, Fe4383, Ca4455, Fe4531, H$\beta$, Mgb, Fe5270, Fe5335, Fe5406, Fe5709, Fe5782). Results are shown in Figure~\ref{SSPfitsToSDSSstacks}.

\subsection{Results for stacked SDSS spectra}
We find smooth trends of increasing stellar population age, increasing metallicity (from sub-solar to super-solar) and increasingly enhanced [$\alpha$/Fe], with $\sigma_0$. Older populations suffer increasing uncertainties from age-metallicity degeneracy, which shows up as a slightly increased scatter in the points at older ages (higher $\sigma_0$).
Whilst the absolute values of these parameters may not be accurate, the relative trends are clear in these plots, with very small systematic changes detectable between adjacent bins in $\sigma_0$. The trends found here are qualitatively similar to those found using different subsets of these data in \cite{LaBarbera14} (their figure 3), where [$\alpha$/Fe] was estimated from four Lick indices using a relative metallicity calibration. In this current work we can access the full spectral shape of the best fitting SSP, which allows for identification of regions that fit well or badly. These results illustrate the use of our new sMILES SSPs for accurately measuring trends in the stellar populations of ETGs. Future developments in stellar population modelling could benefit from incorporating such new SSP spectral libraries, that include abundance variations, into software modelling star formation histories, because we know that different types of galaxies do not follow the abundance patterns of the Milky Way Galaxy (e.g. \citealt{Sen2018,Sen2022}).
\section{Summary}
\label{sec:SummaryConclusions}
Based on our semi-empirical library of stellar spectra (\citealt{Knowles21}) we build new SSPs, covering $3540.5-7409.6\,${\AA}, for a wide range of ages, metallicities and for [$\alpha$/Fe] values of -0.2, 0.0, +0.2, +0.4 and +0.6. SSPs assuming different IMFs are made, using the same methodology and sampling used to build the original (fully empirical) MILES SSPs (\citealt{Vaz2010}), the difference being that our star spectra were modified by theoretical star spectra to sample five specific values of [$\alpha$/Fe] (see Table~\ref{sMILES_SSPmodels_Tab}).
These new sMILES SSPs are intended for use in modelling integrated light from populations of stars. This paper presents the new SSPs and how they were built. We show how some important Lick indices behave in these SSPs, with particular emphasis on their sensitivity to [$\alpha$/Fe] variations (e.g. Figures~\ref{sMILES_SSP_Alpha_Seq} and~\ref{sMILES_SSP_Mgb_vs_Fe}). Figure~\ref{sMILES_SSP_Mgb_vs_Fe} shows that [$\alpha$/Fe] can be distinguished almost independently from effects of age-metallicity degeneracy, making it a valuable tool for probing star formation timescales in integrated light from galaxies.

We compare these new SSPs with previously published ones, particularly \citetalias{Vaz2015}, and find qualitatively similar behaviour, but with some differences (explored in Figures~\ref{sMILES_Vaz_SSP_Age},~\ref{sMILES_V15_HbetaHbetao_Grid} and~\ref{sMILES_SSP_alpha02_isochrone_choice}). These differences are particularly notable for the H$\beta_{\textrm{o}}$  and H$\beta$ index, which arise from differences in the theoretical spectra used to construct SSP models (\citealt{Coelho05,Coelho07} for \citetalias{Vaz2015} SSPs and \citealt{Knowles21} for our new SSPs). More measurements of accurate abundance ratios in stars, from a wider range of star-formation histories, would help to test the accuracy of theoretical star spectra.

To illustrate the applicability of our new SSPs, we fitted them to the high signal-to-noise data of stacked SDSS galaxy spectra from \citetalias{LaBarbera13}. Age, metallicity and [$\alpha$/Fe] trends were measured for galaxy stacks with different stellar velocity dispersions. Figure~\ref{SSPfitsToSDSSstacks} illustrates the fine relative differences that can be distinguished for different classes of galaxies. The variations of these new SSP spectra with [$\alpha$/Fe] provides a useful tool for distinguishing between different star formation histories and timescales of star formation in different galaxy types. These new SSPs will be made publicly available.

\section*{Acknowledgements}
The authors thank the STFC for providing ATK with the studentship for his Ph.D studies and the IAC for providing the support and funds that allowed ATK to visit the institute on two occasions. AES and AV acknowledge travel support from grant AYA2016-77237-C3-1-P from the Spanish Ministry of Economy and Competitiveness (MINECO). AES acknowledges support from the University of Central Lancashire, Jeremiah Horrocks Institute for research visitor funding. ATK and AV acknowledge support from grant PID2019-107427GB-C32 and PID2021-123313NA-I00 from the Spanish Ministry of Science, Innovation and Universities MCIU. This work has also been supported through the IAC project TRACES, which is partially supported through the state budget and the regional budget of the Consejer{\'{i}}a de Econom{\'{i}}a, Industria, Comercio y Conocimiento of the Canary Islands Autonomous Community.
ATK also acknowledges support from the ACIISI, Consejer{\'{i}}a de Econom{\'{i}}a, Conocimiento y Empleo del Gobierno de Canarias and the European Regional Development Fund (ERDF) under grant with reference ProID2021010079. CAP thanks MICINN for grants AYA2017-86389-P and PID2020-117493GBI00. We also thank the referee for their comments and suggestions that have greatly improved the clarity and content of this work.
\section*{Data Availability}
The theoretical stellar spectral library, at fixed spectral sampling, and the semi-empirical stellar library presented in \cite{Knowles21} are publicly available on the UCLanData repository ({\url{https://uclandata.uclan.ac.uk/178/}}) and MILES website ({\url{http://research.iac.es/proyecto/miles/pages/other-predictionsdata.php}}), respectively.
Additional plots and information are available in the Supplementary Materials.
Sets of sMILES SSPs are made available through the MILES website (\url{http://miles.iac.es/}) and UCLanData repository (\url{https://uclandata.uclan.ac.uk/}).

%%%%%%%%%%%%%%%%%%%%%%%%%%%%%%%%%%%%%%%%%%%%%%%%%%



\bibliographystyle{mnras}
\bibliography{sMILES_Paper}% if your bibtex file is called example.bib


%%%%%%%%%%%%%%%%%%%%%%%%%%%%%%%%%%%%%%%%%%%%%%%%%%

%%%%%%%%%%%%%%%%% APPENDICES %%%%%%%%%%%%%%%%%%%%%

\appendix

\section{Stellar Model Predictions of H$\beta$ and H$\beta_{\mathrm{o}}$}
\label{sec:HbetaAppendix}
To supplement the comparison of \cite{Knowles21} and \cite{Coelho05,Coelho07} stellar model predictions of an $[\alpha$/Fe] enhancement in Section~\ref{sMILES_Coelho_StellarMods} and Section 2 of the Supplementary Materials, in Figure~\ref{sMILES_Coelho_Stellar_HbetaHbetao_Indices} we investigate the change of H$\beta$ and H$\beta_{\textrm{o}}$ indices with an increase of $[\alpha$/Fe] from scaled-solar to 0.4 for five different star types. \cite{Knowles21} models consistently predict an increase of both indices for all stars tested, which results in an net increase of indices at the SSP level. \cite{Coelho05,Coelho07} predict a mixture of increasing and decreasing effects on the indices for different stars, which results in a net increase or decrease of the indices on the SSP level, depending on the weighting of the stars in the population model computation. This highlights the model set dependency on the predictions of  H$\beta$ and H$\beta_{\textrm{o}}$ indices, as discussed in \cite{Cervantes09} and \citetalias{Vaz2015}.
\begin{figure*}
\centering
 \includegraphics[width=155mm,angle=0]{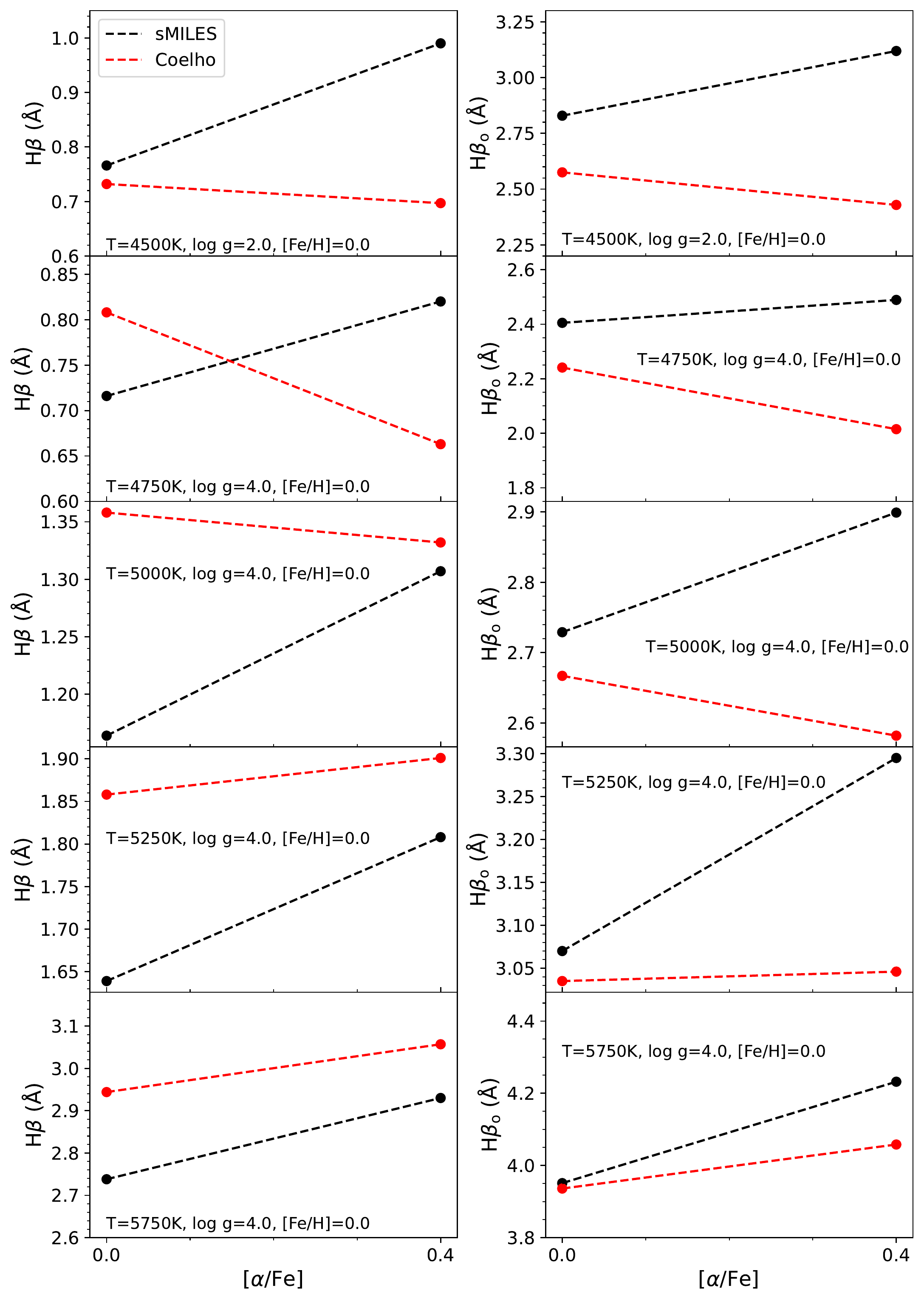}
\caption{Predictions of changes in H$\beta$ (left panel) and H$\beta_{\textrm{o}}$ (right panel) index strength due to an [$\alpha$/Fe] enhancement for \protect\cite{Coelho05,Coelho07} (red lines) and sMILES (\protect\citealt{Knowles21}) (black lines) stellar spectral models, for five star types. Index values are measured at MILES FWHM (2.5{\AA}) resolution.}
\label{sMILES_Coelho_Stellar_HbetaHbetao_Indices}
\end{figure*}

%\counterwithin{figure}{section}

%\counterwithin{figure}{section}

%\newpage

%%%%%%%%%%%%%%%%%%%%%%%%%%%%%%%%%%%%%%%%%%%%%%%%%%

% Don't change these lines
\bsp	% typesetting comment
\label{lastpage}
\end{document}

\title{\textbf{Supplementary Material - Comparisons Between sMILES and Vazdekis et al. (2015) SSP and Stellar Models}}
\date{}
\author{}
\maketitle
\defcitealias{Vaz2015}{V+15}
\noindent Supplementary Material for the article by Knowles et al. 2023, entitled: `sMILES SSPs: A Library of Semi-Empirical MILES Stellar Population Models with Variable [$\alpha$/Fe] Abundances'.
\section{Age and Metallicity Sequences of sMILES SSPs}
Figure \hyperlink{sMILES_SSP_Age_Metal}{S1} shows a sequence of sMILES SSP spectra with ages of 2 Gyr to 14 Gyr for a fixed metallicity and a sequence of sMILES SSPs from [M/H]$_{\textrm{SSP}}$=$-$1.79 to +0.26 for a fixed age (10 Gyr), with a universal Kroupa IMF. The effect of ageing a stellar population can be mimicked by increasing the metallicity of the population at a fixed age (\citealt{Worthey94}), demonstrated by the similarity in the sequences, where old, metal-poor populations look like younger, metal-rich populations. This figure also indicates that metal poor SSPs can be distinguished from their week absorption lines and their overall spectral shape, if well flux calibrated.

\begin{figurenew*}[ht]
    \includegraphics[width=165mm,angle=0]{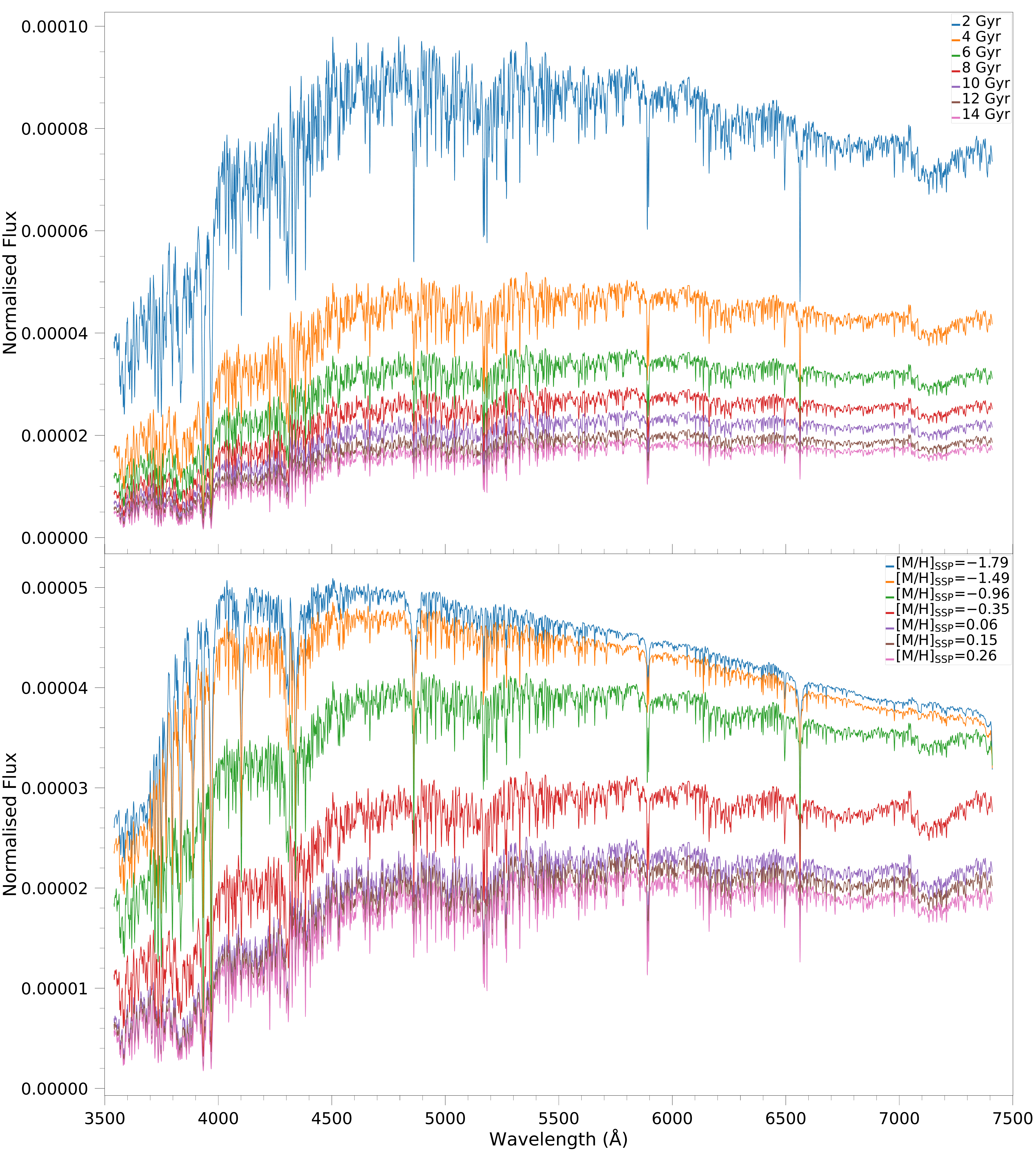}
   \caption{Top panel: sMILES SSP sequence of age for fixed metallicity ([M/H]$_{\textrm{SSP}}$=0.06) and Universal Kroupa IMF in the full MILES wavelength range. Bottom panel: sMILES SSP sequence of metallicity for a fixed age (10 Gyr) and universal Kroupa IMF.}
   \label{sMILES_SSP_Age_Metal}
\end{figurenew*}
\section{H$\beta$ and H$\beta_{o}$ predictions of sMILES and Vazdekis et al. (2015) SSP and Underlying Stellar Models}
To supplement Section 3.2.1 (Figures 8 and 9) of Knowles et al. (2023), in Figure \hyperlink{sMILES_Vaz_HBalmer_regions}{S2} we show the ratio of [$\alpha$/Fe]-enhanced and solar abundance SSP spectra in the region of H$\beta$ and H$\beta_{o}$ indices (Top panel) and over the broader region of Balmer features (Bottom panel) and highlight the Lick index definitions. There are differences between model predictions, particularly seen in spectral features found in the red pseudocontinuum bands of H$\beta$ and H$\beta_{o}$ indices. These differences lead to the differing index predictions shown in Figures 8 and 9 of Knowles et al. (2023). Spectral differences are also found for other Balmer lines but do not appear in the indices, with both models predicting an increase in index strength for increasing [$\alpha$/Fe] abundance ratio at all ages and metallicities tested.

To investigate causes of the different predictions from \citetalias{Vaz2015} and sMILES SSP models in the H$\beta$ region, we compare the underlying theoretical stellar models used in the SSP calculations. \citetalias{Vaz2015} models are constructed using predictions of \cite{Coelho05,Coelho07} stellar models and sMILES SSPs are computed from predictions of the theoretical stellar library presented in \cite{Knowles21}. It has been previously noted that different stellar models can affect predictions of H$\beta$ and H$\beta_{\textrm{o}}$ changes with [$\alpha$/Fe] (e.g. \citealt{Cervantes09}; \citetalias{Vaz2015}), possibly due to differences in the line lists adopted for those computations. In Figure \hyperlink{Coelho_vs_sMILES}{S3}, we plot the ratios of an [$\alpha$/Fe] enhanced and a scaled-solar abundance theoretical stellar spectrum for both \cite{Coelho05,Coelho07} and \cite{Knowles21} models in the region of H$\beta$ and H$\beta_{\textrm{o}}$ and for a broader region of 4000-5000{\AA}. Both sets of models were degraded and rebinned to the MILES FWHM and sampling of 2.5{\AA} and 0.9{{\AA}}, respectively. This was done for three typical stars present within a stellar population. In addition to the SSP level, differences of $\alpha$-enhancement predictions can also be seen at the star level for the three star types tested. These differences are then incorporated into the SSP calculations in different weights, depending on the age, metallicity and abundance pattern of the population.

\begin{figurenew*}[ht]
 \includegraphics[width=\linewidth, angle=0]{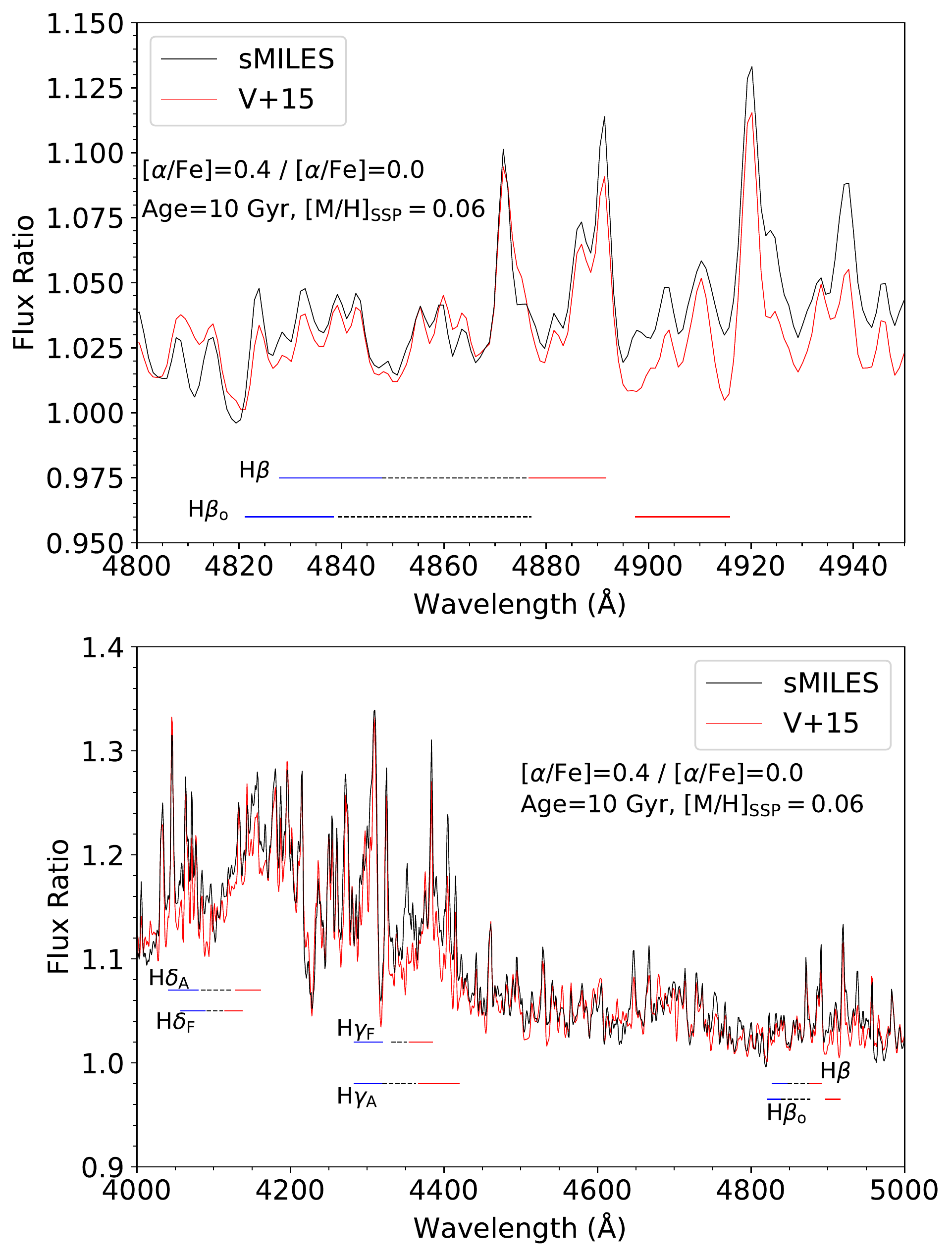}
 \centering
 \caption{sMILES (black lines) and \citetalias{Vaz2015} (red lines) SSP spectral predictions of changes due to $\alpha$-enhancements for solar metallicity and 10 Gyr populations. Top panel: For H$\beta$ and H$\beta_{\textrm{o}}$ index regions. Bottom panel: For 4000-5000{\AA}. Blue, red and feature bands of Balmer Lick index definitions are shown.}
\label{sMILES_Vaz_HBalmer_regions}
\end{figurenew*}
\begin{figurenew*}
\centering
 \includegraphics[width=\linewidth, angle=0]{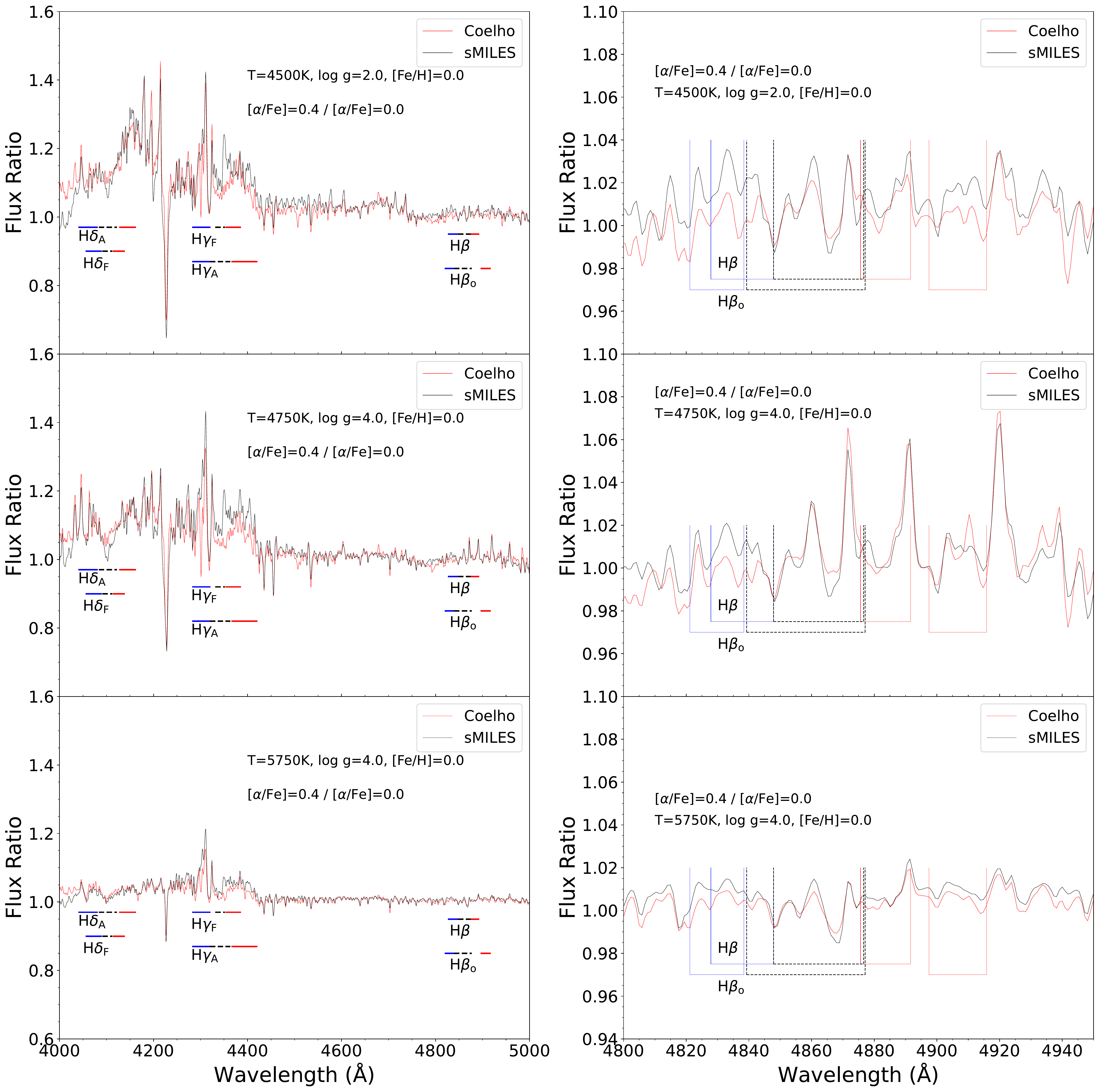}
 \caption{Predictions of changes in flux due to an [$\alpha$/Fe] enhancement for \protect\cite{Coelho05,Coelho07} (red lines) and sMILES (\protect\citealt{Knowles21}) (black lines) stellar spectral models, for three star types at MILES FWHM (2.5{\AA}) resolution. Left panel: For 4000-5000{\AA}. Right panel: For H$\beta$ and H$\beta_{\textrm{o}}$ index regions. Blue, red and feature bands of Balmer Lick index definitions are shown.}
 \label{Coelho_vs_sMILES}
\end{figurenew*}

\section{Fe5270, Fe5335 and Mg$_{\textrm{b}}$ Index Comparisons Between sMILES and Vazdekis et al. (2015) SSP Models}
To supplement Section 3.2.3 of Knowles et al. (2023) in Figure \hyperlink{sMILES_Vaz_SSP_Metal}{S4} we compare sMILES and \citetalias{Vaz2015} SSP model predictions of Fe5270, Fe5335 and Mg$_{\textrm{b}}$ index variations with changes in [M/H]$_{\textrm{SSP}}$. For 10 Gyr old, scaled-solar (star points) and [$\alpha$/Fe]=0.4 (triangular points) populations there is a good agreement in lower metallicity bins and small differences in the highest metallicity bins. For the highest metallicity bins ([M/H]$_{\textrm{SSP}}$=0.06 and [M/H]$_{\textrm{SSP}}$=0.26) sMILES models predict slightly larger index strengths than the \citetalias{Vaz2015} models, with differences of $\sim$0.2-0.3 \AA.

\begin{figurenew*}[ht]
\centering
 \includegraphics[width=120mm, angle=0]{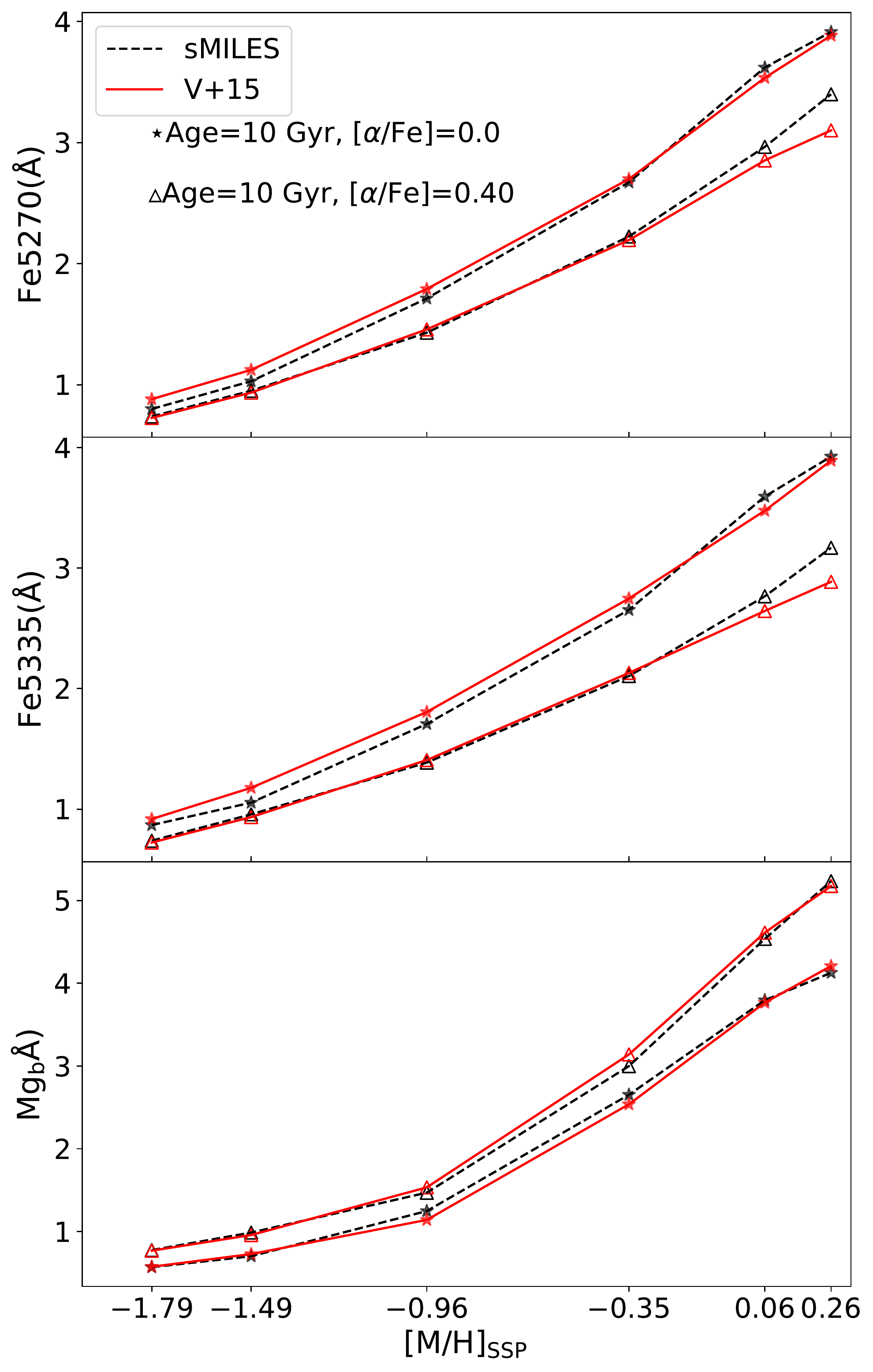}
     \caption{sMILES and \citetalias{Vaz2015} SSP predictions of Fe5270 (Top panel), Fe5333 (Middle panel) and Mg$_{\textrm{b}}$ (Bottom panel) index variations with changing [M/H]$_{\textrm{SSP}}$. The SSPs are 10 Gyr old, with a Universal Kroupa IMF. Index values are measured at MILES FWHM (2.5{\AA}) resolution.}
 \label{sMILES_Vaz_SSP_Metal}
\end{figurenew*}
\section{[MgFe] and [MgFe]' Index Comparisons Between sMILES and Vazdekis et al. (2015) SSP Models}
To supplement Section 3.2.5; Table 3 and Figure 11 of Knowles et al. (2023), we compare the sMILES SSP predictions of [MgFe] and [MgFe]' index changes with [$\alpha$/Fe] variations to those previously calculated with \citetalias{Vaz2015} models, for [$\alpha$/Fe]=0.0 and 0.4 populations, for the same age and metallicity ranged tested in the main article. [MgFe] and [MgFe]' are defined in \cite{Gonzalez93} and \cite{Thomas03Mod} respectively.

In Figures \hyperlink{sMILES_vs_Vaz_MgFe}{S5} and \hyperlink{sMILES_vs_Vaz_MgFe_dash}{S6}
we show the differences in [MgFe] and [MgFe]' indices between sMILES and \citetalias{Vaz2015} models for three ages and various metallicities. Calculated here are the gradients (e.g. how much the index changes with changing [$\alpha$/Fe]), of the line between [$\alpha$/Fe]=0 and 0.4 points in units of {\AA} dex$^{-1}$. As discussed in Knowles et al. (2023), both models generally agree at 7.5 and 14 Gyr old SSPs in absolute [MgFe] and [MgFe]' strengths and in changes of values with metallicity and [$\alpha$/Fe] abundances, with similar gradients found. Differences are found with the younger 2 Gyr old SSP models, with sMILES models predicting a larger change in both indices with increasing [$\alpha$/Fe], particularly for the highest metallicities. These differences originate from both larger Mg$_{\textrm{b}}$ line strength changes and smaller Fe5270 and Fe5335 line strength changes for an increase of [$\alpha$/Fe] in sMILES models, compared to \citetalias{Vaz2015} models (see Table 3 of Knowles et al. 2023).

However, as we note in Knowles et al. (2023), sMILES and \citetalias{Vaz2015} SSP models deviate the most in young, high metallicity and high [$\alpha$/Fe] stellar populations.

We investigate these trends further, for the wider range of [$\alpha$/Fe] values available for sMILES SSPs, in Figure 11 of Knowles et al. (2023), where we uncover non-monotonic behaviour.

\begin{figurenew}[ht]
\centering
\begin{minipage}[b]{0.475\linewidth}
\includegraphics[width=\linewidth, angle=0]{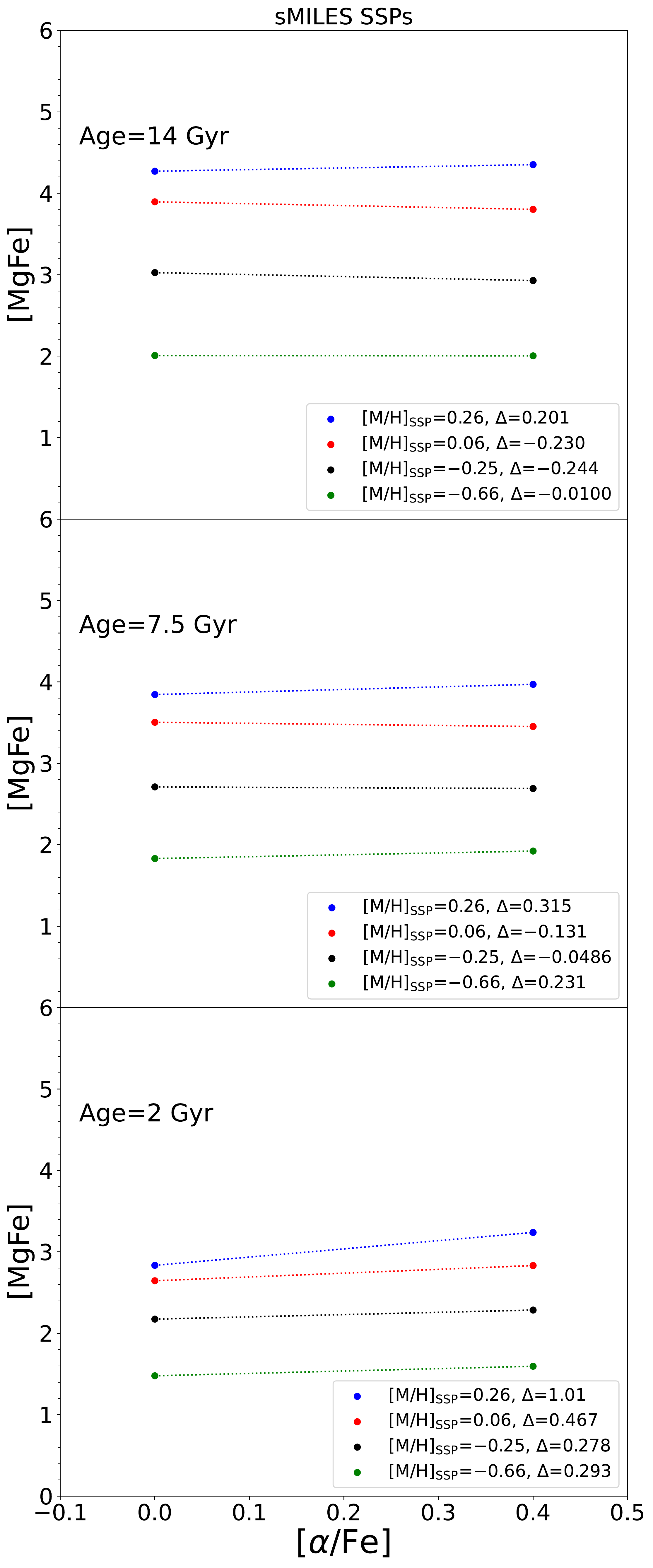}
\end{minipage}
\qquad
\begin{minipage}[b]{0.475\linewidth}
\includegraphics[width=\linewidth, angle=0]{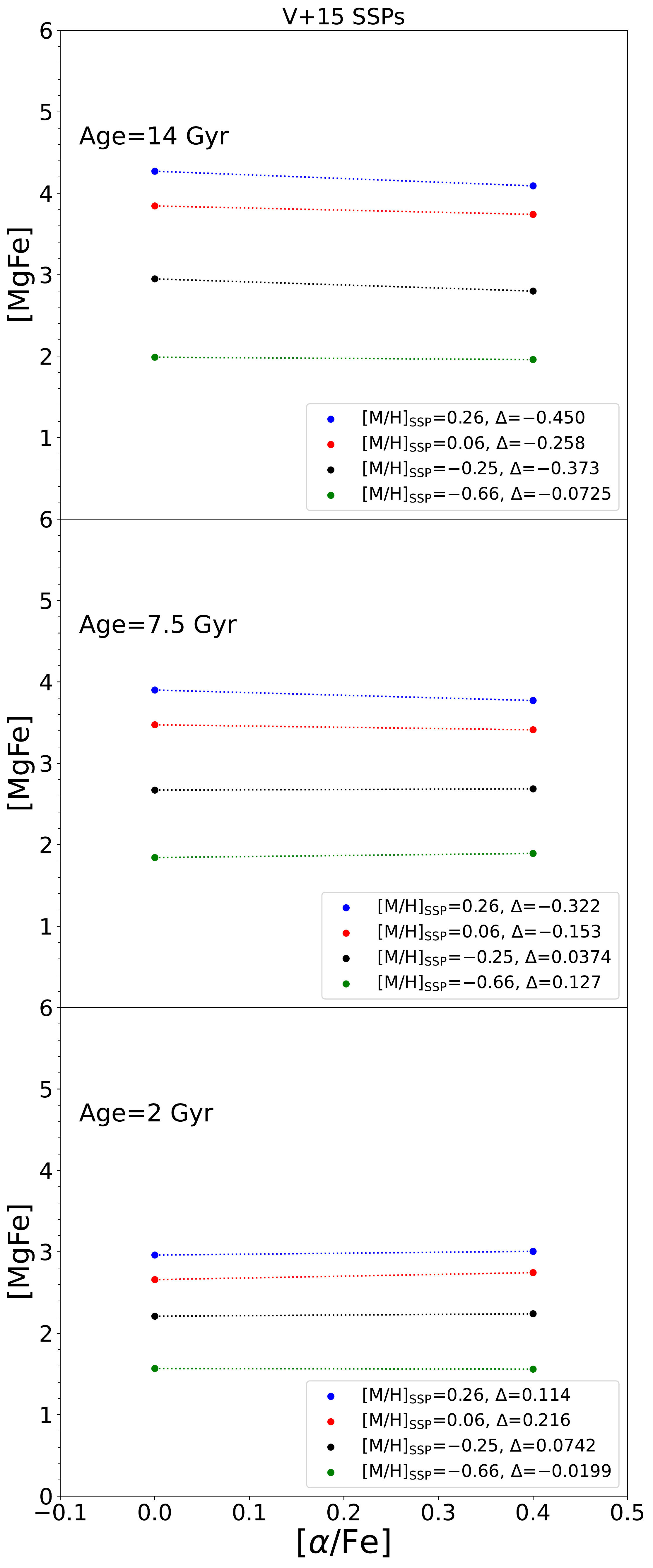}
\end{minipage}
\caption{Difference in [MgFe] strength between sMILES and \citetalias{Vaz2015} SSP models for 3 ages and 4 metallicities at [$\alpha$/Fe]=0.0 and 0.4. Left and right panels show the [MgFe] values of sMILES and \citetalias{Vaz2015} SSP models, respectively. The gradient of the line between [MgFe] values of [$\alpha$/Fe]=0.0 and 0.4 models is given, in units of {\AA} dex$^{-1}$. Index values are measured at MILES FWHM (2.5{\AA}) resolution.}
\label{sMILES_vs_Vaz_MgFe}
\end{figurenew}

\begin{figurenew}[ht]
\centering
\begin{minipage}[b]{0.48\linewidth}
\includegraphics[width=\linewidth, angle=0]{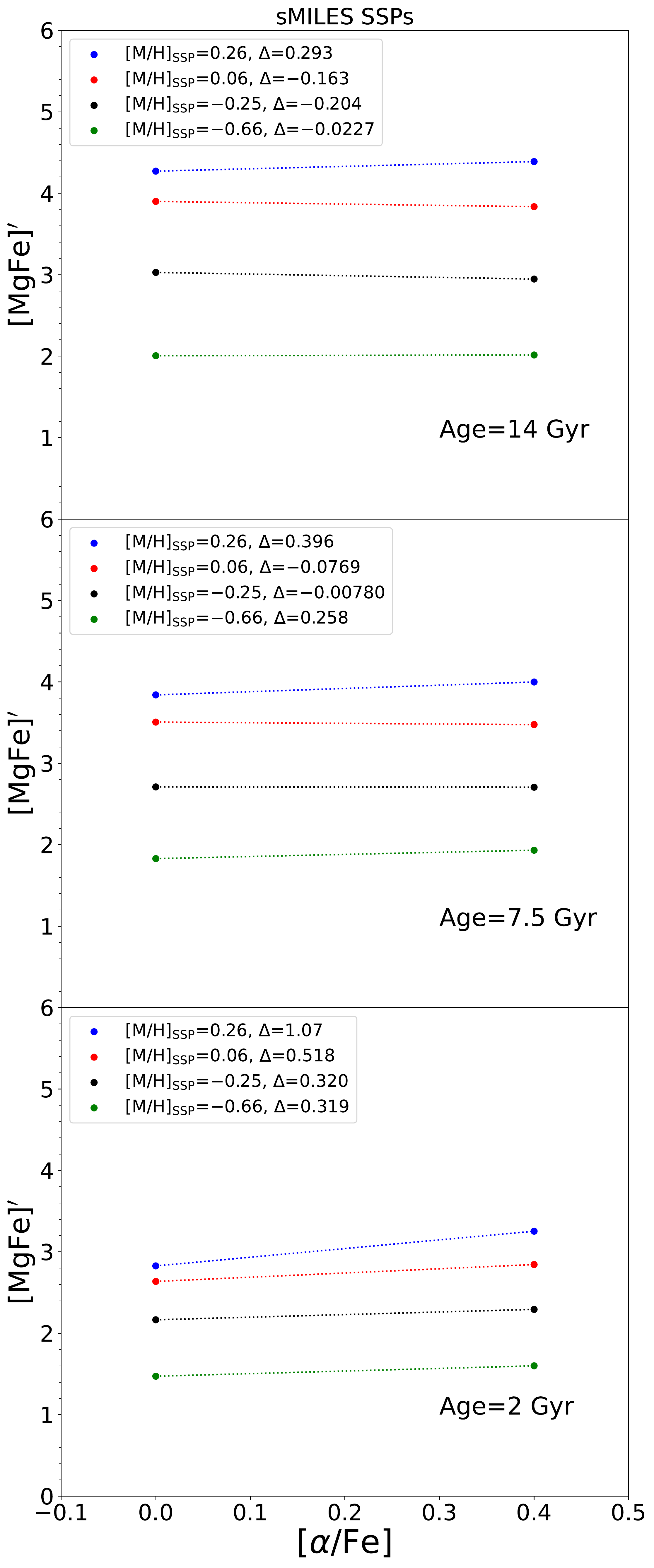}
\end{minipage}
\quad
\begin{minipage}[b]{0.48\linewidth}
\includegraphics[width=\linewidth, angle=0]{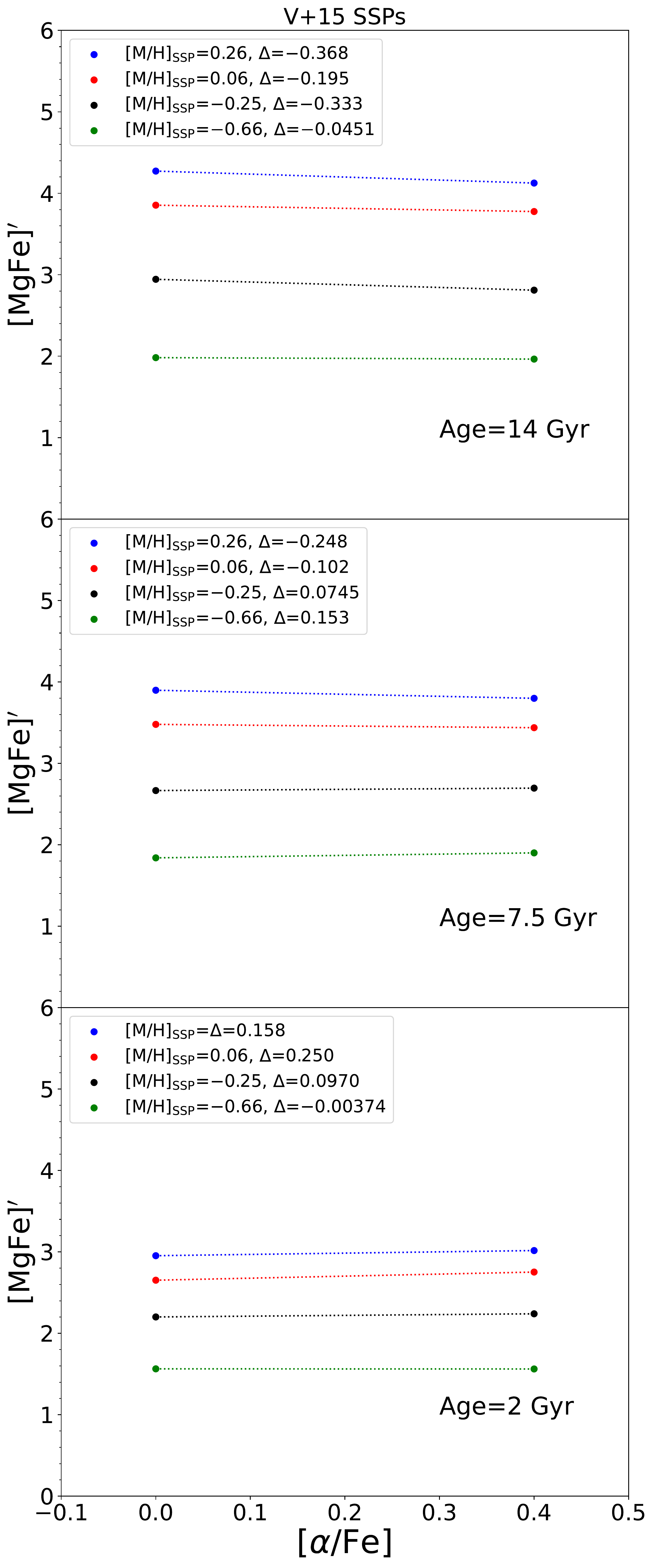}
\end{minipage}
 \centering
 \caption{Difference in [MgFe]' between sMILES and \citetalias{Vaz2015} SSP models for 3 ages and 4 metallicity bins. The left and right panels show the [MgFe]' values of sMILES and \citetalias{Vaz2015} SSP models, respectively. The gradient of the line between [MgFe]' values of [$\alpha$/Fe]=0.0 and 0.4 models is given, in units of {\AA} dex$^{-1}$. Index values are measured at MILES FWHM (2.5{\AA}) resolution.}
 \label{sMILES_vs_Vaz_MgFedash}
\end{figurenew}

\clearpage
\bibliography{Supplementary_Material_sMILES_SSP_Paper.bib}
\bibliographystyle{agsm}